\documentclass[a4paper,11pt]{article}%
\usepackage{amssymb}
\usepackage{graphicx}
\usepackage{amsbsy}
\usepackage{amsmath}
\usepackage{cite}
\usepackage{slashed}
\usepackage{amsfonts}
\usepackage{hyperref}
\hypersetup{
colorlinks   = true,
linkcolor    = blue,
citecolor    = blue,
}

\begin{document}
\thispagestyle{empty} \setcounter{page}{0} \begin{flushright} June 2025\\
\end{flushright}

\vskip3.4 true cm

\begin{center}
{\huge Dark higher-form portals and duality}\\[1.9cm]

\textsc{Cypris Plantier}$^{1}$\textsc{, Christopher Smith}$^{2}$\vspace
{0.5cm}\\[9pt]\smallskip{\small \textsl{\textit{Laboratoire de Physique
Subatomique et de Cosmologie, }}}\linebreak%
{\small \textsl{\textit{Universit\'{e} Grenoble-Alpes, CNRS/IN2P3, Grenoble
INP, 38000 Grenoble, France}.}} \\[1.9cm]\textbf{Abstract}\smallskip
\end{center}

\begin{quote}
\noindent
Light scalar or vector particles are among the most studied dark matter candidates. Yet, those are always described as scalar or vector fields. In this paper, we explore instead the embedding of the scalar particle in an antisymmetric rank-three tensor field, and the dark photon into an antisymmetric rank-two tensor field (a so-called Kalb-Ramond field), and construct minimal bases of effective interactions with Standard Model fields. Then, keeping phenomenological applications as our main objective, a number of theoretical aspects are clarified, in particular related to the impact of existing dualities among the corresponding free theories, and concerning their Stueckelberg representations. Besides, for the rank-two field, we present for the first time its full propagator, accounting for the possible presence of a pseudoscalar mass term. Thanks to these results, and with their different kinematics, gauge-invariant limits, and Lorentz properties, we show that these higher-form fields provide genuine alternative frameworks, with different couplings and expected signatures at low-energy or at colliders.

\let\thefootnote\relax\footnotetext{\newline$^{1}\;$%
cplantier@lpsc.in2p3.fr\newline$^{2}$~chsmith@lpsc.in2p3.fr}
\end{quote}

\newpage

\setcounter{tocdepth}{2}%

\tableofcontents

\newpage

\section{Introduction}

Cosmological and astrophysical evidences for dark matter (DM) have been piling up for a long time now, but despite considerable efforts, its true nature still eludes us. The space of possibilities is \textit{a priori} vast, but has been for quite some time dominated by the so-called weakly-interacting massive particle (WIMP) paradigm, not least thanks to its natural presence in supersymmetric versions of the Standard Model~\cite{Schumann:2019eaa}. The absence of supersymmetric signals at the LHC has weakened this prime candidate, reopening the space of possibilities, but with to some extent the axion taking its place (for recent reviews, see e.g. Refs.~\cite{DiLuzio:2020wdo,Smith:2024uer}). Still, many other light particles, sufficiently long-lived and weakly coupled with normal matter, could make up a sizeable fraction of the observed DM relic abundance. In addition, that particle could be accompanied by others, shorter-lived and of various spins, altogether populating a whole dark sector~\cite{Arbey:2021gdg}. 

The present paper will concentrate on the simple scalar and vector DM candidates, the latter often referred to as a dark photon~\cite{Holdom:1985ag,Fabbrichesi:2020wbt}. However, and contrary to other works, these states will not be introduced via scalar and vector fields,  but will be embedded into so called higher form fields. Those are antisymmetric tensor fields of higher ranks, and in particular, our goal is to use the antisymmetric rank-two tensor $B_{\mu\nu}$ for the dark photon, and the antisymmetric rank-three tensor $C_{\mu\nu\rho}$ for the dark scalar. Theoretically, these fields have been known for a long time, especially in the context of string theory (see e.g. Refs.~\cite{Polchinski:1998rr}), but to our knowledge, they have never been explored in details as true embedding for those dark matter candidates. The only exception is the string theory version of the axion (for reviews, see e.g.\cite{Polchinski:1998rr, Svrcek:2006yi}), originating from a massless $B_{\mu\nu}$, that is most often called a Kalb-Ramond field~\cite{Kalb:1974yc,Lund:1976ze} but sometimes also referred to as the notoph~\cite{Ogievetsky:1966eiu}. Antisymmetric rank-three fields, for their part, have to our knowledge first been described in Refs.~\cite{Curtright:1980yj, Curtright:1980yk}.

There are two main phenomenological motivations to go to the hassle of using higher form fields. First, though not immediately apparent, the massive $B_{\mu\nu}$ and $C_{\mu\nu\rho}$ fields do indeed have the right number of physical degrees of freedom to match that of a massive vector field $A_{\mu}$~\cite{Kemmer:1960,Aurilia:1981xg, Ogievetsky:1966eiu} and a massive scalar field $\phi$~\cite{Aurilia:1969bg, Ogievetsky:1966eiu}, respectively. However, from a Lorentz invariant point of view, it is clear that having a different number of indices changes the way in which those fields can couple to SM matter fields. One immediate question is then to identify which operators exist, and among them, which are of the least mass dimension, the so-called portals.

A second motivation is that we do expect very different scaling behaviors for these portals. Let us take the dark photon to illustrate this point. For a massive vector field, one can always understand its mass as coming from an auxiliary scalar field, in the so-called Stueckelberg construction~\cite{Stueckelberg:1938hvi,Ruegg:2003ps}:
\begin{equation}
A_{\mu} \rightarrow A_{\mu}-\frac{1}{m_V} \partial_{\mu}\phi \ .
\label{Intro1}
\end{equation}
This means that whenever the dark photon is coupled to a nonconserved current, it is its longitudinal degree of freedom represented by $\phi$ that dominates when the typical energy of the process is large compared to the dark photon mass $m_V$, giving it a somewhat axionlike behavior. Now, we will see that the exact opposite happens when the dark photon is introduced via a rank two tensor. Its Stueckelberg construction takes the form
\begin{equation}
B_{\mu\nu} \rightarrow B_{\mu\nu}-\frac{1}{m_V} (\partial_{\mu}A_{\nu}-\partial_{\nu}A_{\mu}) \ ,
\label{Intro2}
\end{equation}
where $A_{\mu}$ represents the transverse polarization states. This reflects the fact that a massless vector is transverse, so it needs to receive one scalar longitudinal degree of freedom to be massive. This is the essence of the Higgs mechanism. On the contrary, a massless $B_{\mu\nu}$ is essentially an axionlike scalar field~\cite{Cremmer:1973mg, Kalb:1974yc}, so it needs to be given a whole transverse vector field to become massive, and those are the modes that get enhanced for small masses. From this, we do expect a quite different phenomenology for those two embeddings of the dark photon.

There is however one profound feature of these higher-rank tensor fields that complicates this program. It stems from the existence of dualities among higher rank tensor field theories (see e.g. Refs.~\cite{Polchinski:1998rr,Hjelmeland:1997eg}). Those are of three types: algebraic, massless, and massive. The first comes from properties of differential forms, while the last two are generalizations of the well-known electromagnetic duality. For instance, the massive $A$ and $B$ fields of Eqs.~(\ref{Intro1}) and~(\ref{Intro2}) are related under~\cite{Takahashi:1970ev, Smailagic:2001ch}
\begin{equation}
B_{\mu\nu} \rightarrow -\frac{1}{2! m} \epsilon_{\mu\nu\rho\sigma}F^{A,\rho\sigma} \ ,
\label{Intro3}
\end{equation}
with $F^A_{\mu\nu} = \partial_{\mu}A_{\nu}-\partial_{\nu}A_{\mu}$, and the massive rank-three form field is related to the scalar field via~\cite{Curtright:1980yj}:
\begin{equation}
C_{\mu\nu\rho} \rightarrow -\frac{1}{m} \epsilon_{\mu\nu\rho\sigma}F^{\phi,\sigma} \ ,
\label{Intro4}
\end{equation}
with $F^{\phi}_{\mu} = \partial_{\mu}\phi$. Such dualities may explain why these fields have not received much phenomenological attention. Yet, there are three important issues to be addressed in an effective framework:
\begin{itemize}
\item First, let us stress that strictly speaking, dualities are valid for free fields only. Typically, they exchange equation of motion and Bianchi identity, so the former better not involve any external current. In practice, duality transformations are still possible in the presence of interactions, but their interpretation changes in that they relate theories that are no longer dynamically identical. 
\item A second caveat is the occurrence of mass scales in Eqs.~(\ref{Intro3}) and~(\ref{Intro4}). Once these dualities are applied to interacting theories, they mix up the effective operators arising at different orders, and what is a portal in one picture is in general no longer so once dualized.
\item A third point is specific to the $B$ field, for which besides the normal mass term $m^2 B_{\mu\nu}B^{\mu\nu}$, there can be a pseudoscalar term $\tilde{m}^2 B_{\mu\nu}\tilde{B}^{\mu\nu}$ where $\tilde{B}_{\mu\nu}=\epsilon_{\mu\nu\rho\sigma}B^{\rho\sigma}/2$. Under the duality transformation of Eqs.~(\ref{Intro3}), this would seem to be equivalent to an irrelevant topological term $\theta F^A_{\mu\nu}\tilde{F}^{A,\mu\nu}$, but we will see this is not the case. Instead, the $\tilde{m}$ term alters even the dynamics of a free $B$ field, changing its polarization states. 
\item On a practical level, though such dualities are very well known (see e.g. Ref.~\cite{Hjelmeland:1997eg,Barbosa:2022zfm} for some recent accounts), they appear scattered in the literature, and are often discussed in abstract field theoretic terms. Indeed, the natural representation of higher rank tensor fields is that of higher rank differential forms. Though it is to some extent necessary to adopt that language, one of our goals will be to review all these aspects and express them back in a form suitable for phenomenological applications. 
\end{itemize}
At the end of the day, we will see that dualities do provide useful dynamical information, but do not make the higher-form embeddings of the dark scalar and photon trivially equivalent to the standard ones. Those turn out to be true alternative frameworks. 

The paper is organized as follows. The first section is intended as an introduction to higher form fields, their action and equations of motion, and their degrees of freedom. The propagators in the massive and massless case are also derived. The only original result in that Section is the non-perturbative treatment of the dual mass term for the rank-two field, leading to a more general polarization sum for these states. In Sec. 2, the effective interactions with all the SM particles are derived for tensor fields of rank between zero and four, with up to two external dark states. We include operators up to rather high dimensions there, to explore the properties of these bases. Then, in Secs. 3, 4, and 5 are discussed the algebraic, massless, and massive dualities, respectively, with emphasis on their impact on the scaling of effective interactions. This is then used in the phenomenological analysis of Sec. 6, in conjunction with the generalized Stueckelberg procedure (which we describe in detail). Finally, our results are summarized in the Conclusion.

\section{Abelian $p$-form fields\label{definitions}}

Higher form gauge fields are particular tensor generalizations of the usual
electromagnetic vector field. To understand their particularities, let us
start by recalling how the vector field arises in QED. First, the gauge field
$A_{\mu}$ is introduced as a connection. It serves to define covariant
derivatives, $\partial_{\mu}\rightarrow\mathcal{D}_{\mu}=\partial_{\mu
}-ieA_{\mu}$ for a field of charge $e$, thereby allowing one to realize the $U(1)$
symmetry locally. Its kinetic term then uses the curvature $[\mathcal{D}_{\mu
},\mathcal{D}_{\nu}]=ieF_{\mu\nu}$, with $F_{\mu\nu}=\partial_{\mu}A_{\nu
}-\partial_{\nu}A_{\mu}$. From this interpretation, it is natural to construct
the Wilson line~\cite{Wilson:1974sk}, which integrates the connection along a path starting at $x$
and going to $y$:%
\begin{equation}
U(x,y)=\exp\left(  -i\int_{P}dz^{\mu}A_{\mu}(z)\right)  \ .
\end{equation}
The quantity $U(x,y)$ is such that $\phi(x)$ and $U(x,y)\phi(y)$ transform
identically. If under the gauge
transformation $A_{\mu}(x)\rightarrow A_{\mu}(x)+\partial_{\mu}\Lambda(x)$, the field undergoes $\phi(x)\rightarrow\exp(-i\Lambda(x))\phi(x)$,
then $U(x,y)\rightarrow\exp(-i\Lambda(x))U(x,y)\exp(i\Lambda(y))$. Though $U(x,y)$ is not
gauge invariant, it becomes so if the path closes into a Wilson loop. In that
case, using Stokes theorem, it is expressible in terms of the flux of the
field strength through the surface enclosed by the loop,%
\begin{equation}
U(x,x)=\exp\left(  -i\int_{\partial\Sigma}dz^{\mu}A_{\mu}(z)\right)
=\exp\left(  -\frac{i}{2}\int_{\Sigma}dn^{\mu\nu}F_{\mu\nu}(z)\right)  \ .
\end{equation}

The idea of higher $p$-form gauge fields is to generalize the Wilson
construction to higher dimensions, to Wilson $p$-dimensional loops enclosing
$p+1$ dimensional surfaces. One peculiarity in this case is that only Abelian
fields can be constructed once $p>1$. Naively, this stems from the liberty in
higher dimensions to move symmetry charge operators past one another, so
that they end up commuting. Those constructions were first encountered in the
context of string theory, with the $p=2$ Kalb-Ramond field as a prototype~\cite{Kalb:1974yc,Lund:1976ze}.
Nowadays, there is a lot of renewed interest in these higher form symmetries,
whether global or local (for reviews, see e.g. Refs.~\cite{Schafer-Nameki:2023jdn,Bhardwaj:2023kri}).

Mathematically, if we want to integrate $p$-dimensional gauge connections
along $p$-dimensional loops, they should be represented by differential $p$-forms. Specifically, a generic $p$-form gauge field corresponds to an
antisymmetric tensor field with $p$ Lorentz indices $A_{\mu_{1}...\mu_{p}}$,
that is
\begin{equation}
A=\frac{1}{p!}A_{\mu_{1}...\mu_{p}}dx^{\mu_{1}}\wedge...\wedge dx^{\mu_{p}%
}\ ,\label{defField}%
\end{equation}
with $C_{p}^{n}=n!/p!(n-p)!$ degrees of freedom (DoF) in $n$ dimensions, and
$p\leqslant n$. This is the natural language to deal with these objects (a
short summary of the main definitions is in the Appendix). Though it is not
compulsory, we will often adopt some aspects of that language to streamline
the notations and calculations. In particular, many developments can be done
keeping $p$ arbitrary instead of painstakingly deriving the results for each
value of $p$. For example, the field strength is the exterior derivative of
the gauge field, i.e., the $p+1$-form $F=dA$. In components, the expression is
less simple%
\begin{equation}
F=dA\Leftrightarrow F_{\mu_{1}...\mu_{p+1}}=(p+1)\partial_{\lbrack\mu_{1}%
}A_{\mu_{2}...\mu_{p+1}]}\ ,\label{FStrength}%
\end{equation}
with the convention that $[...]$ represents the normalized antisymmetrization.
We also immediately get the Bianchi identities $dF=0$ from the fact that
$d^{2}=0$, which corresponds to%
\begin{equation}
\partial_{\lbrack\mu_{1}}F_{\mu_{2}...\mu_{p+2}]}=0\ .
\end{equation}
Nevertheless, in an attempt to provide results of practical phenomenological
use, we will as much as possible fall back to the usual tensorial notation at
important steps. In particular, the naming conventions we shall use are
\begin{subequations}
\label{FieldStrength}%
\begin{align}
p &  =0:\phi\ ,\ F_{\mu}^{\phi}=\partial_{\mu}\phi\ ,\\
p &  =1:A=A_{\mu}dx^{\mu},\ \ F_{\mu\nu}^{A}=\partial_{\mu}A_{\nu}%
-\partial_{\nu}A_{\mu}\ ,\\
p &  =2:B=\frac{1}{2}B_{\mu\nu}dx^{\mu}\wedge dx^{\nu},\ \ F_{\mu\nu\rho}%
^{B}=\partial_{\mu}B_{\nu\rho}+\partial_{\nu}B_{\rho\mu}+\partial_{\rho}%
B_{\mu\nu}\ ,\\
p &  =3:C=\frac{1}{3!}C_{\mu\nu\rho}dx^{\mu}\wedge dx^{\nu}\wedge dx^{\rho
}\ ,\ F_{\mu\nu\rho\sigma}^{C}=\partial_{\mu}C_{\nu\rho\sigma}+\partial_{\nu
}C_{\rho\mu\sigma}+\partial_{\rho}C_{\mu\nu\sigma}+\partial_{\sigma}C_{\nu
\mu\rho}\ ,\\
p &  =4:D=\frac{1}{4!}D_{\mu\nu\rho\sigma}dx^{\mu}\wedge dx^{\nu}\wedge
dx^{\rho}\wedge dx^{\sigma}\ ,\ F_{\mu\nu\rho\sigma\lambda}^{D}=0\ .
\end{align}

Of note is the fact that we will also consider massive $p$-form fields, for
which there is no gauge symmetry. Still, as long as they are represented by
antisymmetric $p$-index tensors, they can be characterized in terms of
differential forms. Said differently, in the present work, it is understood
that free massive $p$-form fields would become $p$-form gauge fields if the
mass term is removed. Specifically, the action for a massive $p$-form field
is,
\end{subequations}
\begin{equation}
\mathcal{S}_{\text{\textrm{p-form}}}=(-1)^{p}\int\frac{1}{2}F\wedge\star
F-\frac{1}{2}m^{2}A\wedge\star A+A\wedge\star j\ ,\label{Action1}%
\end{equation}
where the current $j$ is also a $p$-form. The $\wedge$ operator is the wedge
product, and $\star$ stands for the Hodge dual (see Appendix \ref{sec:appendix}). The reason for
the $(-1)^{p}$ factor comes from our metric signature of $(+1,-1,-1,-1)$, as
will become clear below. In terms of components, this action corresponds to
the Lagrangian%
\begin{equation}
\mathcal{L}_{\text{\textrm{p-form}}}=\frac{(-1)^{p}}{p!}\left(  \frac{1}%
{2}\frac{1}{p+1}F_{\mu_{1}...\mu_{p+1}}F^{\mu_{1}...\mu_{p+1}}-\frac{1}%
{2}m^{2}A_{\mu_{1}...\mu_{p}}A^{\mu_{1}...\mu_{p}}+A_{\mu_{1}...\mu_{p}}%
J^{\mu_{1}...\mu_{p}}\right)  \ .\label{Action2}%
\end{equation}
In both cases, the action for the massless case is simply obtained by setting
$m=0$, while the definitions of $A$ and $F=dA$ stay identical. From this
action, imposing that it is stationary against small variations of the field, the equation of motion (EoM) is found to be:%
\begin{equation}
-\star d\star F+m^{2}A=j\ \Leftrightarrow\partial^{\alpha}F_{\alpha\mu
_{1}...\mu_{p}}+m^{2}A_{\mu_{1}...\mu_{p}}=J_{\mu_{1}...\mu_{p}}\ .
\end{equation}
The similarity is evident with the usual Proca equation describing a massive
vector field, or the inhomogeneous Maxwell equations when $m=0$. To further
explore the consequences, we need to discuss separately the massive and
massless case.

\begin{table}[t]
\centering$%
\begin{tabular}
[c]{c|ccc|cc|cc}\hline
$n=4$ & \multicolumn{3}{|c|}{$\text{Degree of freedom}$} &
\multicolumn{2}{|c|}{Gauge freedom} & Massive & Massless\\
& $\text{Initial}$ & T$\text{emporal}$ & S$\text{patial}$ & Total & Spatial &
& \\\hline
$p$ & $C_{p}^{n}$ & $C_{p-1}^{n-1}$ & $C_{p}^{n-1}$ & $C_{p-1}^{n-1}$ &
$C_{p-1}^{n-2}$ & $C_{p}^{n-1}$ & $C_{p}^{n-2}$\\\hline
$0$ & $1$ & $0$ & $1$ & $0$ & $0$ & $1$ & $1$\\
$1$ & $4$ & $1$ & $3$ & $1$ & $1$ & $3$ & $2$\\
$2$ & $6$ & $3$ & $3$ & $3$ & $2$ & $3$ & $1$\\
$3$ & $4$ & $3$ & $1$ & $3$ & $1$ & $1$ & $0$\\
$4$ & $1$ & $1$ & $0$ & $1$ & $0$ & $0$ & $0$\\\hline
\end{tabular}
\ $\caption{Decomposition of the number of degrees of freedom for the higher form fields in four dimensions. Notice in particular that in the massive case, the 0- and 3-form fields both have one physical DoF, while the 1- and 2-form fields have 3. In the massless case, the 0- and 2-form fields carry one physical DoF. These are the somewhat coincidental correspondences that will be formalized via dualities later on in Secs.~\ref{masslessD} and~\ref{massiveD}.}%
\label{TableDoF}%
\end{table}

\subsection{Massive fields}

When the field is massive, taking an additional derivative shows that
$m^{2}d\star A=d\star j$ since $d^{2}\star F=0$. In other words, $d\star A=0$
for a free field or when the current is conserved, which in components means
that the Lorenz condition $\partial^{\alpha}A_{\alpha\mu_{1}...\mu_{p-1}}=0$
has to be fulfilled. Under this condition, the EoM of the $p$-form field takes
a simpler form. Plugging in $F=dA$,
\begin{equation}
j=-\star d\star dA+m^{2}A=(-\Delta+m^{2})A+d\star d\star A=(-\Delta
+m^{2})A\ ,\label{EoMA}%
\end{equation}
where the differential Laplacian is given by $\Delta=\star d\star d+d\star
d\star$, and in components matches the usual d'Alembertian $\Delta=-\square$
in flat Minkowski space. Despite the heavy use of the differential machinery,
there is nothing special here, and the same result can be found starting from
Eq.~(\ref{Action2}) and integrating by part:%
\begin{equation}
\mathcal{L}_{\text{\textrm{p-form}}}=\frac{(-1)^{p}}{p!}\left(  -\frac{1}%
{2}A_{\mu_{1}...\mu_{p}}(\square+m^{2})A^{\mu_{1}...\mu_{p}}-\frac{p}%
{2}\partial^{\alpha}A_{\alpha\mu_{2}...\mu_{p}}\partial_{\beta}A^{\beta\mu
_{2}...\mu_{p}}+A_{\mu_{1}...\mu_{p}}J^{\mu_{1}...\mu_{p}}\right)
\ ,\label{Action3}%
\end{equation}
where the middle term drops out upon imposing $\partial^{\alpha}A_{\alpha
\mu_{1}...\mu_{p-1}}=0$. The EoM of Eq.~(\ref{EoMA}) is directly obtained by
varying with respect to $A_{\mu_{1}...\mu_{p}}$.

Let us count the number of physical DoF (for an early derivation in four dimensions, see Ref.~\cite{Aurilia:1981xg}). Starting from the $C_{p}%
^{n}=n!/p!(n-p)!$ DoF encoded into the totally antisymmetric tensor
$A_{\mu_{1}...\mu_{p}}$, we can immediately remove all those of the form
$A_{0i_{1}...i_{p-1}}$, where the $i$'s stand for spatial indices. Indeed,
from the definition of the field strength in Eq.~(\ref{FStrength}), it is
clear that only the purely spatial components have time derivatives. In the
Heisenberg picture, the $C_{p-1}^{n-1}$ conjugate momenta of $A_{0i_{1}%
...i_{p-1}}$ vanish, and these $C_{p-1}^{n-1}$ temporal components are
nondynamical. This is consistent since the Lorenz condition, which is built
in the EoM, amounts to $C_{p-1}^{n-1}$ constraints, which is sufficient to fix the
$C_{p-1}^{n-1}$ temporal components $A^{0i_{1}...i_{p-1}}$, leaving no
further constraint on the remaining spatial components. All in all, this
leaves
\begin{equation}
C_{p}^{n}-C_{p-1}^{n-1}=C_{p}^{n-1}\ ,
\end{equation}
propagating DoF, corresponding to the number of ways to pick $p$ spatial
indices in $n$ dimensions, see Table~\ref{TableDoF}.

We can now understand the origin of the $(-1)^{p}$ factor in
Eq.~(\ref{Action1}).\ Going back to $\mathcal{L}_{\text{\textrm{p-form}}}$ in
Eq.~(\ref{Action3}), consider the first term. The $1/p!$ is simply there to
compensate for the summation over permutations of $\mu_{1},...,\mu_{p}$. It
disappears if we identify the $C_{p}^{n}$ fields as $A_{\mu_{1}...\mu_{p}}$
with $\mu_{i}>\mu_{i+1}$. Further, only the purely spatial degrees of freedom
are physical, and bringing down the indices gives $A_{i_{1}...i_{p}}%
=(-1)^{p}A^{i_{1}...i_{p}}$ given our metric $(+1,-1,-1,-1)$. This cancels
with the $(-1)^{p}$ factor in front, leaving the usual Klein-Gordon kinetic
term for each of the $C_{p}^{n-1}$ spatial DoF, $(-1/2)A_{i_{1}...i_{p}%
}(\square+m^{2})A_{i_{1}...i_{p}}$.

\subsection{Massless fields}

Without the mass term, the theory has the gauge symmetry $A\rightarrow
A+d\Lambda$ with $\Lambda$ a $p-1$-form since $F=dA$ is invariant. This
symmetry is preserved by the $A\wedge\star j$ term provided the current is
conserved since under partial integration, $d\Lambda\wedge\star j\rightarrow
(-1)^{p-1}\Lambda\wedge d\star j$ and $d\star j=0$, which is nothing but
$\partial^{\mu_{1}}J_{\mu_{1}...\mu_{p}}=0$. A peculiarity for higher form gauge
fields is that there are fewer gauge DoF than the $C_{p-1}^{n}$ components of
the $p-1$-form $\Lambda$. Indeed, these gauge parameters themselves have gauge
DoF since $\Lambda\rightarrow\Lambda+d\Lambda^{\prime}$ with $\Lambda^{\prime
}$ a $p-2$ form gives the same gauge transformation $A\rightarrow A+d\Lambda$.
This pattern repeats down to a 0-form gauge invariance. Thus, these
recursive invariances mean that there are actually~\cite{Quevedo:1996uu}%
\begin{equation}
C_{p-1}^{n}-C_{p-2}^{n}+C_{p-3}^{n}-...=C_{p-1}^{n-1}\ ,
\label{gaugefree}%
\end{equation}
gauge DoF for an Abelian $p$-form gauge field. Explicitly, these invariances
are
\begin{subequations}
\label{FieldsGauge}%
\begin{align}
\phi &  \rightarrow\phi\ ,\\
A_{\mu} &  \rightarrow A_{\mu}+\partial_{\mu}\Lambda\ ,\\
B_{\mu\nu} &  \rightarrow B_{\mu\nu}+\partial_{\mu}\Lambda_{\nu}-\partial
_{\nu}\Lambda_{\mu}\ ,\\
C_{\mu\nu\rho} &  \rightarrow C_{\mu\nu\rho}+\partial_{\mu}\Lambda_{\nu\rho
}+\partial_{\nu}\Lambda_{\rho\mu}+\partial_{\rho}\Lambda_{\mu\nu}\ ,\\
D_{\mu\nu\rho\sigma} &  \rightarrow D_{\mu\nu\rho\sigma}+\partial_{\mu}%
\Lambda_{\nu\rho\sigma}+\partial_{\nu}\Lambda_{\rho\mu\sigma}+\partial_{\rho
}\Lambda_{\mu\nu\sigma}+\partial_{\sigma}\Lambda_{\mu\rho\nu}\ ,
\end{align}
where $\Lambda_{\mu}$ has the same gauge invariance as $A_{\mu}$, $\Lambda
_{\mu}\rightarrow\Lambda_{\mu}+\partial_{\mu}\Lambda^{\prime}$, $\Lambda
_{\mu\nu}$ the same as $B_{\mu\nu}$, and so on. To quantize these fields, we
generalize the QED Lorenz condition and fix the gauge via $d\star A=0$. Notice
that this leaves some residual gauge invariance corresponding to those
$\Lambda$ that verify $d\star d\Lambda=0$, i.e., $-\Delta\Lambda+d\star
d\star\Lambda=0$. But then, one needs to fix e.g. $d\star\Lambda=0$ to deal
with the gauge-for-gauge invariance under $\Lambda\rightarrow\Lambda
+d\Lambda^{\prime}$, and so on. All in all, the residual invariances are
harmonic at each level, as in QED. In practice, we will see later on how the
propagator for these fields is derived by adding the usual gauge-fixing term
to the Lagrangian.

Let us now count the number of DoF. We cannot simply subtract the total gauge
freedom out of the original $C_{p}^{n}$ DoF since, for the same reason as in
the massive case, the $C_{p-1}^{n-1}$ temporal components are nondynamical.
Said differently, some of the gauge DoF are not relevant for
the counting since they remove nondynamical DoF. To disentangle the
constraints from the gauge DoF, a simple strategy is to first adopt the
temporal gauge and set $A_{0i_{1}...i_{p-1}}=0$, thereby removing the
$C_{p-1}^{n-1}$ temporal components. Then, the remaining gauge DoF are simply
those of a $p$ gauge field living in a $n-1$ dimensional space, since it has
spatial indices only. From Eq.~(\ref{gaugefree}), this corresponds to
$C_{p-1}^{n-2}$ gauge DoF. Altogether then, the number of physical DoF is
\end{subequations}
\begin{equation}
C_{p}^{n}-C_{p-1}^{n-1}-C_{p-1}^{n-2}=C_{p}^{n-2}\ .
\end{equation}
Without surprise, this corresponds to the number of ways to pick $p$ spatial
and transverse indices in $n$ dimensions, see Table~\ref{TableDoF}.

\subsection{Propagators}

Besides the number of DoF, it is necessary to know how to sum over the
polarization states to compute decay rates involving higher-form fields. In
the massive case, these polarization sums can be identified with the numerator
of the corresponding field propagators taken on shell. In turn, those are
immediately obtained by inverting the kinetic terms of Eq.~(\ref{Action3}).
The only delicate point here is that one must properly antisymmetrize these
kinetic term to account for the fact that a $p$ field is fully antisymmetric
in its indices.

Let us define the kinetic kernel $\mathcal{K}$ by writing Eq.~(\ref{Action3})
as $(1/2)A_{\mu_{1}...\mu_{p}}\mathcal{K}^{\mu_{1}...\mu_{p},\nu_{1}...\nu
_{p}}A_{\nu_{1}...\nu_{p}}$. Once properly antisymmetrized and in momentum
space, it becomes
\begin{equation}
\mathcal{K}=(-1)^{p}\left(  \frac{1}{p!}\mathcal{I}_{0}(k^{2}-m^{2})-\frac
{1}{(p-1)!}\mathcal{I}_{2}\right)  \ ,
\end{equation}
where the fully antisymmetric invariants are%
\begin{equation}
\mathcal{I}_{0}^{\mu_{1}...\mu_{p},\nu_{1}...\nu_{p}}=\frac{\delta_{\rho
_{1}...\rho_{p}}^{\mu_{1}...\mu_{p}}}{p!}\frac{\delta_{\sigma_{1}...\sigma
_{p}}^{\nu_{1}...\nu_{p}}}{p!}g^{\rho_{1}\sigma_{1}}...g^{\rho_{p}\sigma_{p}%
}\ ,\ \ \ \mathcal{I}_{2}^{\mu_{1}...\mu_{p},\nu_{1}...\nu_{p}}=\frac
{\delta_{\rho_{1}...\rho_{p}}^{\mu_{1}...\mu_{p}}}{p!}\frac{\delta_{\sigma
_{1}...\sigma_{p}}^{\nu_{1}...\nu_{p}}}{p!}k^{\rho_{1}}k^{\sigma_{1}}%
g^{\rho_{2}\sigma_{2}}...g^{\rho_{p}\sigma_{p}}\ . \label{InvariantDef}%
\end{equation}
The first one is nothing but the identity, $(\mathcal{I}_{0})_{\rho_{1}%
...\rho_{p}}^{\mu_{1}...\mu_{p}}=\delta_{\nu_{1}...\nu_{p}}^{\mu_{1}...\mu
_{p}}/p!$. The propagator in momentum space satisfies $\mathcal{KP}%
=\mathcal{I}_{0}$, hence%
\begin{equation}
\mathcal{P}=i\frac{(-1)^{p}p!}{k^{2}-m^{2}}\left(  \mathcal{I}_{0}-\frac
{p}{m^{2}}\mathcal{I}_{2}\right)  \ . \label{Pmassive}%
\end{equation}
To arrive to that form is immediate using the identities $\mathcal{I}%
_{0}\mathcal{I}_{0}=\mathcal{I}_{0}$, $\mathcal{I}_{0}\mathcal{I}%
_{2}=\mathcal{I}_{2}\mathcal{I}_{0}=\mathcal{I}_{2}$, and $p\mathcal{I}%
_{2}\mathcal{I}_{2}=k^{2}\mathcal{I}_{2}$. This last identity ensures that the
numerator is indeed transverse on shell,
\begin{equation}
\left.  k\cdot\left(  \mathcal{I}_{0}-\frac{p}{m^{2}}\mathcal{I}_{2}\right)
\right\vert _{k^{2}=m^{2}}=\left.  \left(  \mathcal{I}_{0}-\frac{p}{m^{2}%
}\mathcal{I}_{2}\right)  \cdot k\right\vert _{k^{2}=m^{2}}=0\ ,
\label{polsumSTD}
\end{equation}
where $k\cdot\mathcal{I}=k_{\mu_{1}}\mathcal{I}^{\mu_{1}...\mu_{p},\nu
_{1}...\nu_{p}}$, $\mathcal{I}\cdot k=k_{\nu_{1}}\mathcal{I}^{\mu_{1}%
...\mu_{p},\nu_{1}...\nu_{p}}$. Orthogonal invariants could be defined (see
e.g. Ref.~\cite{Malta:2025ydq} for those with $p=2$), but this is not essential and
further, we will see in the following that the $\mathcal{I}_{0} $ and
$\mathcal{I}_{2}$ structure do carry dynamical information.

Explicitly, the invariant functions for $p=0$ are $\mathcal{I}_{0}=1$ and
$\mathcal{I}_{2}=0$, as they should be for a scalar field. For $p=1$, we recover
the usual $\mathcal{I}_{0}^{\mu,\nu}=g^{\mu\nu}$ and $\mathcal{I}_{2}^{\mu
,\nu}=k^{\mu}k^{\nu}$. At higher orders, we get for $p=2$ (in agreement with Ref.~\cite{Tiwary:2021cff}):%
\begin{align}
\mathcal{I}_{0}^{\mu\nu,\alpha\beta} &  =\frac{1}{2}g^{\mu\alpha}g^{\nu\beta
}-\frac{1}{2}g^{\mu\beta}g^{\nu\alpha}\ \ ,\ \\
\mathcal{I}_{2}^{\mu\nu,\alpha\beta} &  =\frac{1}{4}g^{\nu\beta}k^{\mu
}k^{\alpha}-\frac{1}{4}g^{\mu\beta}k^{\nu}k^{\alpha}-\frac{1}{4}g^{\nu\alpha
}k^{\mu}k^{\beta}+\frac{1}{4}g^{\mu\alpha}k^{\nu}k^{\beta}\ ,
\end{align}
in terms of which, for $p=3$,%
\begin{align}
\mathcal{I}_{0}^{\mu\nu\rho,\alpha\beta\gamma} &  =\frac{1}{3}\mathcal{I}%
_{0}^{\mu\alpha,\nu\beta}g^{\rho\gamma}-\frac{1}{3}\mathcal{I}_{0}^{\rho
\alpha,\nu\beta}g^{\mu\gamma}-\frac{1}{3}\mathcal{I}_{0}^{\mu\alpha,\rho\beta
}g^{\nu\gamma}\ ,\\
\mathcal{I}_{2}^{\mu\nu\rho,\alpha\beta\gamma} &  =\frac{1}{9}\mathcal{I}%
_{0}^{\nu\beta,\rho\gamma}k^{\mu}k^{\alpha}-\frac{1}{9}\mathcal{I}_{0}%
^{\mu\beta,\rho\gamma}k^{\nu}k^{\alpha}+\frac{1}{9}\mathcal{I}_{0}^{\mu
\alpha,\rho\gamma}k^{\nu}k^{\beta}-\frac{1}{9}\mathcal{I}_{0}^{\nu\alpha
,\rho\gamma}k^{\mu}k^{\beta}+\frac{1}{9}\mathcal{I}_{0}^{\mu\beta,\nu\gamma
}k^{\rho}k^{\alpha}\nonumber\\
&  \ \ -\frac{1}{9}\mathcal{I}_{0}^{\mu\alpha,\nu\gamma}k^{\rho}k^{\beta
}-\frac{1}{9}\mathcal{I}_{0}^{\nu\beta,\rho\alpha}k^{\mu}k^{\gamma}+\frac
{1}{9}\mathcal{I}_{0}^{\mu\beta,\rho\alpha}k^{\nu}k^{\gamma}+\frac{1}%
{9}\mathcal{I}_{0}^{\mu\alpha,\nu\beta}k^{\rho}k^{\gamma}\ .
\end{align}
Finally, for $p=4$, the propagator is a trivial contact term proportional to
$\mathcal{I}_{0}/m^{2}$ since the field strength vanishes, with $(\mathcal{I}%
_{0})_{\nu_{1}...\nu_{4}}^{\mu_{1}...\mu_{4}}=-\epsilon_{\nu_{1}...\nu_{4}%
}\epsilon^{\mu_{1}...\mu_{4}}/4!$.

For massless states, the kinetic kernel is a projector and cannot be inverted.
This situation is well-known in QED: the gauge has to be fixed to quantize the
theory. This can be done by adding to the Lagrangian a term quadratic in the
Lorenz condition:%
\begin{equation}
\mathcal{S}_{\text{\textrm{p-form}}}=(-1)^{p}\int\frac{1}{2}F\wedge\star
F-\frac{1}{2\xi}d\star A\wedge\star d\star A\ ,
\end{equation}
in which case the gauge-dependent propagator becomes%
\begin{equation}
\mathcal{P}=i\frac{(-1)^{p}p!}{k^{2}}\left(  \mathcal{I}_{0}-\left(
1-\xi\right)  \frac{p}{k^{2}}\mathcal{I}_{2}\right)  \ . \label{Pmassless}%
\end{equation}
The cancelation of the $\mathcal{I}_{2}$ term is then ensured by the
conservation of the currents to which the propagator couples.

Finally, there are a few peculiar one- and two-point vertices specific to four
dimensions: $dA\wedge dA$ for $p=1$, $A\wedge A$, $d\star A\wedge\star dA$ for
$p=2$, and $dA$ for $p=3$. All of them are 4-forms and can thus enter in
the action. Now, the former is the well-known theta term $F_{\mu\nu}^{A}%
\tilde{F}^{A,\mu\nu}$ that can be discarded for a topologically trivial
$U(1)$, while the latter is a pure boundary term that can also be discarded.
This leaves only the $p=2$ terms: the pseudoscalar mass term $\tilde{m}%
^{2}B\wedge B\rightarrow\tilde{m}^{2}B_{\mu\nu}\tilde{B}^{\mu\nu}$ and the
mixed kinetic term $d\star B\wedge\star dB\rightarrow\epsilon_{\mu\nu
\rho\sigma}\partial_{\alpha}B^{\alpha\mu}F^{B,\nu\rho\sigma}$. For simplicity
here, we will keep the standard kinetic term only (except briefly in
Sec.~\ref{AlgDualities}). Let us thus concentrate on $\tilde{m}^{2}B_{\mu\nu
}\tilde{B}^{\mu\nu}$. It could be dealt with perturbatively, but since
$\tilde{m}$ could be as large as $m$, a better way to proceed is to
immediately resum all $\tilde{m}$ mass insertions by encoding the $\tilde{m}$
term directly into the $B$ propagator. This requires one to extend the basis of
antisymmetric invariants to include%
\begin{align}
\mathcal{I}_{3}^{\mu_{1}\mu_{2},\nu_{1}\nu_{2}} &  =\frac{1}{2}\epsilon
^{\mu_{1}\mu_{2}\nu_{1}\nu_{2}}\ ,\\
\mathcal{I}_{41}^{\mu_{1}\mu_{2},\nu_{1}\nu_{2}} &  =\frac{1}{4}(k^{\mu_{1}%
}k_{\alpha}\epsilon^{\alpha\mu_{2}\nu_{1}\nu_{2}}-k^{\mu_{2}}k_{\alpha
}\epsilon^{\alpha\mu_{1}\nu_{1}\nu_{2}})\ ,\\
\mathcal{I}_{42}^{\mu_{1}\mu_{2},\nu_{1}\nu_{2}} &  =\frac{1}{4}(k^{\nu_{1}%
}k_{\alpha}\epsilon^{\mu_{1}\mu_{2}\alpha\nu_{2}}-k^{\nu_{2}}k_{\alpha
}\epsilon^{\mu_{1}\mu_{2}\alpha\nu_{1}})\ ,
\end{align}
which obey simple relations with the other ones like e.g. $\mathcal{I}%
_{3}\mathcal{I}_{3}=-\mathcal{I}_{0}$, $\mathcal{I}_{3}\mathcal{I}%
_{2}=-\mathcal{I}_{42}$, $\mathcal{I}_{2}\mathcal{I}_{3}=-\mathcal{I}_{41}$,
$\mathcal{I}_{2}\mathcal{I}_{42}=\mathcal{I}_{41}\mathcal{I}_{2}=0$,
$2\mathcal{I}_{42}\mathcal{I}_{2}=k^{2}\mathcal{I}_{42}$, $2\mathcal{I}%
_{2}\mathcal{I}_{41}=k^{2}\mathcal{I}_{41}$, and so on. Inverting the kinetic
term, we then find%
\begin{equation}
\mathcal{P}^{\text{2-form}}=\frac{2}{k^{2}-m^{2}(1+\tilde{m}^{4}/m^{4}%
)}\left(  \mathcal{I}_{0}-\frac{2}{m^{2}}\mathcal{I}_{2}-\frac{\tilde{m}^{2}%
}{k^{2}-m^{2}}\left(  \mathcal{I}_{3}+\frac{2}{m^{2}}(\mathcal{I}%
_{41}+\mathcal{I}_{42})\right)  \right)  \ ,
\end{equation}
where it is understood that the Lagrangian terms are normalized as $-\tilde
{m}^{2}B_{\mu\nu}B^{\mu\nu}/4+\tilde{m}^{2}B_{\mu\nu}\tilde{B}^{\mu\nu}/4$,
with $\tilde{B}_{\mu\nu}=\epsilon_{\mu\nu\rho\sigma}B^{\rho\sigma}/2$. Notice
that if $m=0$ but $\tilde{m}\neq0$, the kinetic term is not invertible,
so the situation is a bit pathological in that case. Here, taking both mass
terms as nonzero, the true pole mass of the $B$ field becomes\footnote{A similar result
was obtained in Ref.~\cite{Quevedo:1996uu}, but in the context of compact QED with monopole condensates.}%
\begin{equation}
m_{B}^{2}=\frac{m^{4}+\tilde{m}^{4}}{m^{2}}\ .\label{Bmass}%
\end{equation}
Yet, given this mass, the presence of a second pole at $k^{2}=m^{2}$ in
$\mathcal{P}^{\text{2-form}}$ cannot be physical. To see that it is spurious,
one should remember that $\mathcal{P}^{\text{2-form}}$ will always be
sandwiched as $J_{1,\mu_{1}\mu_{2}}\mathcal{P}_{\text{2-form}}^{\mu_{1}%
\mu_{2},\nu_{1}\nu_{2}}J_{2,\nu_{1}\nu_{2}}$ for some vertices $J_{1}$ and
$J_{2}$ that are antisymmetric in their indices. As a result, one can check
that $J_{1}\cdot(\mathcal{I}_{41}+\mathcal{I}_{42})\cdot J_{2}=-(k^{2}%
/2)J_{1}\cdot\mathcal{I}_{3}\cdot J_{2}$, such that effectively, the full
propagator can be taken as%
\begin{equation}
\mathcal{P}^{\text{2-form}}=\frac{2}{k^{2}-(m^{4}+\tilde{m}^{4})/m^{2}}\left(
\mathcal{I}_{0}-\frac{2}{m^{2}}\mathcal{I}_{2}+\frac{\tilde{m}^{2}}{m^{2}%
}\mathcal{I}_{3}\right)  \ ,\label{PropMtilde}%
\end{equation}
which now has its pole at the physical mass $k^{2}=m_{B}^{2}$. The singularity as $m\rightarrow 0$ have the same physical interpretation as that in Eq.~(\ref{polsumSTD}). In both cases, the physics abruptly changes when $m\rightarrow0$ since the kinetic term ceased to be invertible. We will see later on how to rederive these results using duality, along with
the polarization sum in the numerator of the full propagator. At this stage,
one may be surprised to notice that this polarization sum is no longer
transverse on shell.\ However, this should be expected. In the presence of the
$\tilde{m}$ term, the EoM becomes
\begin{equation}
\partial^{\alpha}F_{\alpha\mu_{1}\mu_{2}}^{B}+m^{2}B_{\mu_{1}\mu_{2}}%
-\tilde{m}^{2}\tilde{B}_{\mu_{1}\mu_{2}}=J_{\mu_{1}\mu_{2}}\ .
\end{equation}
Thus, for free fields, the Lorenz condition should read (why a term
proportional to the field strength appears will become obvious in
Sec.~\ref{AlgDualities}):%
\begin{equation}
m^{2}\partial^{\nu}B_{\mu\nu}-\frac{1}{3!}\tilde{m}^{2}\epsilon_{\mu\nu
\rho\sigma}F^{B,\nu\rho\sigma}=0\ ,\label{GenLorenz}%
\end{equation}
and the polarization matrices thus have to satisfy $k_{\mu}(m^{2}%
\mathcal{I}_{0}^{\mu\nu,\rho\sigma}-\tilde{m}^{2}\mathcal{I}_{3}^{\mu\nu
,\rho\sigma})\varepsilon_{\rho\sigma}^{(\lambda)}=0$. One can check that this
is consistent with the sum in the numerator of Eq.~(\ref{PropMtilde}):%
\begin{equation}
\left.  k\cdot(m^{2}\mathcal{I}_{0}-\tilde{m}^{2}\mathcal{I}_{3})\cdot\left(
\mathcal{I}_{0}-\frac{2}{m^{2}}\mathcal{I}_{2}+\frac{\tilde{m}^{2}}{m^{2}%
}\mathcal{I}_{3}\right)  \right\vert _{k^{2}=m_{B}^{2}}=0\ .
\end{equation}

\section{Effective couplings for higher fields}

The goal here is to construct the leading operators coupling $p$-form fields
to SM matter fields, with $p=0,...,4$. Neither the Lorenz condition
$\partial^{\mu}A_{\mu\nu...}=0$ nor the free EoM $\partial^{\mu}F_{\mu\nu
...}=0$ (or $\partial^{2}A_{\mu\nu...}=0$) are imposed. Only the Bianchi
identities for $p=0,1,2$ are enforced. Importantly, no dark gauge symmetry (or shift symmetry for $p=0$) will be imposed here. Instead, invariant and noninvariant operators will be constructed and put in separate classes. This will permit one to compare phenomenological scenarios in which the dark gauge symmetry is totally broken at the low scale, or only broken by the dark state mass term.

Also, operators with three or more $p$ fields will not be constructed. Though one could introduce some quantum number
to allow no fewer than $n$ of them to be produced, for $n$ any given integer, one rarely encounters $n$ larger than $2$ since this would conflict with the kinetic terms. Phenomenologically, producing more
than two dark states is strongly but trivially phase-space suppressed, and
thus less likely to end up as observable. Another argument is that one of our
goals will be to analyze the relationships between these bases, comparing the
situation when a gauge (or shift) symmetry is active or not. For three or more
$p$ fields, the leading symmetric operators would involve three or more field
strengths, and thus be of prohibitively high dimension.

The language of differential form is not well suited to the construction of
all these effective interactions. Instead, we will follow the pedestrian path
of taking products of antisymmetric tensor fields together with SM fields,
contracting all the Lorentz indices to form invariants. Mathematically, this
corresponds to taking various products of interior products of differential
forms with suitably constructed vectors made of combinations of SM fields, but
there appears to be no advantage in adopting that formalism. The same is true
concerning higher form field self-interactions. Generic renormalizable
self-interactions could be constructed from $A\wedge A\wedge\star A$,
$A\wedge\star A\wedge\star(d\star A)$,$\ A\wedge\star A\wedge\star dA$,
$(A\wedge\star A)\wedge\star(A\wedge\star A)$, etc, but this is not
particularly efficient since only some of these combinations correspond to
4-form for a given $p$. It is actually much easier to directly construct
the suitable self-interactions for each $p$. Though neither those
renormalizable couplings nor their extensions to higher orders will concern us
here (see Ref.~\cite{Hell:2021wzm} for a discussion in the $p=1$ and $p=2$ case), for completeness, the full list up to dimension four can easily be written down and is actually quite
limited,
\begin{subequations}
\begin{align}
p &  =0:\phi^{3}\ ,\ \phi^{4}\ ,\\
p &  =1:A_{\mu}A^{\mu}\partial^{\nu}A_{\nu}\ ,\ A_{\mu}A^{\mu}A_{\nu}A^{\nu
}\ ,\\
p &  =2:B_{\mu\nu}B^{\mu\nu}B_{\rho\sigma}B^{\rho\sigma}\ ,\ B_{\mu\nu}%
B^{\mu\nu}B_{\rho\sigma}\tilde{B}^{\rho\sigma}\ ,\ B_{\mu\nu}B^{\mu\rho
}B_{\rho\sigma}B^{\sigma\nu}\ ,\ B_{\mu\nu}B^{\mu\rho}B_{\rho\sigma}\tilde
{B}^{\sigma\nu}\ ,\\
p &  =3:C_{\mu\nu\rho}C^{\mu\nu\rho}\epsilon^{\alpha\beta\gamma\delta
}F_{\alpha\beta\gamma\delta}^{C}\ ,\ C_{\mu\nu\rho}C^{\mu\nu\rho}%
C_{\alpha\beta\gamma}C^{\alpha\beta\gamma}\ ,\\
p &  =4:D_{\mu\nu\rho\sigma}D^{\mu\nu\rho\sigma}D_{\alpha\beta\gamma\delta
}\epsilon^{\alpha\beta\gamma\delta}\ ,\ D_{\mu\nu\rho\sigma}D^{\mu\nu
\rho\sigma}D_{\alpha\beta\gamma\delta}D^{\alpha\beta\gamma\delta}\ .
\end{align}
All these extra operators break the dark gauge symmetry. Their presence or impact could thus be restricted phenomenologically by assuming that this symmetry remains active in the dark sector, except of course for the dark state mass term. Notice also that only $\phi$ and $D_{\mu\nu\rho\sigma}$ can have a dimensionful self-interaction, while none of the others can.

In the following, most of the work will concentrate on fermionic operators,
and for them, various identities are useful. Those are totally standard, but
it is worth collecting them here because they are not that often encountered
in practice. Indeed, a specificity of higher form fields is to have many
indices, and thus the effective operators involve many contractions, including
with the epsilon tensor. Further, since in this paper we are concerned by the
properties of these bases, we will push the construction to effective
operators involving up to two extra derivatives, further extending the Lorentz
index counts. Yet, the properties of the Dirac matrices makes reducing these
operators nearly always possible, drastically reducing the number of
independent operators at each order.

First, we define $2\sigma^{\mu\nu}\equiv i\left[  \gamma^{\mu},\gamma^{\nu
}\right]  $, and reduce any string of more than two Dirac matrices using the
Chisholm identity:%
\end{subequations}
\begin{equation}
\gamma^{\mu}\gamma^{\nu}\gamma^{\rho}=g^{\mu\nu}\gamma^{\rho}-g^{\mu\rho
}\gamma^{\nu}+g^{\nu\rho}\gamma^{\mu}+i\epsilon^{\mu\nu\rho\alpha}%
\gamma_{\alpha}\gamma_{5}\Rightarrow\epsilon_{\mu\nu\rho\sigma}\gamma^{\mu
}\gamma^{\nu}\gamma^{\rho}=-(3!)i\gamma_{\sigma}\gamma_{5}\ .\label{Id1}%
\end{equation}
This also shows that whenever a $\sigma^{\mu\nu}$ appears, it cannot be
accompanied by a epsilon tensor since $2\sigma^{\mu\nu}\gamma_{5}%
=i\epsilon^{\mu\nu\alpha\beta}\sigma_{\alpha\beta}$ implies%
\begin{equation}
\epsilon^{\alpha\beta\gamma\delta}\sigma_{\mu\nu}\psi_{R}=\frac{i}{2}%
\epsilon^{\alpha\beta\gamma\delta}\epsilon_{\mu\nu\rho\sigma}\sigma
^{\rho\sigma}\psi_{R}=-\frac{i}{2}\delta_{\mu\nu\rho\sigma}^{\alpha\beta
\gamma\delta}\sigma^{\rho\sigma}\psi_{R}\ ,\label{Id2}%
\end{equation}
and similarly with $\psi_{L}$. For fermionic fields, we do use the fermion
equation of motion whenever possible\footnote{Covariant derivatives are always meant to include the SM gauge fields only, and not any of the dark fields.}, $i\gamma^{\mu}\mathcal{D}_{\mu}%
\psi_{L,R}\rightarrow m\psi_{R,L}$. This implies in particular%
\begin{equation}
\sigma^{\mu\nu}\mathcal{D}_{\nu}\psi_{R}=-i(g^{\mu\nu}-\gamma^{\mu}\gamma
^{\nu})\mathcal{D}_{\nu}\psi_{R}=-i\mathcal{D}^{\mu}\psi_{R}+m\gamma^{\mu}%
\psi_{L}\ .\label{Id3}%
\end{equation}
Within a spinor contraction, derivative can be assumed to always act on the
right, since this is equivalent to $\overleftrightarrow{\mathcal{D}}$ up to a
total derivative of the whole spinor contraction which can then be made to act
on the other fields by partial integration. By consistency once operators with
two derivatives are included, those with SM field strengths have to be present
since $[\mathcal{D}_{\mu},\mathcal{D}_{\nu}]=ieF_{\mu\nu}$ when acting on a
charged fermion field, leading to identities like%
\begin{equation}
\mathcal{D}_{\mu}\mathcal{D}^{\mu}\psi_{L}=m^{2}\psi_{L}+\frac{i}{2}%
\sigma^{\mu\nu}[\mathcal{D}_{\mu},\mathcal{D}_{\nu}]\psi_{L}=m^{2}\psi
_{L}-\frac{e}{2}\sigma^{\mu\nu}F_{\mu\nu}\psi_{L}\ .
\end{equation}
Finally, additional identities that prove useful are
\begin{subequations}
\label{Id4}%
\begin{align}
\partial_{\mu}(\bar{\psi}_{L}\gamma^{\mu}\psi_{L}) &  =m\bar{\psi}_{L}\psi
_{R}-m\bar{\psi}_{R}\psi_{L}\ ,\\
\partial_{\mu}(\bar{\psi}_{L}\sigma^{\mu\nu}\psi_{R}) &  =i\bar{\psi}%
_{L}\overleftrightarrow{\mathcal{D}}^{\nu}\psi_{R}-m\bar{\psi}_{L}\gamma^{\nu
}\psi_{L}-m\bar{\psi}_{R}\gamma^{\nu}\psi_{R}\ ,\\
\partial_{\mu}(\bar{\psi}\sigma^{\mu\nu}\mathcal{D}^{\rho}\psi) &
=-i\partial^{\nu}(\bar{\psi}\mathcal{D}^{\rho}\psi)+2i\bar{\psi}%
_{L}\mathcal{D}^{\nu}\mathcal{D}^{\rho}\psi_{R}+e\bar{\psi}_{L}\gamma^{\nu
}\gamma^{\mu}\psi_{R}F_{\mu}^{\rho}\nonumber\\
&  \ \ \ -m\bar{\psi}_{R}\gamma^{\nu}\mathcal{D}^{\rho}\psi_{R}-m\bar{\psi
}_{L}\gamma^{\nu}\mathcal{D}^{\rho}\psi_{L}\ .
\end{align}
By convention in the present section, $F$ always denotes SM field strengths,
while those associated with higher form fields receive that field as a
superscript, as $F^{A}$, $F^{B}$, and so on, see Eq.~(\ref{FieldStrength}).

Finally, it should be noted that in the SM before symmetry breaking, the
fermion EoM would involve the Higgs boson and not simply the fermion mass.
Yet, generalizing our bases to that case amounts to add Higgs fields for all
$\psi_{L}\Gamma\psi_{R}$ contractions, plus operators involving derivatives of
the Higgs fields, and that can easily be done separately.

\subsection{0-form field effective operators\label{EffOp0}}

These operators are well-known (see e.g. Ref.~\cite{Kamenik:2011vy}) since a 0-form field is simply a generic neutral scalar field. The leading operators to SM fermion fields are%
\end{subequations}
\begin{equation}%
\begin{tabular}
[c]{ccc}\hline
$d$ & Type & \\\hline
4 & II & $\bar{\psi}_{L}\psi_{R}\phi$\\
5 & I & $(\bar{\psi}_{L}\gamma^{\mu}\psi_{L}F_{\mu}^{\phi})$\\
& II & --\\
6 & I & $\bar{\psi}_{L}\psi_{R}\partial^{\mu}F_{\mu}^{\phi}$\\
& II & --\\
& III & $\bar{\psi}_{L}\sigma_{\mu\nu}\psi_{R}\phi F^{\mu\nu}$\\\hline
\end{tabular}
\ \label{FPhi1}%
\end{equation}
for one dark field, and
\begin{equation}%
\begin{tabular}
[c]{cccc}\hline
$d$ & Type &  & \\\hline
5 & II & $\bar{\psi}_{L}\psi_{R}\phi^{2}$ & \\
6 & II & $\bar{\psi}_{L}\gamma^{\mu}\psi_{L}\phi F_{\mu}^{\phi}$ & \\
7 & I & $\bar{\psi}_{L}\psi_{R}F_{\mu}^{\phi}F^{\phi,\mu}$ & \\
& II & $\bar{\psi}_{L}\psi_{R}\phi\partial^{2}\phi$ & $\bar{\psi}%
_{L}\mathcal{D}^{\mu}\psi_{R}\phi F_{\mu}^{\phi}$\\
& III & $\bar{\psi}_{L}\sigma_{\mu\nu}\psi_{R}\phi^{2}F^{\mu\nu}$ & \\\hline
\end{tabular}
\ \label{FPhi2}%
\end{equation}
for two dark fields. In those Tables, type I and II distinguish
shift-symmetric and non-shift-symmetric operators, while type III involves an
extra photon field. For simplicity, all the operators with $L\leftrightarrow
R$ are understood. Those are simply the Hermitian conjugate operators for
those involving scalar or tensor spinor structures. Remember that $F_{\mu
}^{\phi}=\partial_{\mu}\phi$, and the Bianchi identity holds (so an operator
involving $F^{\mu\nu}\partial_{\mu}F_{\nu}^{\phi}\rightarrow0$ is absent). The
operator $\bar{\psi}_{L}\gamma^{\mu}\psi_{L}F_{\mu}^{\phi}$ is in parentheses
because it is reducible, $\bar{\psi}_{L}\gamma^{\mu}\psi_{L}F_{\mu}^{\phi
}\rightarrow\partial_{\mu}(\bar{\psi}_{L}\gamma^{\mu}\psi_{L})\phi$ using the
Dirac equation for on-shell fermions. Yet, it is important for $\bar{\psi}%
_{L}\gamma^{\mu}\psi_{L}F_{\mu}^{\phi}$ to appear explicitly since it is the
leading shift-invariant effective interaction. This situation is ubiquitous
in axion effective theories. By contrast, all the other operators that could
be eliminated by partial integration and the use of the Dirac equation do not
bring anything special and have not been kept.

If the effective operators are constructed at the SM scale, we must add a
Higgs field for all the $LR$ operators of Eqs.~(\ref{FPhi1}) and~(\ref{FPhi2})
to make them symmetric under $SU(2)_{L}\otimes U(1)_{Y}$ since $\psi_{R}$ and
$\psi_{L}$ have different gauge charges. This increases their dimension by
one. Then, restricting our attention to operators up to dimension five, the
non-fermionic operators that can be constructed are (omitting Wilson
coefficients for clarity)%
\begin{align}
\mathcal{L}_{Int}^{\text{0-Form}} &  =\Lambda\Phi^{\dagger}\Phi\phi
+\Phi^{\dagger}\Phi\phi^{2}\nonumber\\
&  +\frac{1}{\Lambda}\Phi^{\dagger}\overleftrightarrow{\mathcal{D}}_{\mu}\Phi
F^{\phi,\mu}+\frac{1}{\Lambda}\phi\mathcal{D}^{\mu}\Phi^{\dagger}%
\mathcal{D}_{\mu}\Phi+\frac{1}{\Lambda}\phi(\Phi^{\dagger}\Phi)^{2}+\frac
{1}{\Lambda}\phi F_{\mu\nu}F^{\mu\nu}+\frac{1}{\Lambda}\phi F_{\mu\nu}%
\tilde{F}^{\mu\nu}\nonumber\\
&  +\mathcal{O}(\Lambda^{-2})\ ,\label{SMPhi}%
\end{align}
where $\Phi$ stands for the Higgs boson doublet and $F$ for either the
electroweak or gluon field strength, $B_{\mu\nu}$, $W_{\mu\nu}^{i}$, or
$G_{\mu\nu}^{a}$.

\subsection{1-form field effective operators\label{EffOp1}}

The operators for a Proca field together with a fermion and up to two
derivatives are (operators with $L\leftrightarrow R$ are understood):%
\begin{equation}%
\begin{tabular}
[c]{cccc}\hline
$d$ & Type &  & \\\hline
4 & II & $\bar{\psi}_{L}\gamma^{\mu}\psi_{L}A_{\mu}$ & \\
5 & I & $\bar{\psi}_{L}\sigma^{\mu\nu}\psi_{R}F_{\mu\nu}^{A}$ & \\
& IV & $\bar{\psi}_{L}\psi_{R}\partial^{\mu}A_{\mu}$ & \\
6 & I & $\bar{\psi}_{L}\gamma^{\mu}\mathcal{D}^{\nu}\psi_{L}F_{\mu\nu}^{A}$ &
$\bar{\psi}_{L}\gamma^{\nu}\psi_{L}\partial^{\mu}F_{\mu\nu}^{A}$\\
& II & $-$ & \\
& III & $\bar{\psi}_{L}\gamma_{\nu}\psi_{L}A_{\mu}F^{\mu\nu}$ & $\bar{\psi
}_{L}\gamma_{\nu}\psi_{L}A_{\mu}\tilde{F}^{\mu\nu}$\\\hline
\end{tabular}
\ \label{FA1}%
\end{equation}
Remember here that the covariant derivative acting on $\psi$ may include the SM gauge fields only. For reasons that will become clear later on, we introduce as a fourth class of
operators all those that vanish upon enforcing the Lorenz condition. Notice
that no other operator appears at $\mathcal{O}(\Lambda^{-1})$ because
Eq.~(\ref{Id4}) is used to express $\bar{\psi}_{L}\mathcal{D}^{\nu}\psi
_{R}A_{\nu}$ in terms of $\bar{\psi}_{L}\sigma^{\mu\nu}\psi_{R}F_{\mu\nu}^{A}%
$. Similarly, at the dimension-six level, $\bar{\psi}_{L}\gamma^{\mu
}\mathcal{D}^{\nu}\psi_{L}\partial_{\nu}A_{\mu}$ is reduced to $\bar{\psi}%
_{L}\gamma^{\mu}\mathcal{D}^{\nu}\psi_{L}F_{\mu\nu}^{A}$ by first subtracting
$\bar{\psi}_{L}\gamma^{\mu}\mathcal{D}^{\nu}\psi_{L}\partial_{\mu}A_{\nu}$,
reducible via the Dirac equation, and then by using Eq.~(\ref{Id1}) and the
antisymmetry under $\mu\leftrightarrow\nu$. Also, rewriting Eq.~(\ref{Id1})
as
\begin{equation}
\gamma^{\mu}\mathcal{D}^{\nu}-\gamma^{\nu}\mathcal{D}^{\mu}=\gamma^{\mu}%
\gamma^{\nu}\gamma^{\rho}\mathcal{D}_{\rho}-g^{\mu\nu}\gamma^{\rho}%
\mathcal{D}_{\rho}-i\epsilon^{\mu\nu\rho\alpha}\gamma_{\alpha}\gamma
_{5}\mathcal{D}_{\rho}\ ,\label{Id5}%
\end{equation}
allows one to relate $\bar{\psi}_{L}\gamma^{\mu}\mathcal{D}^{\nu}\psi_{L}F_{\mu
\nu}^{A}$ and $\bar{\psi}_{L}\gamma^{\mu}\mathcal{D}^{\nu}\psi_{L}\tilde
{F}_{\mu\nu}^{A}$, up to $\mathcal{O}(m)$ corrections to $\bar{\psi}_{L}%
\sigma^{\mu\nu}\psi_{R}F_{\mu\nu}^{A}$. So, in the absence of the photon
field, there are no gauge-dependent dimension-six operators because they all
collapse to $\mathcal{O}(m)$ contributions to the dimension-five $\bar{\psi
}_{L}\psi_{R}\partial^{\mu}A_{\mu}$ operator and $\mathcal{O}(m^{2})$
contribution to the dimension four $\bar{\psi}_{L}\gamma^{\mu}\psi_{L}A_{\mu}$
operator. Since in the absence of the gauge symmetry, those should be present
anyway, there is no need to include separately the gauge-dependent
dimension-six ones. Finally, concerning the leading nongauge invariant
couplings $\bar{\psi}_{L}\gamma^{\mu}\psi_{L}A_{\mu}$ and $\bar{\psi}%
_{R}\gamma^{\mu}\psi_{R}A_{\mu}$, it should be remembered that the vector
combination is actually gauge invariant since the vector current is conserved.

With two Proca fields, the leading operators are%
\begin{equation}%
\begin{tabular}
[c]{cccccc}\hline
$d$ & Type &  &  &  & \\\hline
5 & II & $\bar{\psi}_{L}\psi_{R}A_{\mu}A^{\mu}$ &  &  & \\
6 & II & $\bar{\psi}_{L}\gamma^{\nu}\psi_{L}A^{\mu}F_{\mu\nu}^{A}$ &
$\bar{\psi}_{L}\gamma^{\nu}\psi_{L}A^{\mu}\tilde{F}_{\mu\nu}^{A}$ & $\bar
{\psi}_{L}\gamma^{\mu}\mathcal{D}^{\nu}\psi_{L}A_{\mu}A_{\nu}$ & \\
& IV & $\bar{\psi}_{L}\gamma^{\nu}\psi_{L}A_{\nu}\partial^{\mu}A_{\mu} $ &  &
& \\
7 & I & $\bar{\psi}_{L}\psi_{R}F_{\mu\nu}^{A}F^{A,\mu\nu}$ & $\bar{\psi}%
_{L}\psi_{R}F_{\mu\nu}^{A}\tilde{F}^{A,\mu\nu}$ &  & \\
& II & $\bar{\psi}_{L}\psi_{R}\partial_{\nu}A_{\mu}\partial^{\nu}A^{\mu}$ &
$\bar{\psi}_{L}\psi_{R}A^{\mu}\partial^{\nu}F_{\mu\nu}$ & $\bar{\psi}%
_{L}\sigma^{\mu\nu}\psi_{R}A_{\mu}\partial^{\rho}F_{\nu\rho}$ & $\bar{\psi
}_{L}\sigma^{\mu\nu}\psi_{R}A_{\rho}\partial^{\rho}F_{\mu\nu}^{A}$\\
&  & $\bar{\psi}_{L}\sigma^{\mu\nu}\mathcal{D}^{\rho}\psi_{R}A_{\mu}%
\partial_{\rho}A_{\nu}$ & $\bar{\psi}_{L}\sigma^{\nu\rho}\mathcal{D}^{\mu}%
\psi_{R}A_{\mu}F_{\nu\rho}^{A}$ & $\bar{\psi}_{L}\mathcal{D}^{\mu}%
\mathcal{D}^{\nu}\psi_{R}A_{\mu}A_{\nu}$ & \\
& III & $\bar{\psi}_{L}\sigma^{\mu\nu}\psi_{R}A_{\rho}A^{\rho}F_{\mu\nu}$ &
$\bar{\psi}_{L}\sigma^{\mu\nu}\psi_{R}A_{\mu}A^{\rho}F_{\rho\nu}$ &  & \\
& IV & $\bar{\psi}_{L}\psi_{R}A_{\mu}\partial^{\mu}\partial^{\nu}A_{\nu}$ &
$\bar{\psi}_{L}\psi_{R}\partial^{\mu}A_{\mu}\partial^{\nu}A_{\nu}$ &
$\bar{\psi}_{L}\sigma^{\mu\nu}\psi_{R}F_{\mu\nu}^{A}\partial^{\rho}A_{\rho}$ &
\\\hline
\end{tabular}
\label{FA2}%
\end{equation}
Again, many identities have been exploited to reduce the number of independent
operators, and this basis is minimal.

Applied at the electroweak scale, the same provision as in the scalar case
applies for adding Higgs boson fields for all $LR$ operators. Then, the
operators involving only SM bosonic fields are, up to dimension five,%
\begin{equation}
\mathcal{L}_{Int}^{\text{1-Form}}=F_{\mu\nu}^{A}F_{Y}^{\mu\nu}+\Phi^{\dagger
}\Phi\partial^{\mu}A_{\mu}+\Phi^{\dagger}\overleftrightarrow{\mathcal{D}}%
_{\mu}\Phi A^{\mu}+\Phi^{\dagger}\Phi A_{\mu}A^{\mu}+\mathcal{O}(\Lambda
^{-2})\ ,\label{SMA}%
\end{equation}
where $\Phi$ stands for the Higgs boson doublet and we denote as $F_{Y}%
^{\mu\nu}$ the $U(1)_{Y}$ field strength. The first operator is the well-known
kinetic mixing operator. We do not include the topological $F_{\mu\nu}%
^{A}\tilde{F}_{Y}^{\mu\nu}$ coupling, which vanishes upon partial integration
thanks to the Bianchi identity. Notice that a specific combination of the last
two operators is gauge invariant. It can be absorbed into the covariant
derivative acting on the Higgs field, thereby giving it a dark charge. Once
diagonalized, the kinetic mixing has a similar impact. This is described in
detail in many places, see e.g. Ref.~\cite{Williams:2011qb, Fabbrichesi:2020wbt}, but for completeness, let us
briefly summarize the main feature. We take the case of a low-energy mixing
with the photon field alone, which we denote as $A_{\mu}^{\gamma}$ and
$F_{\mu\nu}^{\gamma}$, and start from%
\begin{equation}
L_{kin}=-\frac{1}{4}F_{\mu\nu}^{\gamma}F^{\gamma,\mu\nu}-\frac{1}{4}F_{\mu\nu
}^{A}F^{A,\mu\nu}+m_{A}^{2}A_{\mu}A^{\mu}+\frac{\chi}{2}F_{\mu\nu}%
^{A}F^{\gamma,\mu\nu}\ .
\end{equation}
To diagonalize the kinetic terms, one performs%
\begin{equation}
A_{\mu}^{\gamma}\rightarrow A_{\mu}^{\gamma}+\sinh\eta A_{\mu}\ ,\ \ A_{\mu
}\rightarrow\cosh\eta A_{\mu}\ ,\ \ \tanh\eta=\chi\ .\label{KmixReparam}%
\end{equation}
This is really a reparametrization of the fields, not a mere rotation, and as
such it need not be unitary. All that matters is to produce canonical kinetic
terms. Phenomenologically, the two-point effective coupling rescales the dark
vector mass as $m_{A}\rightarrow m_{A}\cosh\eta$, and adds gauge invariant
couplings with strength $q\times\sinh\eta$ for all the fermions of electric
charge $q$ to the dark vector. Those do not alter the basis constructed above,
and can be absorbed either directly in its gauge-invariant effective
couplings, or into gauge-invariant combinations of its effective couplings
like $\bar{\psi}_{L}\gamma^{\mu}\psi_{L}A_{\mu}+\bar{\psi}_{R}\gamma^{\mu}%
\psi_{R}A_{\mu}$ for a charged fermion, or $\Phi^{\dagger}\overleftrightarrow
{\mathcal{D}}_{\mu}\Phi A^{\mu}+\Phi^{\dagger}\Phi A_{\mu}A^{\mu}$ for a
charged scalar. The description of the full $SU(2)_{L}\otimes U(1)$ case,
including the mixing with the $Z$ boson, does not alter this picture, see Ref.~\cite{Williams:2011qb, Fabbrichesi:2020wbt}.

\subsection{2-form field effective operators\label{EffOp2}}

For the 2-form field, the dominant operators are found to be:%
\begin{equation}%
\begin{tabular}
[c]{ccccc}\hline
$d$ & Type &  &  & \\\hline
$4$ & II & $\bar{\psi}_{L}\sigma^{\mu\nu}\psi_{R}B_{\mu\nu}$ &  & \\
$5$ & I & $\bar{\psi}_{L}\gamma_{\sigma}\psi_{L}\epsilon^{\mu\nu\rho\sigma
}F_{\mu\nu\rho}^{B}$ &  & \\
& II & $\bar{\psi}_{L}\gamma^{\mu}\mathcal{D}^{\nu}\psi_{L}B_{\mu\nu}$ &  & \\
& IV & $\bar{\psi}_{L}\gamma^{\nu}\psi_{L}\partial^{\mu}B_{\mu\nu}$ &  & \\
$6$ & I & $\bar{\psi}_{L}\sigma^{\mu\nu}\psi_{R}\partial^{\rho}F_{\mu\nu\rho
}^{B}$ &  & \\
& II & $-$ &  & \\
& III & $\bar{\psi}_{L}\psi_{R}B_{\mu\nu}F^{\mu\nu}$ & $\bar{\psi}_{L}\psi
_{R}B_{\mu\nu}\tilde{F}^{\mu\nu}$ & $\bar{\psi}_{L}\sigma^{\mu\rho}\psi
_{R}B_{\mu\nu}F^{\nu}{}_{\rho}$\\
& IV & $\bar{\psi}_{L}\sigma^{\mu\nu}\psi_{R}\partial_{\mu}\partial^{\rho
}B_{\nu\rho}$ &  & \\\hline
\end{tabular}
\ \label{FB1}%
\end{equation}
The dimension-four and the first dimension-five operator were already identified in Ref.~\cite{Malta:2025ydq}, while the leading tensor interaction was considered in Refs.~\cite{Magnus:2024sdp} and~\cite{Tiwary:2021cff} in the context of DM searches and Bhabha scattering, respectively. To derive all the others, partial integration was used as much as possible and the Bianchi identity $\epsilon^{\alpha\mu\nu\rho}\partial_{\alpha}F_{\mu\nu\rho
}^{B}=0$ was enforced. Besides the fermionic identities discussed before, we
also used Eq.~(\ref{AppId}) to simplify any partial contraction between an
epsilon tensor and $B$ or $F^{B}$, e.g. as%
\begin{equation}
\epsilon^{\alpha\beta\gamma\mu}B_{\mu\nu}=-\frac{1}{4}\epsilon^{\alpha
\beta\gamma\mu}\epsilon_{\mu\nu\rho\sigma}\epsilon^{\lambda\kappa\rho\sigma
}B_{\lambda\kappa}=-\frac{1}{4}\delta_{\nu\rho\sigma}^{\alpha\beta\gamma
}\epsilon^{\lambda\kappa\rho\sigma}B_{\lambda\kappa}=-\frac{1}{2}(\delta_{\nu
}^{\alpha}\tilde{B}^{\sigma\rho}+\delta_{\nu}^{\beta}\tilde{B}^{\rho\alpha
}+\partial_{\nu}^{\gamma}\tilde{B}^{\alpha\sigma})\ ,\label{AntiB}%
\end{equation}
where $\tilde{B}_{\mu\nu}=(1/2)\epsilon_{\mu\nu\rho\sigma}B^{\rho\sigma}$ is
the dual of the $B$ field. For two $B$ fields, the leading operators are%
\begin{equation}%
\begin{tabular}
[c]{ccccc}\hline
$d$ & Type &  &  & \\\hline
5 & II & $\bar{\psi}_{L}\psi_{R}B_{\mu\nu}B^{\mu\nu}$ & $\bar{\psi}_{L}%
\psi_{R}B_{\mu\nu}\tilde{B}^{\mu\nu}$ & \\
6 & II & $\bar{\psi}\gamma_{\rho}\mathcal{D}^{\mu}\psi B_{\mu\nu}B^{\nu\rho}$
& $\bar{\psi}\gamma_{\rho}\mathcal{D}^{\mu}\psi B_{\mu\nu}\tilde{B}^{\nu\rho}$
& $\bar{\psi}\gamma_{\rho}\mathcal{D}^{\mu}\psi\tilde{B}_{\mu\nu}B^{\nu\rho}%
$\\
&  & $\bar{\psi}_{L}\gamma^{\mu}\psi_{L}B_{\mu\nu}\partial_{\rho}\tilde
{B}^{\nu\rho}$ & $\bar{\psi}_{L}\gamma^{\mu}\psi_{L}\tilde{B}_{\mu\nu}%
\partial_{\rho}\tilde{B}^{\rho\nu}$ & \\
& IV & $\bar{\psi}_{L}\gamma^{\mu}\psi_{L}\tilde{B}_{\mu\nu}\partial_{\rho
}B^{\nu\rho}$ & $\bar{\psi}_{L}\gamma^{\mu}\psi_{L}B_{\mu\nu}\partial_{\rho
}B^{\rho\nu}$ & \\
7 & I & $\bar{\psi}_{L}\psi_{R}F_{\mu\nu\rho}^{B}F^{B,\mu\nu\rho}$ &  & \\
& II,III,IV & ... &  & \\\hline
\end{tabular}
\ \label{FB2}%
\end{equation}
At the dimension-seven level, we keep only the gauge-invariant operator
because it is the leading one being so. The gauge-dependent operators are too
numerous to be useful, so their quite tricky reduction to a minimal basis does
not appear worth the effort.

The couplings with the other SM fields are%
\begin{align}
\mathcal{L}_{Int}^{\text{2-Form}} &  =\Lambda B_{\mu\nu}F_{Y}^{\mu\nu}+\Lambda
B_{\mu\nu}\tilde{F}_{Y}^{\mu\nu}\nonumber\\
&  \ \ \ \ +\Phi^{\dagger}\Phi B_{\mu\nu}B^{\mu\nu}+\Phi^{\dagger}\Phi
B_{\mu\nu}\tilde{B}^{\mu\nu}\nonumber\\
&  \ \ \ \ +\frac{1}{\Lambda}\Phi^{\dagger}\overleftrightarrow{\mathcal{D}%
}_{\alpha}\Phi\epsilon^{\mu\nu\rho\alpha}F_{\mu\nu\rho}^{B}+\frac
{1}{\Lambda}\partial^{\alpha}F_{\alpha\beta}\epsilon^{\mu\nu\rho\beta
}F_{\mu\nu\rho}^{B}+\frac{1}{\Lambda}H^{\dagger}HF^{\mu\nu}B_{\mu\nu
}\nonumber\\
&  \ \ \ +\frac{1}{\Lambda}\Phi^{\dagger}\overleftrightarrow{\mathcal{D}%
}_{\alpha}\Phi\partial_{\mu}B^{\mu\alpha}+\frac{1}{\Lambda}\partial^{\alpha
}F_{\alpha\nu}\partial_{\mu}B^{\mu\nu}+\frac{1}{\Lambda}H^{\dagger}H\tilde
{F}^{\mu\nu}B_{\mu\nu}\nonumber\\
&  \ \ \ \ +\mathcal{O}(\Lambda^{-2})\label{SMB}%
\end{align}
Notice that the $B_{\mu\nu}F_{Y}^{\mu\nu}$ term at $\mathcal{O}(\Lambda)$ is
actually a surface term upon enforcing the Lorentz condition $\partial^{\mu
}B_{\mu\nu}=0$ since%
\begin{equation}
B_{\mu\nu}F_{Y}^{\mu\nu}=2B_{\mu\nu}\partial^{\mu}A_{Y}^{\nu}=2\partial^{\mu
}(B_{\mu\nu}A_{Y}^{\nu})-\partial^{\mu}B_{\mu\nu}A^{\nu}\ .
\end{equation}
It is thus essentially topological (in the absence of kinetic terms for $A$
and $B$, this term alone is known as the topological BF theory~\cite{Blau:1989bq}) and can be discarded. The second term is not topological since $B_{\mu\nu}\tilde{F}_{Y}^{\mu\nu}=\epsilon^{\mu\nu\rho\sigma}B_{\mu\nu}\partial_{\rho}A_{Y,\sigma}=\epsilon^{\mu\nu\rho\sigma}A_{Y,\sigma}F_{\mu\nu\rho}^{B}$. As for the kinetic mixing, this two-point vertex must be eliminated to get canonical kinetic terms, and this can be done again via a reparametrization. 

Specifically, let us consider the low-energy situation of a mixing with the
photon field strength%
\begin{equation}
\mathcal{L}_{eff}^{\text{2-Form}}\supset\frac{1}{12}F_{\mu\nu\rho}^{B}%
F^{B,\mu\nu\rho}-\frac{1}{4}m^{2}B_{\mu\nu}B^{\mu\nu}+\frac{1}{4}\tilde{m}%
^{2}B_{\mu\nu}\tilde{B}^{\mu\nu}-\frac{1}{4}F_{\mu\nu}^{\gamma}F^{\gamma
,\mu\nu}+\frac{\Lambda_{\gamma}}{2}B_{\mu\nu}\tilde{F}^{\gamma,\mu\nu}\ .
\end{equation}
We can perform the reparametrization%
\begin{equation}
A_{\mu}^{\gamma}\rightarrow A_{\mu}^{\gamma}\left(  1-\eta^{2}\right)
^{-1/2},\ \ B_{\mu\nu}\rightarrow B_{\mu\nu}-\eta F_{\mu\nu}^{\gamma
}\ ,\ \ \ \eta=\frac{\Lambda_{\gamma}}{\tilde{m}^{2}}\ ,
\end{equation}
to eliminate the $B_{\mu\nu}\tilde{F}^{\gamma,\mu\nu}$ coupling. This leaves
the field strength $F^{B}=dB$ invariant, but regenerates a topological
$B_{\mu\nu}F_{Y}^{\mu\nu}$ coupling that needs again to be discarded (along
with a topologically trivial $F_{\mu\nu}^{\gamma}\tilde{F}^{\gamma,\mu\nu}$ term).

A crucial difference with the dark vector kinetic mixing though is that it is
now the dark field that is shifted, not the photon field. As a result, the $B$
field does not inherit any new coupling. Instead, it is the photon that does
if $B$ is coupled to SM fields. For instance, shifting the $\bar{\psi}%
_{L}\sigma^{\mu\nu}\psi_{R}B_{\mu\nu}$ operator generates new contributions to
the electric and/or magnetic dipole operators (EDM and MDM, respectively),%
\begin{equation}
c_{1}\bar{\psi}_{L}\sigma^{\mu\nu}\psi_{R}B_{\mu\nu}\rightarrow c_{1}\bar
{\psi}_{L}\sigma^{\mu\nu}\psi_{R}B_{\mu\nu}-c_{1}\eta\bar{\psi}_{L}\sigma
^{\mu\nu}\psi_{R}F_{\mu\nu}^{\gamma}\ ,
\end{equation}
which are tightly bounded. Specifically, adding the $L\leftrightarrow R$
operator, which is simply the Hermitian conjugate, this becomes%
\begin{equation}
c_{1}\bar{\psi}_{L}\sigma^{\mu\nu}\psi_{R}B_{\mu\nu}\rightarrow-\frac
{\operatorname{Re}(c_{1}\eta)}{2}\bar{\psi}\sigma^{\mu\nu}\psi F_{\mu\nu
}^{\gamma}+i\frac{\operatorname{Im}(c_{1}\eta)}{2}\bar{\psi}\sigma^{\mu\nu
}\gamma^{5}\psi F_{\mu\nu}^{\gamma}\ ,
\end{equation}
with then the $\psi$ MDM and EDM given by $\operatorname{Re}(c_{1}%
\eta)=ea_{\psi}/2m_{\psi}$ and $d_{\psi}=\operatorname{Im}(c_{1}\eta)$,
respectively. Overall, assuming $c_{1}$ is $\mathcal{O}(1)$ and real,%
\begin{equation}
a_{\psi}\sim\frac{m_{\psi}}{m}\frac{\operatorname{Re}\Lambda_{\gamma}}%
{\tilde{m}}\ ,\ \ d_{\psi}\sim\frac{\operatorname{Im}\Lambda_{\gamma}}%
{\tilde{m}^{2}}\ .
\end{equation}
If we ask that the contribution to $a_{\mu}$ is less than $10^{-12}$, and that
to $d_{e}$ less than $10^{-30}~$e$cm$ (based on the PDG data~\cite{ParticleDataGroup:2024cfk}), this imposes $\operatorname{Re}%
\Lambda_{\gamma}/\tilde{m}^{2}<10^{-20}$~eV and $\operatorname{Im}%
\Lambda_{\gamma}/\tilde{m}^{2}<10^{-25}$~eV. If $\tilde{m}_{B}$ is to be below
the electroweak scale, then $\Lambda_{\gamma}$ needs to be extremely
suppressed. If $\tilde{m}_{B}$ is larger, these $LR$ operators have to first
involve the Higgs boson, and $c_{0}\sim\mathcal{O}(v_{ew}/\Lambda)$, with
$v_{ew}$ the electroweak vacuum. Yet, in this case also, a large scaling
$\Lambda_{\gamma}\ll\Lambda$ is required if $\tilde{m}_{B}$ is to be
relatively light since otherwise, with $\Lambda_{\gamma}\approx\Lambda$,
$a_{\psi}$ and $d_{\psi}$ would depend only on $\tilde{m}_{B}^{2}$ which would
then have to be well above the TeV scale.

For completeness, had we kept the topological $\Lambda^{\prime}B_{\mu\nu}%
F_{Y}^{\mu\nu}$ term, or in the absence of the $\tilde{m}_{B}$ term, an
additional $\eta^{\prime}\tilde{F}_{\mu\nu}$ shift of $B_{\mu\nu}$ would be
needed, with $\eta^{\prime}$ either $\Lambda^{\prime}/\tilde{m}$ or $\Lambda
/m$, respectively. Under this shift, besides the EDM and MDM operators, new
corrections arise because the field strength $F^{B,\mu\nu\rho}$ ceased to be
invariant. The shift then produces a contribution to the $\epsilon
_{\alpha\beta\gamma\sigma}F^{B,\alpha\beta\gamma}\partial_{\mu}F^{\gamma
,\mu\sigma}$ operator of Eq.~(\ref{SMB}), along with a Uehling correction in
$\eta^{\prime2}F_{\mu\nu}\partial^{2}F^{\mu\nu}$. These corrections are much
less strict than that coming from EDM and MDM, with e.g. $|\Lambda
|/m^{2}<10^{-10}$~eV if we require the Uehling correction to be less than a
thousandth of the QED contribution of $\alpha/(60\pi m_{e}^{2})$.

To close this Section, it should be stressed that having a strong suppression
of the $B_{\mu\nu}\tilde{F}^{\gamma,\mu\nu}$ mixing term is not unrealistic.
Indeed, contrary to the dark vector kinetic mixing, $B_{\mu\nu}\tilde
{F}^{\gamma,\mu\nu}$ and $B_{\mu\nu}F^{\gamma,\mu\nu}$ break the $B$ gauge
invariance. That is a crucial feature since one could imagine models in which
a breaking of that symmetry first generates the $B$ mass term, and only
subsequently induces $B_{\mu\nu}\tilde{F}^{\gamma,\mu\nu}$, $B_{\mu\nu
}F^{\gamma,\mu\nu}$, $\bar{\psi}_{L}\sigma^{\mu\nu}\psi_{R}B_{\mu\nu}$,....
Further, a strong scaling $m\gg\Lambda$ could be relatively stable as the
coupling $\bar{\psi}_{L}\sigma^{\mu\nu}\psi_{R}B_{\mu\nu}$ cannot induce a
$B_{\mu\nu}\tilde{F}^{\gamma,\mu\nu}$ interaction at one loop, and the
contribution from the subleading coupling $\bar{\psi}_{L}\gamma_{\sigma}%
\psi_{L}\epsilon^{\mu\nu\rho\sigma}F_{\mu\nu\rho}^{B}$ is protected by gauge
invariance. So, in our opinion, a scenario with a light $B$ state that does
not significantly mix with the photon, and is dominantly coupled to SM
matter via either $\bar{\psi}_{L}\sigma^{\mu\nu}\psi_{R}B_{\mu\nu}$ or
$\bar{\psi}_{L}\gamma_{\sigma}\psi_{L}\epsilon^{\mu\nu\rho\sigma}F_{\mu\nu
\rho}^{B}$, is viable and phenomenologically relevant. We will come back to
this point in Sec.~\ref{Pheno}.

\subsection{3-form field effective operators\label{EffOp3}}

For a three-form field, the leading fermionic operators are
\begin{equation}%
\begin{tabular}
[c]{cccc}\hline
$d$ & Type &  & \\\hline
4 & II & $\bar{\psi}_{L}\gamma_{\sigma}\psi_{L}\epsilon^{\mu\nu\rho\sigma
}C_{\mu\nu\rho}$ & \\
5 & I & $\bar{\psi}_{L}\psi_{R}\epsilon^{\mu\nu\rho\sigma}F_{\mu\nu\rho\sigma
}^{C}$ & \\
& IV & $\bar{\psi}_{L}\sigma^{\mu\nu}\psi_{R}\partial^{\rho}C_{\mu\nu\rho}$ &
\\
6 & I & $-$ & \\
& II & $-$ & \\
& III & $\bar{\psi}_{L}\gamma^{\rho}\psi_{L}C_{\mu\nu\rho}F^{\mu\nu}$ &
$\bar{\psi}_{L}\gamma^{\rho}\psi_{L}C_{\mu\nu\rho}\tilde{F}^{\mu\nu}$\\
& IV & $\bar{\psi}_{L}\gamma^{\mu}\mathcal{D}^{\nu}\psi_{R}\partial^{\rho
}C_{\mu\nu\rho}$ & $\bar{\psi}_{L}\gamma_{\sigma}\psi_{R}\epsilon
^{\sigma\alpha\mu\nu}\partial_{\alpha}\partial^{\rho}C_{\mu\nu\rho}$\\\hline
\end{tabular}
\ \label{FC1}%
\end{equation}
while for two 3-form fields, we find%
\begin{equation}%
\begin{tabular}
[c]{ccccc}\hline
$d$ & Type &  &  & \\\hline
5 & II & $\bar{\psi}_{L}\psi_{R}C^{\mu\nu\rho}C_{\mu\nu\rho}$ &  & \\
6 & II & $\bar{\psi}_{L}\gamma^{\mu}\psi_{L}C^{\nu\rho\sigma}F_{\mu\nu
\rho\sigma}^{C}$ & $\bar{\psi}_{L}\gamma^{\mu}\mathcal{D}_{\nu}\psi_{L}%
C_{\mu\rho\sigma}C^{\nu\rho\sigma}$ & \\
& IV & $\bar{\psi}_{L}\gamma^{\sigma}\psi_{L}C_{\sigma\mu\nu}\partial_{\rho
}C^{\mu\nu\rho}$ & $\bar{\psi}_{L}\gamma^{\mu}\psi_{L}\epsilon^{\nu\alpha
\beta\gamma}C_{\alpha\beta\gamma}\partial^{\rho}C_{\mu\nu\rho}$ & \\
7 & I & $\bar{\psi}_{L}\psi_{R}F_{\mu\nu\rho\sigma}^{C}F^{C,\mu\nu\rho\sigma}$
&  & \\
& II & $\bar{\psi}_{L}\psi_{R}\partial_{\sigma}C^{\mu\nu\rho}\partial^{\sigma
}C_{\mu\nu\rho}$ & $\bar{\psi}_{L}\psi C^{\mu\nu\rho}\partial^{\sigma}%
F_{\mu\nu\rho\sigma}^{C}$ & \\
&  & $\bar{\psi}_{L}\sigma^{\mu\nu}\mathcal{D}^{\rho}\psi_{R}C^{\alpha\beta}%
{}_{\nu}\partial_{\rho}C_{\mu\alpha\beta}$ & $\bar{\psi}_{L}\mathcal{D}^{\mu
}\mathcal{D}_{\nu}\psi_{R}C_{\mu\rho\sigma}C^{\nu\rho\sigma}$ & \\
& III & $\bar{\psi}_{L}\sigma^{\alpha\beta}\psi_{R}C^{\mu\nu\rho}C_{\mu\nu
\rho}F_{\alpha\beta}$ & $\bar{\psi}_{L}\sigma^{\mu\nu}\psi_{R}C_{\mu
\alpha\beta}C^{\rho\alpha\beta}F_{\rho\nu}$ & \\
& IV & $\bar{\psi}_{L}\psi_{R}C^{\mu\nu\sigma}\partial_{\sigma}\partial^{\rho
}C_{\mu\nu\rho}$ & $\bar{\psi}_{L}\sigma^{\mu\nu}\psi_{R}C^{\alpha\beta}%
{}_{\nu}\partial_{\alpha}\partial^{\rho}C_{\mu\beta\rho}$ & $\bar{\psi}%
_{L}\sigma^{\alpha\beta}\psi_{R}\partial_{\rho}C^{\mu\nu\rho}F_{\mu\nu
\alpha\beta}^{C}$\\
&  & $\bar{\psi}_{L}\psi_{R}\epsilon^{\mu\nu\alpha\beta}\partial^{\rho
}C_{\alpha\beta\rho}\partial^{\sigma}C_{\mu\nu\sigma}$ & $\bar{\psi}_{L}%
\sigma_{\alpha\beta}\psi_{R}C^{\mu\nu\alpha}\partial^{\beta}\partial^{\rho
}C_{\mu\nu\rho}$ & \\
&  & $\bar{\psi}_{L}\psi_{R}\partial_{\rho}C^{\mu\nu\rho}\partial^{\sigma
}C_{\mu\nu\sigma}$ & $\bar{\psi}_{L}\sigma^{\nu\rho}\mathcal{D}^{\mu}\psi
_{R}C^{\alpha}{}_{\rho\nu}\partial^{\sigma}C_{\mu\alpha\sigma}$ & \\\hline
\end{tabular}
\ \ \ \label{FC2}%
\end{equation}
Many reductions are quite subtle and require not only the various spinor
identities, but also exploiting the antisymmetry of $C_{\mu\nu\rho}$ or
$F_{\mu\nu\rho\sigma}^{C}$ together with Eq.~(\ref{AppId}), in a way similar
as in Eq.~(\ref{AntiB}). For example, the operator $\bar{\psi}_{L}%
\gamma_{\alpha}\psi_{L}\epsilon^{\mu\nu\rho\alpha}\partial^{\sigma}F_{\mu
\nu\rho\sigma}^{C}$ is absent because it can be written as $\bar{\psi}%
_{L}\gamma_{\alpha}\psi_{L}\partial^{\alpha}(\epsilon^{\mu\nu\rho\sigma}%
F_{\mu\nu\rho\sigma}^{C})$, which is reducible by integrating by part and
using the Dirac equation. The situation is to be contrasted to that of the
$\bar{\psi}_{L}\gamma^{\mu}\psi_{L}F_{\mu}^{\phi}$ operator in the scalar
basis. In that case, we choose to add it in parentheses because it gets reduced
to $\bar{\psi}_{L}\psi_{R}\phi$, hiding the shift symmetry. By contrast here,
$\bar{\psi}_{L}\gamma_{\alpha}\psi_{L}\epsilon^{\mu\nu\rho\alpha}%
\partial^{\sigma}F_{\mu\nu\rho\sigma}^{C}$ sum up to $\mathcal{O}(m)$
contributions to $\bar{\psi}_{L}\psi_{R}\epsilon^{\mu\nu\rho\sigma}F_{\mu
\nu\rho\sigma}^{C}$, which is still gauge invariant, so there is no need to
keep track of $\bar{\psi}_{L}\gamma_{\alpha}\psi_{L}\epsilon^{\mu\nu\rho
\alpha}\partial^{\sigma}F_{\mu\nu\rho\sigma}^{C}$.

All in all, it is quite remarkable that so few operators survive. At first
glance, with many Lorentz indices at our disposal hence many alternative ways
to contract them, one could have expected the number of operators to be quite
large, especially with two $C$ fields. The reason why this is not the case
resides in the existence of Hodge dualities relating this basis to that for
the $p=1$ field. This will be explored in detailed in Sec.~\ref{AlgDualities},
but we can already state that provided all the operators vanishing under the
Lorenz conditions are included, there are precisely as many operators for a
one and a 3-form field. In practice, we nevertheless derived all the above
operators from scratch except for those of dimension seven, which were
directly constructed from their Proca field counterparts. Indeed, at that
level, the number of ways to contract all the indices is simply too large for
a brute force method.

With SM fields, we can construct%
\begin{equation}
\mathcal{L}_{Int}^{\text{3-Form}}=\partial^{\rho}C_{\rho\mu\nu}\tilde{F}%
_{Y}^{\mu\nu}+\Phi^{\dagger}\Phi\epsilon^{\mu\nu\rho\sigma}F_{\mu\nu\rho
\sigma}^{C}+\Phi^{\dagger}\overleftrightarrow{\mathcal{D}}_{\alpha}%
\Phi\epsilon^{\mu\nu\rho\sigma}C_{\mu\nu\rho}+\Phi^{\dagger}\Phi C^{\mu\nu
\rho}C_{\mu\nu\rho}+\mathcal{O}(\Lambda^{-2})\ .\label{SMC}%
\end{equation}
Notice that $\partial^{\rho}C_{\rho\mu\nu}\tilde{F}_{Y}^{\mu\nu}\sim
\partial^{\tau}F_{Y,\tau\sigma}\epsilon^{\mu\nu\rho\sigma}C_{\mu\nu\rho}$,
while $\partial^{\rho}C_{\rho\mu\nu}F_{Y}^{\mu\nu}$ is not included because it
vanishes upon partial integration thanks to the Bianchi identity for
$F_{Y}^{\mu\nu}$.

\subsection{4-form field effective operators\label{EffOp4}}

With four indices, the number of possible contractions becomes very limited since no more than two Dirac matrices can appear. In practice, a Levi-Civita tensor is always needed to bring down the number of indices. At the same time, for a 4-form field, there is a unique way to contract its indices with the antisymmetric tensor because%
\begin{equation}
\epsilon^{\alpha\beta\gamma\delta}D_{\mu\nu\rho\sigma}=-\frac{1}{4!}%
\epsilon^{\alpha\beta\gamma\delta}D_{\lambda\kappa\pi\sigma}\epsilon
^{\lambda\kappa\pi\sigma}\epsilon_{\mu\nu\rho\sigma}=\frac{1}{4!}\delta
_{\mu\nu\rho\sigma}^{\alpha\beta\gamma\delta}(\epsilon^{\lambda\kappa
\pi\sigma}D_{\lambda\kappa\pi\sigma})\ ,
\end{equation}
from which identities for $\epsilon^{\alpha\beta\gamma\sigma}D_{\mu\nu
\rho\sigma}$, $\epsilon^{\alpha\beta\rho\sigma}D_{\mu\nu\rho\sigma}$, and
$\epsilon^{\alpha\nu\rho\sigma}D_{\mu\nu\rho\sigma}$ can be deduced. As a
result, only operators involving the scalar combination $\epsilon^{\mu\nu
\rho\sigma}D_{\mu\nu\rho\sigma}$ need to be considered. In practice, the construction of all the possible operators then follows exactly the same step as for the scalar fields since $\epsilon^{\mu\nu
\rho\sigma}D_{\mu\nu\rho\sigma}$ and $\phi$ have the same Lorentz properties. The final list of operators need not be repeated here; it is simply obtained from that in Eqs.~(\ref{FPhi1}) and~(\ref{FPhi2}) by substituting $\phi\rightarrow\epsilon^{\mu\nu\rho\sigma}D_{\mu\nu\rho\sigma}$. 

\section{Equivalences via algebraic dualities\label{AlgDualities}}

When constructing the basis for the $p$-form fields in the previous Section, we only paid attention to the Lorentz structure of the fields, not to their dynamics. In Sec.~\ref{definitions}, we describe how kinetic terms can be constructed for $p$-form fields by generalizing the Proca or Maxwell Lagrangian. We also show how this dynamics, supplemented or not by a gauge symmetry, reduces the number of degrees of freedom down to the physical ones.

Once dealing with higher-form fields, there is another route to prescribe the dynamics, thereby leading to a different number of physical degrees of freedom. Indeed, a gauge-fixing Lagrangian term built on the Lorenz condition is nothing but a kinetic term for the Hodge dual field:%
\begin{equation}
\mathcal{L}_{gf}(A)=\frac{1}{2}d\star A\wedge\star d\star A=\frac{1}%
{2}F^{\star A}\wedge\star F^{\star A}=\mathcal{L}_{kin}(\star A)\ .
\end{equation}
The converse is obviously true: the usual kinetic term for a field $A$ can be written as a gauge fixing term for the dual field $\star A$.\ To be more explicit, consider a $p$-form field $A$. By definition, its Hodge dual $\star
A$ is the $n-p$-form field given by%
\begin{equation}
(\star A)_{\mu_{1}...\mu_{n-p}}=\frac{1}{p!}\epsilon_{\mu_{1}...\mu_{n-p}%
\nu_{1}...\nu_{p}}A^{\nu_{1}...\nu_{p}}\ ,
\end{equation}
and one can check that the Lorenz condition for $\star A$ gives back the field
strength for $A$:%
\begin{equation}
\partial^{\mu_{1}}(\star A)_{\mu_{1}...\mu_{n-p}}=\frac{1}{(p+1)!}%
\epsilon_{\mu_{1}...\mu_{n-p}\nu_{1}...\nu_{p}}F^{A,\mu_{1}\nu_{1}...\nu_{p}%
}\ .\label{LorenzKin}%
\end{equation}
In practice, this means that the kinetic terms of $p$-form field theories have two algebraically equivalent realizations:
\begin{subequations}
\begin{align}
\text{3-form theory} &  :\mathcal{L}_{kin}=\frac{1}{2}\partial_{\mu}(\star
C)^{\mu}\partial_{\nu}(\star C)^{\nu}=-\frac{1}{2}\frac{1}{4!}F_{\mu\nu
\rho\sigma}^{C}F^{C,\mu\nu\rho\sigma}\ \label{Equiv3},\\
\text{2-form theory} &  :\mathcal{L}_{kin}=-\frac{1}{2}\partial_{\mu}(\star
B)^{\mu\nu}\partial^{\rho}(\star B)_{\rho\nu}=\frac{1}{2}\frac{1}{3!}F_{\mu
\nu\rho}^{B}F^{B,\mu\nu\rho}\ \label{Equiv2},\\
\text{1-form theory} &  :\mathcal{L}_{kin}=\frac{1}{2}\frac{1}{2!}%
\partial_{\mu}(\star A)^{\mu\nu\rho}\partial^{\sigma}(\star A)_{\sigma\nu\rho
}=-\frac{1}{2}\frac{1}{2!}F_{\mu\nu}^{A}F^{A,\mu\nu}\ ,\\
\text{0-form theory} &  :\mathcal{L}_{kin}=-\frac{1}{2}\frac{1}{4!}%
\partial_{\mu}(\star\phi)^{\nu\rho\sigma\lambda}\partial^{\mu}(\star\phi
)_{\nu\rho\sigma\lambda}=\frac{1}{2}F_{\mu}^{\phi}F^{\phi,\mu}\ .
\end{align}
Adding mass terms for either $A$ or $\star A$ is totally equivalent since mass terms are invariant (up to the sign) under dualization, $A\wedge\star
A=-(\star A)\wedge\star(\star A)$. Let us stress that these relationships are purely mathematical. They do not truly represent dualities between theories but rather alternative notational representations. It is however important to keep these in mind as in the literature, the 3-form theory is often prescribed via its dual vector field as in Eq.~{\ref{Equiv3}}, or the 2-form field via Eq.~{\ref{Equiv2}}.

In the absence of mass terms, it must be remarked that the Lorenz kinetic term does have precisely the required gauge symmetry to match that manifest in the dual field-strength form. For example, imagine starting with a 1-form field $A^{\mu}$, with its four DoF. If we prescribe the usual Maxwell kinetic term, only two physical DoFs survive. This is not the case if we instead set its kinetic term as,
\end{subequations}
\begin{equation}
\mathcal{L}_{kin}=\frac{1}{2}\partial_{\mu}A^{\mu}\partial_{\nu}A^{\nu}\ .
\end{equation}
It is obviously not invariant under the usual gauge invariance $A_{\mu
}\rightarrow A_{\mu}+\partial_{\mu}\Lambda$ for generic $\Lambda$, but rather under
\begin{equation}
A_{\mu}\rightarrow A_{\mu}+\frac{1}{2!}\epsilon_{\mu\nu\rho\sigma}%
\partial^{\nu}\Lambda^{\rho\sigma}\ ,
\end{equation}
for any 2-form $\Lambda$. This is precisely the larger gauge invariance expected for a 3-form field since%
\begin{equation}
C_{\mu\nu\rho}\rightarrow C_{\mu\nu\rho}+\partial_{\mu}\Lambda_{\nu\rho
}+\partial_{\nu}\Lambda_{\rho\mu}+\partial_{\rho}\Lambda_{\mu\nu}%
\rightarrow(\star C)_{\mu}\rightarrow(\star C)_{\mu}+\frac{1}{2!}\epsilon
_{\nu\rho\sigma\mu}\partial^{\nu}\Lambda^{\rho\sigma}\ .
\end{equation}
Thus, the Lorenz kinetic term for a vector field actually describes a 3-form gauge field~\cite{Curtright:1980yj}, which has no remaining physical DoF. The converse is of course also true: a Lorenz kinetic term for a 3-form field
has the $C_{\mu\nu\rho}\rightarrow C_{\mu\nu\rho}+\epsilon_{\sigma\mu\nu\rho
}\partial^{\sigma}\Lambda$ symmetry. This is a smaller invariance, matching that of the 1-form gauge field $A_{\mu}\rightarrow A_{\mu}+\partial_{\mu}\Lambda$ with the usual Maxwell kinetic term. Since $A^{\mu}$ and $C^{\mu\nu\rho}$ both start with four degrees of freedom, this invariance leaves precisely the usual two transverse degrees of freedom, so the 3-form theory with the Lorenz kinetic term indeed corresponds to the 1-form theory with the usual Maxwell kinetic term.

Notice that this dualization is equally valid at the level of the propagator.
Under%
\begin{equation}
\mathcal{I}_{i}^{\mu_{1}...\mu_{p},\alpha_{1}...\alpha_{p}}\rightarrow
\mathcal{I}_{i,\rho_{1}...\rho_{n-p},\gamma_{1}...\gamma_{n-p}}=\frac
{\epsilon_{\mu_{1}...\mu_{p}\rho_{1}...\rho_{n-p}}}{p!}\frac{\epsilon
_{\alpha_{1}...\alpha_{p}\gamma_{1}...\gamma_{n-p}}}{p!}\mathcal{I}_{i}%
^{\mu_{1}...\mu_{p},\alpha_{1}...\alpha_{p}}\ ,
\end{equation}
we get
\begin{equation}
\mathcal{I}_{0}^{p}\rightarrow-\frac{(n-p)!}{p!}\mathcal{I}_{0}^{n-p}%
\ ,\ \mathcal{I}_{2}^{p}\rightarrow\frac{(n-p)!}{p!}\left(  \frac{n-p}%
{p}\mathcal{I}_{2}^{n-p}-\frac{k^{2}}{p}\mathcal{I}_{0}^{n-p}\right)  \ ,
\end{equation}
where the superscripts on the $\mathcal{I}_{i}$ invariants indicate their
dimensionality. With this, the propagator of a field is dualized as%
\begin{equation}
\mathcal{P}(A)=i\frac{(-1)^{p}p!}{k^{2}-m^{2}}\left(  \mathcal{I}_{0}%
^{p}-\frac{p}{m^{2}}\mathcal{I}_{2}^{p}\right)  \rightarrow\mathcal{P}(\star
A)=i\frac{(-1)^{p}(n-p)!}{m^{2}}\left(  \mathcal{I}_{0}^{n-p}-\frac{n-p}%
{k^{2}-m^{2}}\mathcal{I}_{2}^{n-p}\right)  \ ,
\end{equation}
which is precisely what one could derive directly starting from the Lorenz
kinetic term. Similarly for the massless propagator, dualizing the invariants
in Eq.~(\ref{Pmassless}) reproduces that of the Lorenz kinetic term
gauge-fixed via a $F^{\star A}\wedge\star F^{\star A}/2\xi$ term\footnote{In the earliest work on $p=2$ fields~\cite{Ogievetsky:1966eiu} (see also Ref.~\cite{Papaloucas:1989ku}), the kinetic term is actually written in the dual form $\partial_{\mu}B^{\mu\rho}\partial^{\nu}B_{\nu\rho}$, and the gauge is fixed by enforcing $\epsilon^{\mu\nu\rho\sigma}F^B_{\nu\rho\sigma} = 0$.}.

Returning to our effective operator bases, it is now clear that they are not
all independent of each other. For the scalar field, dualization is kind of
automatic, and we already showed that the basis for $D^{\mu\nu\rho\sigma}$ is
in one-to-one correspondence with that for $\phi$. The situation is more
interesting for 3-form fields. For instance, if we add to the operators
involving $C^{\mu\nu\rho}$ a Lorenz kinetic term $\partial_{\mu}C^{\mu\nu\rho
}\partial^{\sigma}C_{\sigma\nu\rho}$, and dualize the $C$ field into a
1-form field $C_{\mu\nu\rho}\rightarrow\epsilon_{\mu\nu\rho\sigma}A^{\sigma
}$, the whole effective theory matches onto that of the vector field. This
provides a powerful check of these operator bases, provided of course that all
the operators involving the Lorenz condition are kept. That is the reason why
we did so in the previous Section. For example, at the dimension five level,
\begin{align}
\bar{\psi}_{L}\psi_{R}\epsilon^{\mu\nu\rho\sigma}F_{\mu\nu\rho\sigma}^{C} &
\leftrightarrow\bar{\psi}_{L}\psi_{R}\partial^{\mu}A_{\mu}\ ,\\
\bar{\psi}_{L}\sigma^{\mu\nu}\psi_{R}\partial^{\rho}C_{\mu\nu\rho} &
\leftrightarrow\bar{\psi}_{L}\sigma^{\mu\nu}\psi_{R}F_{\mu\nu}^{A}\ ,
\end{align}
and similarly for the other couplings to SM fields in Eqs.~(\ref{SMA})
and~(\ref{SMC}). This also explains why the $C$ field has no dimension-six
gauge invariant operator (apart from an extra photon field). It is a
manifestation of the fact that only gauge invariant operators exist for the
1-form field, and those all get mapped onto Lorenz operators for the
3-form field.

Similarly, for the 2-form field, the basis must be self-dual under
$B^{\mu\nu}\leftrightarrow\tilde{B}^{\mu\nu}$ in the sense that the operators
of a given dimension can only get reorganized. Most operators are by
themselves self-dual thanks to Dirac matrices identities like Eq.~(\ref{Id2}) or
Eq.~(\ref{Id4}), while field strength and Lorenz condition operators get
interchanged,%
\begin{align}
\bar{\psi}_{L}\gamma_{\sigma}\psi_{L}\epsilon^{\mu\nu\rho\sigma}F_{\mu\nu\rho
}^{B} &  \leftrightarrow\bar{\psi}_{L}\gamma^{\nu}\psi_{L}\partial^{\mu}%
B_{\mu\nu}\ ,\\
\bar{\psi}_{L}\sigma^{\mu\nu}\psi_{R}\partial^{\rho}F_{\mu\nu\rho}^{B} &
\leftrightarrow\bar{\psi}_{L}\sigma^{\mu\nu}\psi_{R}\partial_{\mu}%
\partial^{\rho}B_{\nu\rho}\ .
\end{align}
A peculiarity of the $B$ field is to have both a $m^{2}B_{\mu\nu}B^{\mu\nu}$
and a $\tilde{m}^{2}B_{\mu\nu}\tilde{B}^{\mu\nu}$ mass terms, and we have seen
before that when both are present, the Lorenz condition must be generalized to
include a term proportional to the dual of the field strength, see
Eq.~(\ref{GenLorenz}). This can be easily understood on the basis of
Eq.~(\ref{LorenzKin}). Indeed, the $\tilde{m}^{2}$ term could be absorbed
entirely into the $m^{2}$ term upon a reparametrization $B\rightarrow
B+\lambda\star B$ for some $\lambda$, but this would split the kinetic term
into a combination of $F^{B}\wedge\star F^{B}$ and $d\star B\wedge\star d\star
B$, from which the generalized Lorenz condition Eq.~(\ref{GenLorenz}) would emerge.

While these dualities are very useful as cross-checks for the operator bases,
they do not bring much phenomenologically. In the following, we will always
assume that the operators are accompanied by the usual field strength kinetic terms.

\section{Equivalences via massless dualities}
\label{masslessD}

For massless theories, dualities between a $p$-gauge field $A$ and a $n-p-2$
gauge field $A^{\star}$ can be obtained by dualizing their field strengths.
Indeed, using the properties of the wedge product, the usual kinetic term can
be rewritten as%
\begin{equation}
\mathcal{L}_{kin}(F=dA)=F\wedge\star F=-(\star F)\wedge\star(\star
F)=\mathcal{L}_{kin}^{\star}(\star F=dA^{\star})\ .
\end{equation}
Beware that here, $A^{\star}$ is not the dual to $A$, which would be a $n-p$
gauge field. Instead, dualizing the field strength in the absence of sources
interchanges EoM and Bianchi identity. Specifically, the EoM derived from
$\mathcal{L}_{kin}$ is $d\star F=0$ and the Bianchi identity is $dF=0$, while
the dual theory $\mathcal{L}_{kin}^{\star}$ has the EoM $d\star(\star
F)=dF=0$, and the Bianchi identity $d(\star F)=d\star F=0$. This ensures that
there exists a gauge field $A^{\star}$ such that $\star F=dA^{\star}$, but it
does not tell us how it is related to $A$. In practice, the only nontrivial
dualizations in $n=4$ dimensions are that interchanging $F_{\mu\nu}^{A}$ and
$\tilde{F}_{\mu\nu}^{A}$, which corresponds to the well-known electromagnetic
duality of the Maxwell theory in vacuum, and that relating the massless scalar
$F_{\mu}^{\phi}$ to the tensor $F_{\mu\nu\rho}^{B}$, which has the same
number of degrees of freedom, see Table~\ref{TableDoF}. These cases are
detailed in Sec.~\ref{massless1} and~\ref{massless2} below, while in
Sec.~\ref{massless3}, we show what happens if one tries to dualize the
3-form field strength.

\subsection{Equivalence in the massless 1-form model\label{massless1}}

To set the stage, let us discuss the duality $F_{\mu\nu}^{A}\leftrightarrow
\tilde{F}_{\mu\nu}^{A}$ in a way that can immediately be generalized to the
$0$- and $2$-form duality. The idea is to start from a parent Lagrangian (see e.g. App. B.4 in Ref.~\cite{Polchinski:1998rr})%
\begin{equation}
\mathcal{L}_{\mathrm{parent}}(F,\tilde{A})=-\frac{1}{2}\frac{1}{2!}F_{\mu\nu
}F^{\mu\nu}+\frac{1}{2!}\frac{1}{2!}\epsilon_{\mu\nu\rho\sigma}\tilde
{A}^{\sigma}\partial^{\mu}F^{\nu\rho}\ \ ,
\end{equation}
where $F$ and $\tilde{A}$ are fundamental fields of mass dimension 2 and
1, respectively (the tilde on $A$ has no particular meaning, it is just a
notation). The key feature is the mixing term, which can make either $\tilde
{A}$ or $F$ auxiliary under partial integration since $\epsilon_{\mu\nu
\rho\sigma}\tilde{A}^{\sigma}\partial^{\mu}F^{\nu\rho}\leftrightarrow
-\tilde{F}_{\mu\nu}^{\tilde{A}}F^{\mu\nu}$.

A first point of view on this Lagrangian is that $\tilde{A}$ is a Lagrange
multiplier. It is not propagating since it has no kinetic term. All it does is imposing $\epsilon_{\mu\nu\rho\sigma}\partial^{\mu}F^{\nu\rho}=0$ via its
EoM, that is, a Bianchi identity for $F$. Barring topological obstructions, if
$F$ is closed it is also exact, and $F$ must be the field strength of some
vector field, $F=F^{A}=dA$. Then,
\begin{equation}
\mathcal{L}_{\mathrm{parent}}(F=F^{A},0)=-\frac{1}{2}\frac{1}{2!}F_{\mu\nu
}^{A}F^{A,\mu\nu}\ ,\ \ F_{\mu\nu}^{A}=\partial_{\mu}A_{\nu}-\partial_{\nu
}A_{\mu}\ ,\label{dualA}%
\end{equation}
which is the usual Maxwell Lagrangian for the gauge field $A$. Alternatively,
treating $F$ as auxiliary after integrating by part, its EoM is
$F=-\tilde{F}^{\tilde{A}}$. With this, we find again the Maxwell Lagrangian,
but in terms of $F^{\tilde{A}}$:%
\begin{equation}
\mathcal{L}_{\mathrm{parent}}(F(\tilde{A}),\tilde{A})=-\frac{1}{2}\frac{1}%
{2!}F_{\mu\nu}^{\tilde{A}}F^{\tilde{A},\mu\nu}\ ,\ \ F_{\mu\nu}^{\tilde{A}%
}=\partial_{\mu}\tilde{A}_{\nu}-\partial_{\nu}\tilde{A}_{\mu}\ .\label{dualG}%
\end{equation}
The $A$ and $\tilde{A}$ formulations are totally equivalent. In this case,
their Lagrangian even have the same form because both $A$ and $\tilde{A}$ are
1-form fields. All this is nothing but the usual electromagnetic duality,
i.e., the fact that Maxwell's equations in vacuum are symmetric under the
exchange of electric and magnetic fields, $E\rightarrow B$, $B\rightarrow-E$,
that is, under $F_{\mu\nu}^{\gamma}\leftrightarrow\tilde{F}_{\mu\nu}^{\gamma}$
for the photon field strength.

In the presence of interactions, the situation is more complicated. Duality
interchanges Bianchi identity and EoM, with $\partial_{\mu}F^{\mu\nu}%
=\epsilon_{\mu\nu\rho\sigma}\partial^{\nu}\tilde{F}^{\rho\sigma}=0$ and
$\partial_{\mu}\tilde{F}^{\mu\nu}=\epsilon_{\mu\nu\rho\sigma}\partial^{\nu
}F^{\rho\sigma}=0$. Interactions break this pattern since $\partial_{\mu
}F^{\mu\nu}=J^{\nu}$. Yet, the above formulation can accommodate for some
effective interactions, so let us see what happens in that case. First, since
it is $F$ that starts as fundamental, only interactions of the form $F_{\mu
\nu}J^{\mu\nu}/2!$ can be added, with $J^{\mu\nu}$ encoding effective
interactions like $F^{\gamma,\mu\nu}$, $\bar{\psi}_{L}\sigma^{\mu\nu}\psi_{R}%
$, $\bar{\psi}_{L}\gamma^{\mu}\mathcal{D}^{\nu}\psi_{L}$, $\partial^{\mu}%
(\bar{\psi}_{L}\gamma^{\nu}\psi_{L})$, etc , i.e., all those present in the
basis of Eq.~(\ref{FB1}). This is in accordance with the fact that SM fields
would all be neutral under the dark gauge symmetry\footnote{Consequently, remember that the covariant derivative acting on $\psi$ may contain only SM gauge fields.}. Then, eliminating $G$
proceeds as before, simply replacing $F_{\mu\nu}J^{\mu\nu}\rightarrow
F_{\mu\nu}^{A}J^{\mu\nu}$. Eliminating $F$, on the other hand, is affected by
the presence of $J$. The EoM becomes $F=-\tilde{F}^{\tilde{A}}+J$, which when
plugged back into $\mathcal{L}_{\mathrm{parent}}(F=-\tilde{F}^{\tilde{A}%
}+J,\tilde{A})$, generates the dual interactions $\tilde{F}_{\mu\nu}%
^{\tilde{A}}J^{\mu\nu}$ together with a whole series of contact terms
$J_{\mu\nu}J^{\mu\nu}$~\cite{Dalmazi:2011df}.

Explicitly, the effective interactions when $F=F^{A}$, $\tilde{A}=0$ and those
when $F=-\tilde{F}^{\tilde{A}}+J$ are related as
\begin{subequations}
\begin{align}
F_{\mu\nu}^{A}F_{Y}^{\mu\nu} &  \rightarrow F_{\mu\nu}^{\tilde{A}}\tilde
{F}_{Y}^{\mu\nu}\ ,\\
F_{\mu\nu}^{A}\tilde{F}_{Y}^{\mu\nu} &  \rightarrow F_{\mu\nu}^{\tilde{A}%
}F_{Y}^{\mu\nu}\ ,\\
\bar{\psi}_{L}\sigma^{\mu\nu}\psi_{R}F_{\mu\nu}^{A} &  \rightarrow i\bar{\psi
}_{L}\sigma^{\mu\nu}\psi_{R}F_{\mu\nu}^{\tilde{A}}\ ,\\
\bar{\psi}_{L}\gamma^{\mu}\mathcal{D}^{\nu}\psi_{L}F_{\mu\nu}^{A} &
\rightarrow\bar{\psi}_{L}\gamma^{\mu}\mathcal{D}^{\nu}\psi_{L}\tilde{F}%
_{\mu\nu}^{\tilde{A}}\ \label{opcovD},\\
\bar{\psi}_{L}\gamma^{\nu}\psi_{L}\partial^{\mu}\tilde{F}_{\mu\nu}^{A} &
\rightarrow\bar{\psi}_{L}\gamma^{\nu}\psi_{L}\partial^{\mu}F_{\mu\nu}%
^{\tilde{A}}\ ,\\
\bar{\psi}_{L}\gamma^{\nu}\psi_{L}\partial^{\mu}F_{\mu\nu}^{A} &
\rightarrow\bar{\psi}_{L}\gamma^{\nu}\psi_{L}\partial^{\mu}\tilde{F}_{\mu\nu
}^{\tilde{A}}\ .
\end{align}
For the third operator, the appearance of the $i$ factor comes
from spinor identities, see Eqs.~(\ref{Id1}) to~(\ref{Id4}). In practice, this
swaps magnetic and electric interactions, including the magnetic and electric
dipole operators. For the other operators, remember that $\partial^{\mu}%
\tilde{F}_{\mu\nu}=0$ when $F=dA$ and $\partial^{\mu}\tilde{F}_{\mu\nu
}^{\tilde{A}}=0$ when $F^{\tilde{A}}=d\tilde{A}$, while $\partial^{\mu}%
F_{\mu\nu}=\partial^{\mu}F_{\mu\nu}^{\tilde{A}}=0$ on shell. This means in
particular that the kinetic mixings are unrelated in those two formulations.

In the effective basis, we also derived many operators involving pairs of dark
states. Those operators cannot be put in the form $F_{\mu\nu}J^{\mu\nu}/2!$.
Yet, when the dark vector field is external, whether one computes physical
observables in terms of $A$ or $\tilde{A}$ always gives the same results. For
example, consider the process $A\rightarrow J$ derived from $F_{\mu\nu}%
^{A}J^{\mu\nu}$ and $\tilde{A}\rightarrow J$ from $\tilde{F}_{\mu\nu}%
^{\tilde{A}}J^{\mu\nu}$. The amplitudes have the forms
\end{subequations}
\begin{align}
\mathcal{M}(A &  \rightarrow J)=\varepsilon_{\alpha}^{(\lambda)}(k^{\mu}%
g^{\nu\alpha}-k^{\nu}g^{\mu\alpha})J_{\mu\nu}\ ,\\
\mathcal{M}(\tilde{A} &  \rightarrow J)=\tilde{\varepsilon}_{\alpha}%
^{(\lambda)}k_{\gamma}\epsilon^{\alpha\gamma\mu\nu}J_{\mu\nu}\ ,
\end{align}
and, using $\sum_{\lambda}\varepsilon_{\alpha}^{(\lambda)}\varepsilon_{\beta
}^{\ast(\lambda)}=\sum_{\lambda}\tilde{\varepsilon}_{\alpha}^{(\lambda)}%
\tilde{\varepsilon}_{\beta}^{\ast(\lambda)}=-g_{\alpha\beta}$ since the
$k_{\alpha}k_{\beta}$ part cancels out,
\begin{equation}
\sum_{\lambda}|\mathcal{M}(\tilde{A}\rightarrow J)|^{2}=\sum_{\lambda
}|\mathcal{M}(A\rightarrow J)|^{2}+2k^{2}J_{\mu\nu}J^{\mu\nu}\ .
\end{equation}
On shell, $k^{2}=0$ and both expressions coincide. Clearly, this remains true
even if more than one dark photon is present. However, dividing this equation
by $k^{2}+i\varepsilon$, it relates the $J\rightarrow\tilde{A}\rightarrow J$
and $J\rightarrow A\rightarrow J$ amplitudes since the $\xi$ dependent part of
the propagator cancels out when the couplings to $J^{\mu\nu}$ are gauge
invariant. Off shell, both then differ by the $J_{\mu\nu}J^{\mu\nu}$ contact
term, in accordance with our earlier finding.

It should be understood that duality no longer produces equivalent theories in
the presence of interactions. Phenomenologically, this has several
consequences. First, whether the effective interactions are understood in
terms of $F^{A}$ or its dual does not change their forms, but mixes up their
$CP$ properties. This is particularly relevant for the kinetic mixing term.
Indeed, if one could justify that $F_{\mu\nu}\tilde{F}_{Y}^{\mu\nu}$ is
initially absent by imposing some $CP$ properties on $F_{\mu\nu}J^{\mu\nu}$,
while still allowing for the antisymmetric mixing term, then the dual theory
ends up with no kinetic mixing at all. This offers an alternative realization
of the dark photon scenario in which the dominant couplings would be the
$\bar{\psi}_{L}\sigma^{\mu\nu}\psi_{R}F_{\mu\nu}^{A}$ couplings. Notice that
the kinetic mixing is by essence off shell, so it is consistent for it to be
intrinsically different in both realizations.

A second consequence is that once nonrenormalizable effective interactions
with SM fields are present, consistency requires the presence of effective
interactions among SM fields only. For example, if the dimension-five
$\bar{\psi}_{L}\sigma^{\mu\nu}\psi_{R}F_{\mu\nu}^{A}/\Lambda$ coupling is
present, the fact that we do not know whether it should instead be interpreted
as $\bar{\psi}_{L}\sigma^{\mu\nu}\psi_{R}F_{\mu\nu}^{\tilde{A}}/\Lambda$ means
that we should also include the dimension-six contact interaction $\bar{\psi
}_{L}\sigma^{\mu\nu}\psi_{R}\bar{\psi}_{L}\sigma_{\mu\nu}\psi_{R}/\Lambda^{2}%
$. In some sense, this is expected since once a complete UV theory at some
scale $\Lambda$ is able to produce $\bar{\psi}_{L}\sigma^{\mu\nu}\psi
_{R}F_{\mu\nu}^{\tilde{A}}$ interactions, there is no reason not to expect it
to also generate this contact term.

\subsection{Equivalence between the massless 0- and 2-form
models\label{massless2}}

There are many ways to express the massless $p=2$ gauge theory in terms of a
scalar field. Here, let us derive it following the same steps as in the
previous Section, starting with the parent Lagrangian
\begin{equation}
\mathcal{L}_{\mathrm{parent}}(\phi,F)=\frac{1}{2}\frac{1}{3!}F_{\mu\nu\rho
}F^{\mu\nu\rho}-\frac{1}{3!}\epsilon_{\mu\nu\rho\sigma}\phi\partial^{\mu
}F^{\nu\rho\sigma}\ \ .\label{ParentMassless}%
\end{equation}
At this level, $F_{\mu\nu\rho}$ is a fundamental antisymmetric tensor field of
mass dimension 2, while $\phi$ is a scalar field. Treating $\phi$ as a
Lagrange multiplier, its equation of motion enforces $\epsilon_{\mu\nu
\rho\sigma}\partial^{\mu}F^{\nu\rho\sigma}=0$, i.e., that $F$ has to satisfy
the Bianchi identity $dF=0$. Being closed and barring topological issues, $F$
is also exact and $F\rightarrow F^{B}=dB$ with $B$ a 2-form field. Thus,
after eliminating $\phi$,
\begin{equation}
\mathcal{L}_{\mathrm{parent}}(0,F=dB)=\frac{1}{2}\frac{1}{3!}F_{\mu\nu\rho
}^{B}F^{B,\mu\nu\rho}\ ,\ \ F_{\mu\nu\rho}^{B}=\partial_{\mu}B_{\nu\rho
}+\partial_{\nu}B_{\rho\mu}+\partial_{\rho}B_{\mu\nu}\ ,
\end{equation}
which is the massless $p=2$ gauge theory. The opposite situation follows from
first integrating by part the second term, upon which $F^{\nu\rho\sigma}$
becomes an auxiliary field. Integrating it out gives%
\begin{equation}
\mathcal{L}_{\mathrm{parent}}(\phi,F(\phi))=\frac{1}{2}F_{\mu}^{\phi}%
F^{\phi,\mu}\ ,\ \ F_{\mu}^{\phi}=\partial_{\mu}\phi\ ,\label{EoMmassless}%
\end{equation}
which is the usual massless scalar theory. This proves the duality between the
massless 0- and 2-form theories, in the same sense as the duality between
Eqs.~(\ref{dualA}) and~(\ref{dualG}).

Let us proceed by adding effective operators to the parent Lagrangian. As
discussed in the previous section, the EoM and the Bianchi identity can be
dual only provided the fields are massless and free. Yet, there is no real
obstruction to simply adding effective operators to the parent Lagrangian of Eq.~(\ref{ParentMassless}), provided we work to a given order, stick to
operators involving only one dark field, and integrate all these effective
interactions by part to put them in a suitable algebraic form. Further, at the
parent level, $F_{\mu\nu\rho}$ is a generic 3-form field (of mass
dimension two) and not the $B$ field strength yet. Thus, the relevant
operators to add are those of the $C$ field basis, with $C_{\mu\nu\rho
}\rightarrow F_{\mu\nu\rho}$, so that
\begin{equation}
\mathcal{L}_{\mathrm{parent}}^{eff}(\phi,F)=\mathcal{L}_{\mathrm{parent}}%
(\phi,F)+\frac{1}{3!}F_{\mu\nu\rho}J^{\mu\nu\rho}\ ,\label{parentmassless1}%
\end{equation}
with (in the present Section, for simplicity, we keep only the operators
appearing in Eq.~(\ref{FC1}), and leave out all those involving
$L\leftrightarrow R$ fields)
\begin{equation}
F_{\mu\nu\rho}J^{\mu\nu\rho}=\frac{c_{1}}{\Lambda}\bar{\psi}_{L}\gamma
_{\sigma}\psi_{L}\epsilon^{\mu\nu\rho\sigma}F_{\mu\nu\rho}-\frac{3c_{2}%
}{\Lambda^{2}}\bar{\psi}_{L}\sigma^{\mu\nu}\psi_{R}\partial^{\rho}F_{\mu
\nu\rho}-\frac{c_{3}}{\Lambda^{2}}\bar{\psi}_{L}\psi_{R}\epsilon^{\sigma\mu
\nu\rho}\partial_{\sigma}F_{\mu\nu\rho}+\mathcal{O}(\Lambda^{-3})\ .
\end{equation}
The last two operators have to be integrated by part to keep $F$ auxiliary.
Notice that compared to simply taking the operators involving $F^{B}$ in the
$B$ field basis, there is the additional operator with $c_{3}$, which can be
written $\bar{\psi}_{L}\psi_{R}\epsilon^{\sigma\mu\nu\rho}\partial_{\sigma
}F_{\mu\nu\rho}$. This is irrelevant if we integrate $\phi$ out since we then
get back the Bianchi identity $\epsilon_{\mu\nu\rho\sigma}\partial^{\mu}%
F^{\nu\rho\sigma}=0$ and $F\rightarrow F^{B}$. This extra operator cancels out
and we recover exactly the type I operators of Eq.~(\ref{FB1}):%
\begin{equation}
\mathcal{L}_{\mathrm{parent}}^{eff}(0,F=dB)=\frac{1}{2}\frac{1}{3!}F_{\mu
\nu\rho}^{B}F^{B,\mu\nu\rho}+\frac{1}{3!}\frac{c_{1}}{\Lambda}\bar{\psi}%
_{L}\gamma_{\sigma}\psi_{L}\epsilon^{\mu\nu\rho\sigma}F_{\mu\nu\rho}^{B}%
-\frac{1}{2!}\frac{c_{2}}{\Lambda^{2}}\bar{\psi}_{L}\sigma^{\mu\nu}\psi
_{R}\partial^{\rho}F_{\mu\nu\rho}^{B}\ .\label{BpDual1}%
\end{equation}
If instead we integrate $F^{\alpha\beta\gamma}$ out, the EoM in
Eq.~(\ref{EoMmassless}) receives an extra term $J^{\alpha\beta\gamma}$, and
the effective Lagrangian becomes%
\begin{equation}
\mathcal{L}_{\mathrm{parent}}^{eff}(\phi,F(\phi))=\frac{1}{2}\partial^{\alpha
}\phi\partial_{\alpha}\phi+\frac{c_{1}}{\Lambda}\bar{\psi}_{L}\gamma^{\mu}%
\psi_{L}F_{\mu}^{\phi}+\frac{c_{3}}{\Lambda^{2}}\bar{\psi}_{L}\psi_{R}%
\partial^{\mu}F_{\mu}^{\phi}+\frac{1}{2}\frac{c_{1}^{2}}{\Lambda^{2}}\bar
{\psi}_{L}\gamma_{\mu}\psi_{L}\bar{\psi}_{L}\gamma^{\mu}\psi_{L}%
+\mathcal{O}(\Lambda^{-3})\ .\label{BpDual2}%
\end{equation}
This time, it is the $c_{2}$ operator that cancels out trivially because it
ends up proportional to $\bar{\psi}_{L}\sigma^{\mu\nu}\psi_{R}\partial_{\mu
}\partial_{\nu}\phi$, and we recover the same two type I effective operators
as in the $\phi$ basis of Eq.~(\ref{FPhi1}). The presence of the contact term
is to be noted though: Duality cannot work without a complete basis of
operators, including those that do not involve the dark state. Notice finally
that $c_{2}$ and $c_{3}$ disappear on shell since $\phi$ and $B$ are massless,
leaving only $c_{1}$ in both cases.

To describe the opposite situation of starting with the scalar effective
operators, we must consider the parent Lagrangian%
\begin{equation}
\mathcal{L}_{\mathrm{parent}}^{eff}(F,B)=\frac{1}{2}F^{\mu}F_{\mu}-\frac
{1}{3!}\epsilon_{\mu\nu\rho\sigma}F^{\mu}F^{B,\nu\rho\sigma}+F_{\mu}J^{\mu
}\ \ ,\label{parentmassless2}%
\end{equation}
where $F^{\mu}$ is a generic vector field of mass dimension two, thus with the
effective interactions taken from the $A$ field basis,%
\begin{equation}
F_{\mu}J^{\mu}=\frac{c_{1}}{\Lambda}\bar{\psi}_{L}\gamma^{\mu}\psi_{L}F_{\mu
}+\frac{c_{2}}{\Lambda^{2}}\bar{\psi}_{L}\sigma^{\mu\nu}\psi_{R}(\partial
_{\mu}F_{\nu}-\partial_{\nu}F_{\mu})+\frac{c_{3}}{\Lambda^{2}}\bar{\psi}%
_{L}\psi_{R}\partial^{\mu}F_{\mu}+\mathcal{O}(\Lambda^{-3})\ .
\end{equation}
Compared to the scalar effective operators, we again have one more operator in
the form of $c_{2}$. Integrating $B$ out, its EoM implies $\epsilon_{\mu
\nu\rho\sigma}\partial^{\mu}F^{\nu}=0$, hence $F_{\mu}=\partial_{\mu}%
\phi=F_{\mu}^{\phi}$ (barring topological obstruction). The extra operator
$c_{2}$ disappears thanks to the Bianchi identity $\epsilon^{\mu\nu
\rho\sigma}\partial_{\nu}F_{\mu}^{\phi}$, and the 0-form field effective
theory is correctly reproduced%
\begin{equation}
\mathcal{L}_{\mathrm{parent}}(F=d\phi,0)=\frac{1}{2}\partial^{\alpha}%
\phi\partial_{\alpha}\phi+\frac{c_{1}}{\Lambda}\bar{\psi}_{L}\gamma^{\mu}%
\psi_{L}F_{\mu}^{\phi}+\frac{c_{3}}{\Lambda^{2}}\bar{\psi}_{L}\psi_{R}%
\partial^{\mu}F_{\mu}^{\phi}+\mathcal{O}(\Lambda^{-3})\ .\label{BpDual3}%
\end{equation}
If instead we integrate $F_{\mu}$ out, its EoM is algebraic and once plugged
back in the Lagrangian, we find%
\begin{align}
\mathcal{L}_{\mathrm{parent}}^{eff}(F(B),B) &  =\frac{1}{2}\frac{1}{3!}%
F_{\mu\nu\rho}^{B}F^{B,\mu\nu\rho}+\frac{1}{3!}\frac{c_{1}}{\Lambda}\bar{\psi
}_{L}\gamma_{\mu}\psi_{L}\epsilon^{\mu\nu\rho\sigma}F_{\nu\rho\sigma}%
^{B}-\frac{1}{2!}\frac{c_{2}}{\Lambda^{2}}\bar{\psi}_{L}\sigma^{\mu\nu}%
\psi_{R}\partial^{\rho}F_{\mu\nu\rho}^{B}\nonumber\\
&  \ \ \ -\frac{1}{2}\frac{c_{1}^{2}}{\Lambda^{2}}\bar{\psi}_{L}\gamma_{\mu
}\psi_{L}\bar{\psi}_{L}\gamma^{\mu}\psi_{L}+\mathcal{O}(\Lambda^{-3}%
)\ ,\label{BpDual4}%
\end{align}
where the $c_{3}$ operator disappears upon enforcing the Bianchi identity
$\epsilon^{\mu\nu\rho\sigma}\partial_{\mu}F_{\nu\rho\sigma}^{B}$. The results
Eqs.~(\ref{BpDual1}) and~(\ref{BpDual2}) are manifestly consistent with
Eqs.~(\ref{BpDual3}) and~(\ref{BpDual4}), as could have been expected since
the effective currents of the parent Lagrangian Eq.~(\ref{parentmassless1})
and~(\ref{parentmassless2}) are dual to each other, $J_{\mu\nu\rho
}=\epsilon_{\mu\nu\rho\sigma}J^{\sigma}$. Also, if we imagine that the
four-fermion operator is initially already present in either Lagrangian, both
treatments shift its coupling strength in opposite directions, making the
final realizations consistent with each other.

As for the dark photon, the duality between the scalar and 2-form fields holds
more generally provided these states are kept on shell and external since they
are then essentially free. For instance, if we dualize $F_{\mu\nu\rho}%
^{B}J^{\mu\nu\rho}=3\partial_{\mu}B_{\nu\rho}J^{\mu\nu\rho}$ into $F^{\phi
,\mu}\epsilon_{\mu\nu\rho\sigma}J^{\nu\rho\sigma}$, the $B\rightarrow J$
and $\phi\rightarrow J$ amplitudes%
\begin{align}
\mathcal{M}(B &  \rightarrow J)=3k^{\mu}\varepsilon_{(\lambda)}^{\nu\rho
}J_{\mu\nu\rho}\ ,\\
\mathcal{M}(\phi &  \rightarrow J)=-k^{\sigma}\varepsilon_{\mu\nu\rho\sigma
}J^{\mu\nu\rho}\ ,
\end{align}
are related as
\begin{equation}
\sum_{\lambda}|\mathcal{M}(B\rightarrow J)|^{2}=|\mathcal{M}(\phi\rightarrow
J)|^{2}+3!k^{2}J_{\mu\nu\rho}J^{\mu\nu\rho}\ ,\label{SpinBphi}%
\end{equation}
where we used that $\sum_{\lambda}\varepsilon_{(\lambda)}^{\ast\alpha\beta
}\varepsilon_{(\lambda)}^{\mu\nu}=2\mathcal{I}_{0}^{\alpha\beta,\mu\nu}$ since
the $\mathcal{I}_{2}^{\alpha\beta,\mu\nu}$ component cancels out by gauge
invariance. Exactly like in the previous Section, the scalar and 2-form
duality $\epsilon_{\mu\nu\rho\sigma}F^{\phi,\sigma}\leftrightarrow
F_{\mu\nu\rho}^{B}$ holds on shell, where $k^{2}=0$, even for interactions
involving more than one dark state, but contact terms are necessary off shell.

These contact terms also take on a new role compared to that in the previous
Section. There, the current was defined at the level of the field strength,
with $F_{\mu\nu}^{A}J^{\mu\nu}$ associated to $F_{\mu\nu}^{\tilde{A}}\tilde
{J}^{\mu\nu}$. These currents need not be conserved for gauge invariance.
Instead, after partial integration, we get $A_{\nu}\partial_{\mu}J^{\mu\nu}$
associated to $\tilde{A}_{\nu}\partial_{\mu}\tilde{J}^{\mu\nu}$. It is these
divergences $\partial_{\mu}J^{\mu\nu}$ and $\partial_{\mu}\tilde{J}^{\mu\nu}$
that need to be conserved to preserve gauge invariance, and they trivially are
since $\partial_{\mu}\partial_{\nu}J^{\mu\nu}=\partial_{\mu}\partial_{\nu
}\tilde{J}^{\mu\nu}=0$. The same happens starting with the $F_{\mu\nu\rho}%
^{B}J^{\mu\nu\rho}$ coupling, which is equivalent to $-3B_{\nu\rho}%
\partial_{\mu}J^{\mu\nu\rho}$ after partial integration, but the scalar field
breaks this pattern. While $p$-form fields are built on gauge-invariance, all
that remains for $p=0$ is a constant shift symmetry.\ But dualizing as
$J_{\mu\nu\rho}=\epsilon_{\mu\nu\rho\sigma}J^{\sigma}$, the shift symmetry of
$F_{\mu}^{\phi}\epsilon^{\mu\nu\rho\sigma}J_{\nu\rho\sigma}$ is lost after
partial integration into $\phi\epsilon^{\mu\nu\rho\sigma}\partial_{\mu}%
J_{\nu\rho\sigma}$. Intuitively, a constant shift of $\phi$ prevents it from
vanishing at infinity, so it is not really surprising that partial integration
can be problematic.

The only way for the shift symmetry to remain active is for $\epsilon^{\mu
\nu\rho\sigma}\partial_{\mu}J_{\nu\rho\sigma}$ to vanish, in which case $\phi$
decouples entirely. Then, setting $\mathcal{M}(\phi\rightarrow J)=0$ in
Eq.~(\ref{SpinBphi}), it predicts that $B$ must also decouple entirely at
$k^{2}=0$ even though the coupling $B_{\nu\rho}\partial_{\mu}J^{\mu\nu\rho}$
does not vanish! Let us illustrate this phenomenon with an example. Imagine
that $\bar{\psi}_{L}\gamma^{\mu}\psi_{L}F_{\mu}^{\phi}$ is accompanied by
$\bar{\psi}_{R}\gamma^{\mu}\psi_{R}F_{\mu}^{\phi}$, summing up in the
vectorial combination $\bar{\psi}\gamma^{\mu}\psi F_{\mu}^{\phi}$. Such a
coupling is spurious since after partial integration, $\phi$ decouples as
$\partial_{\mu}(\bar{\psi}\gamma^{\mu}\psi)=0$. Yet, the dual $B$ amplitude is
induced by $\bar{\psi}\gamma_{\mu}\psi\epsilon^{\mu\nu\rho\sigma}F_{\nu
\rho\sigma}^{B}$ which does not vanish. An explicit calculation shows that
Eq.~(\ref{SpinBphi}) still holds, with the amplitude squared for $B\rightarrow
J$ proportional to $k^{2}$. In other words, the occurrence of the contact term
times $k^{2}$ in Eq.~(\ref{SpinBphi}) is here needed to maintain consistency
when the $\phi\rightarrow J$ process is trivially absent.

\subsection{Triviality of the massless 3-form model\label{massless3}}

As said at the beginning, dualizing field strengths relates $p$ and
$n-p-2$-gauge field models. In $d=4$, neither could be 3-form fields, so
one may think that model is simply independent of all the others. In fact,
this impossibility is a manifestation of the triviality of massless three-form
fields. To see this, notice that it is actually possible to write down a
parent Lagrangian for a 3-form field as%
\begin{equation}
\mathcal{L}_{\mathrm{parent}}^{eff}(F,C)=-\frac{1}{2}FF+\frac{1}{4!}%
\epsilon_{\mu\nu\rho\sigma}FF^{C,\mu\nu\rho\sigma}\ ,
\end{equation}
where $F$ is of mass dimension two, and $F^{C}=dC$ with $C$ a 3-form field
of mass dimension one. Integrating $F$ out gives back the 3-form kinetic
term%
\begin{equation}
\mathcal{L}_{\mathrm{parent}}^{eff}(F(C),C)=-\frac{1}{2}\frac{1}{4!}F_{\mu
\nu\rho\sigma}^{C}F^{C,\mu\nu\rho\sigma}\ .
\end{equation}
However, writing $\epsilon_{\mu\nu\rho\sigma}FF^{C,\mu\nu\rho\sigma}%
=-\epsilon_{\mu\nu\rho\sigma}\partial^{\mu}FC^{\nu\rho\sigma}$, $C$ becomes a
Lagrange multiplier and imposes $\partial^{\mu}F=0$, i.e., $F=F_{0}$ must be
constant,
\begin{equation}
\mathcal{L}_{\mathrm{parent}}^{eff}(F=F_{0},0)=-\frac{1}{2}F_{0}^{2}\ .
\end{equation}
Thus, the fact that a 3-form field is not paired with any other field
under field-strength dualization implies that it does not have any dynamics at
all, in agreement with the counting done in Table~\ref{TableDoF}.

This conclusion remains true if we add interactions in the form of $FJ$.
Integrating $C$ out still imposes $F=F_{0}$, but integrating $F$ out now
produces%
\begin{equation}
\mathcal{L}_{\mathrm{parent}}^{eff}(F(C),C)=-\frac{1}{2}\frac{1}{4!}F_{\mu
\nu\rho\sigma}^{C}F^{C,\mu\nu\rho\sigma}+\frac{1}{4!}J\epsilon_{\mu\nu
\rho\sigma}F^{C,\mu\nu\rho\sigma}+\frac{J^{2}}{2}\ .
\end{equation}
Notice that the interaction term is actually fully general for a
gauge-invariant $C$ field, because any effective interaction of the form
$J_{\mu\nu\rho\sigma}F^{C,\mu\nu\rho\sigma}$ can be rewritten as
$J\epsilon_{\mu\nu\rho\sigma}F^{C,\mu\nu\rho\sigma}$ with $J=\epsilon_{\mu
\nu\rho\sigma}J^{\mu\nu\rho\sigma}$ using Eq.~(\ref{AppId}). In practice, the
$C\rightarrow J$ amplitude is thus
\begin{equation}
\mathcal{M}(C\rightarrow J)=\frac{1}{3!}\epsilon_{\mu\nu\rho\sigma}%
\varepsilon_{(\lambda)}^{\mu\nu\rho}k^{\sigma}J~~~\rightarrow\sum_{\lambda
}|\mathcal{M}(C\rightarrow J)|^{2}=-k^{2}J^{2}\ .
\end{equation}
The amplitude squared vanishes on shell, while off shell, only a trivial
contact term survives.

\section{Equivalences via massive dualities}
\label{massiveD}

For massive states, dualities have to be established at the level of fields,
not field strengths. Generically, those arise between a $p$-form field $A$ and
a $n-p-1$-form-field $A^{\prime}$, and follow from the parent Lagrangian~\cite{Casini:2002jm}%
\begin{equation}
\mathcal{L}_{\mathrm{parent}}(A,A^{\prime})=-(-1)^{p}\frac{m_{1}^{2}}%
{2}A\wedge\star A+m_{2}(-1)^{p}A\wedge dA^{\prime}-(-1)^{n-p-1}\frac{m_{3}%
^{2}}{2}A^{\prime}\wedge\star A^{\prime}\ .
\end{equation}
Fundamentally, this parent Lagrangian includes those discussed for massless
states as special cases. For instance, if we set one of the mass term to zero,
say $m_{1}=0$, then the EoM of $A$ asks for $dA^{\prime}=0$, which means that
$A^{\prime}$ is not a field but rather the field strength of a $n-p-2$ field,
$A^{\prime}=dA^{\prime\prime}$. It is then massless since the $m_{3}$ term
becomes the usual kinetic term for $A^{\prime\prime}$.

Let us thus assume that neither $m_{1}$ nor $m_{3}$ vanishes. We also need
$m_{2}\neq0$ otherwise there is no dynamics. Then, either $A$ or $A^{\prime}$
can be made algebraic since $dA\wedge A^{\prime}=(-1)^{p}A\wedge dA^{\prime}$,
up to a total derivative, giving
\begin{equation}%
\begin{array}
[c]{rl}%
A\overset{}{=}(-1)^{p(n-p)+1}\dfrac{m_{2}}{m_{1}^{2}}\star F^{A^{\prime}}: &
\mathcal{L}_{\mathrm{parent}}(A^{\prime})=\dfrac{(-1)^{p-1}}{2}F^{A^{\prime}%
}\wedge\star F^{A^{\prime}}-(-1)^{n-p-1}\dfrac{m_{1}^{2}m_{3}^{2}}{2m_{2}^{2}%
}A^{\prime}\wedge\star A^{\prime}\ ,\\
A^{\prime}\overset{}{=}(-1)^{n-p}\dfrac{m_{2}}{m_{3}^{2}}\star F^{A}: &
\mathcal{L}_{\mathrm{parent}}(A)=\dfrac{(-1)^{n-p}}{2}F^{A}\wedge\star
F^{A}-(-1)^{p}\dfrac{m_{1}^{2}m_{3}^{2}}{2m_{2}^{2}}A\wedge\star A\ ,
\end{array}
\end{equation}
where the fields have been rescaled as either $A^{\prime}\rightarrow
A^{\prime}\times m_{1}/m_{2}$ or $A\rightarrow A\times m_{3}/m_{2}$ to bring
the kinetic term in its canonical form. The signs are consistent with our
Lorentzian metric, but could be adapted to other cases. In practice, we get a
nontrivial duality if $p=0$, between a massive scalar and a massive
3-form tensor field, or if $p=1$, between a massive vector and a massive
2-form tensor field, see Table~\ref{TableDoF}. Dualities can also be
obtained starting with a $m_{2}\star A\wedge dA^{\prime}$ mixing term, but
those produce Lorenz-type kinetic terms and are simply the Hodge dual of those
obtained from $m_{2}A\wedge dA^{\prime}$, see Sec.~\ref{AlgDualities}.

\subsection{Equivalence between the massive 0- and 3-form models}

Let us start with the duality between the $p=0$- and $p=3$-form fields with
parent Lagrangian%
\begin{equation}
\mathcal{L}_{\mathrm{parent}}(\phi,C)=-\frac{m_{1}^{2}}{2}\phi^{2}+\frac
{m_{2}}{3!}\epsilon_{\mu\nu\rho\sigma}C^{\mu\nu\rho}\partial^{\sigma}%
\phi+\frac{1}{2}\frac{m_{3}^{2}}{3!}C_{\mu\nu\rho}C^{\mu\nu\rho}\ \ .
\end{equation}
By partial integration, either the $C$ or $\phi$ field can be made algebraic
and integrated out:%
\begin{equation}%
\begin{array}
[c]{rl}%
C_{\mu\nu\rho}=-\dfrac{m_{2}}{m_{3}^{2}}\epsilon_{\mu\nu\rho\sigma}%
\partial^{\sigma}\phi: & \mathcal{L}_{\mathrm{parent}}(\phi,C(\phi))=\dfrac
{1}{2}\partial_{\mu}\phi\partial^{\mu}\phi-\dfrac{m_{1}^{2}m_{3}^{2}}%
{2m_{2}^{2}}\phi^{2}\ ,\\
\phi=\dfrac{1}{4!}\dfrac{m_{2}}{m_{1}^{2}}\epsilon_{\mu\nu\rho\sigma}%
F^{C,\mu\nu\rho\sigma}: & \mathcal{L}_{\mathrm{parent}}(\phi(C),C)=-\dfrac
{1}{2}\dfrac{1}{4!}F_{\mu\nu\rho\sigma}^{C}F^{C,\mu\nu\rho\sigma}+\dfrac{1}%
{2}\dfrac{1}{3!}\dfrac{m_{1}^{2}m_{3}^{2}}{m_{2}^{2}}C_{\mu\nu\rho}C^{\mu
\nu\rho}\ .
\end{array}
\label{CmassEoM}%
\end{equation}
After the rescaling $\phi\rightarrow\phi\times m_{3}/m_{2}$ or $C\rightarrow
C\times m_{1}/m_{2}$, these states end up with the same mass $m_{S}=m_{1}%
m_{3}/m_{2}$.

As in the massless case, if we add generic effective couplings, the EoM for
the selected auxiliary field may no longer be algebraic, and even if it is, it
is in general impossible to solve. If only effective operators linear in the
dark field are kept, then that field EoM can remain algebraic, but contact
interactions will necessarily appear.

For phenomenological purposes, it is not that interesting to express the
scalar theory in terms of a 3-form field theory, but the converse offers a
genuine alternative description for a dark scalar field. So, let us add the
effective operators that are linear in the three form $C$ by defining an
effective $J_{eff}^{\mu\nu\rho}$ current, and take the parent effective
Lagrangian as $\mathcal{L}_{\mathrm{parent}}^{eff}(\phi,C)=\mathcal{L}%
_{\mathrm{parent}}(\phi,C)+C_{\mu\nu\rho}J_{eff}^{\mu\nu\rho}/3!$. If we
integrate $\phi$ out, we simply get back the effective Lagrangian for the
massive $C$ field, but with all the effective interactions rescaled by
$m_{1}/m_{2}$. Integrating $C$ out instead, $J_{eff}^{\mu\nu\rho}$ now appears
in its EoM in Eq.~(\ref{CmassEoM}), and we find%
\begin{equation}
\mathcal{L}_{\mathrm{parent}}^{eff}(\phi,C(\phi))=\frac{1}{2}\partial_{\alpha
}\phi\partial^{\alpha}\phi-\dfrac{m_{1}^{2}m_{3}^{2}}{2m_{2}^{2}}\phi
^{2}-\frac{1}{3!}\dfrac{1}{m_{3}}\epsilon_{\mu\nu\rho\alpha}\partial^{\alpha
}\phi J_{eff}^{\mu\nu\rho}-\frac{1}{2}\frac{1}{3!}\dfrac{1}{m_{3}^{2}%
}J_{eff,\mu\nu\rho}J_{eff}^{\mu\nu\rho}\ ,
\end{equation}
where we have rescaled $\phi\rightarrow\phi\times m_{3}/m_{2}$. In going from
the effective interactions in terms of $C$ to that in terms of $\phi$, they
all increase by one dimension, with $m_{3}$ acting as a compensating scale.
This parameter is essentially free since the values of $m_{1}$ and $m_{2}$ can
be adapted to keep $m_{S}=m_{1}m_{3}/m_{2}$ fixed at some chosen value.
Explicitly, the effective $C$ and $\phi$ couplings of dimensions up to six are
then related as%
\begin{align}
\frac{m_{1}}{m_{2}}\bar{\psi}_{L}\gamma_{\sigma}\psi_{L}\epsilon^{\mu\nu
\rho\sigma}C_{\mu\nu\rho} &  \leftrightarrow\frac{1}{m_{3}}\bar{\psi}%
_{L}\gamma_{\alpha}\psi_{L}\partial^{\alpha}\phi\ ,\\
\frac{m_{1}}{m_{2}}\bar{\psi}_{L}\psi_{R}\epsilon^{\mu\nu\rho\sigma}F_{\mu
\nu\rho\sigma}^{C} &  \leftrightarrow\frac{1}{m_{3}}(\bar{\psi}_{L}\psi
_{R})\partial^{2}\phi\ ,\\
\frac{m_{1}}{m_{2}}\bar{\psi}_{L}\sigma^{\mu\nu}\psi_{R}\partial^{\rho}%
C_{\mu\nu\rho} &  \leftrightarrow0\ ,
\end{align}
plus the same relations for $L\leftrightarrow R$. Note that these effective
interactions could have been obtained directly by substituting $C_{\mu\nu\rho
}\rightarrow(m_{1}/m_{2})C_{\mu\nu\rho}$ on one hand, and $C_{\mu\nu\rho
}=(1/m_{3})\epsilon_{\mu\nu\rho\sigma}\partial^{\sigma}\phi$ on the other. In
particular, the effective couplings initially involving the $C$ field Lorenz
condition do not have an equivalent in the $\phi$ description because of the
$\phi$ Bianchi identity, $\partial^{\rho}C_{\mu\nu\rho}\rightarrow
\epsilon_{\mu\nu\rho\sigma}\partial^{\rho}\partial^{\sigma}\phi=0$. This is
expected since in the $C$ description, the Lorenz condition holds so they
would also disappear. Another feature is to generate only shift-invariant effective scalar interactions, all involving the
$\phi$ field strength $F_{\mu}^{\phi}=\partial_{\mu}\phi$. If this pattern
holds also for operators involving pairs of $C$ field (and we will see below
when it indeed does), the leading operator would end up being the
dimension-seven operator%
\begin{equation}
\frac{m_{1}^{2}}{m_{2}^{2}}\bar{\psi}_{L}\psi_{R}C^{\mu\nu\rho}C_{\mu\nu\rho
}\leftrightarrow\frac{1}{m_{3}^{2}}\bar{\psi}_{L}\psi_{R}F_{\mu}^{\phi}%
F^{\phi,\mu}\ .
\end{equation}

For the contact interactions, there are only four-fermion couplings at the
dimension-six level,
\begin{equation}
\dfrac{1}{m_{3}^{2}}J_{eff,\mu\nu\rho}J_{eff}^{\mu\nu\rho}=\dfrac{c_{1,L}^{2}%
}{m_{3}^{2}}\bar{\psi}_{L}\gamma_{\mu}\psi_{L}\bar{\psi}_{L}\gamma^{\mu}%
\psi_{L}+\dfrac{c_{1,R}^{2}}{m_{3}^{2}}\bar{\psi}_{L}\gamma_{\mu}\psi_{L}%
\bar{\psi}_{R}\gamma^{\mu}\psi_{R}+\dfrac{c_{1,L}c_{1,R}}{m_{3}^{2}}\bar{\psi
}_{L}\gamma_{\mu}\psi_{L}\bar{\psi}_{R}\gamma^{\mu}\psi_{R}\ .
\end{equation}
For consistency, duality thus requires all these effective operators to be
present. Though dimension-seven operators are not relevant phenomenologically,
one feature is worth mentioning. At that level, contact interactions arising
from operators that would vanish upon enforcing the Lorenz condition start to
appear, with for example%
\begin{equation}
c_{1}\bar{\psi}_{L}\gamma_{\sigma}\psi_{L}\epsilon^{\mu\nu\rho\sigma}%
C_{\mu\nu\rho}\otimes\frac{c_{3}}{\Lambda}\bar{\psi}_{L}\sigma^{\mu\nu}%
\psi_{R}\partial^{\rho}C_{\mu\nu\rho}\rightarrow\dfrac{c_{1}c_{3}}{m_{3}%
^{2}\Lambda}\bar{\psi}_{L}\gamma_{\mu}\psi_{L}\partial_{\nu}(\bar{\psi}%
_{L}\sigma^{\mu\nu}\psi_{R})\ .
\end{equation}
Had we set $\partial^{\rho}C_{\mu\nu\rho}\rightarrow0$ too soon, this kind of
contact interaction would be missed.

Including now the SM operators of Eq.~(\ref{SMC}), we find%
\begin{align}
\partial^{\rho}C_{\rho\mu\nu}\tilde{F}_{Y}^{\mu\nu} &  \leftrightarrow0\ ,\\
\frac{m_{1}}{m_{2}}\Phi^{\dagger}\Phi\epsilon^{\mu\nu\rho\sigma}F_{\mu
\nu\rho\sigma}^{C} &  \leftrightarrow\dfrac{1}{m_{3}}\Phi^{\dagger}%
\Phi\partial^{2}\phi\ ,\\
\frac{m_{1}}{m_{2}}\Phi^{\dagger}\overleftrightarrow{\mathcal{D}}_{\alpha}%
\Phi\epsilon^{\mu\nu\rho\sigma}C_{\mu\nu\rho} &  \leftrightarrow\dfrac
{1}{m_{3}}\Phi^{\dagger}\overleftrightarrow{\mathcal{D}}_{\mu}\Phi
\partial^{\mu}\phi\ ,\\
\frac{m_{1}^{2}}{m_{2}^{2}}\Phi^{\dagger}\Phi C^{\mu\nu\rho}C_{\mu\nu\rho} &
\leftrightarrow\dfrac{1}{m_{3}^{2}}\Phi^{\dagger}\Phi\partial_{\mu}%
\phi\partial^{\mu}\phi\ .
\end{align}
From the point of view of the stability of the scalar theory, introducing it
via the parent Lagrangian thus appears quite desirable. First, we can suppress
all the couplings with the Higgs doublet by setting $m_{3}$ and $m_{2}$ to
rather large values, with $m_{1}$ alone setting the scale of the dark scalar
mass via $m_{S}=m_{1}\times m_{3}/m_{2}$. Second, all the $\phi$ operators
have to be shift invariant, and one avoids the dangerous renormalizable
$\phi\Phi^{\dagger}\Phi$ or $\phi^{2}\Phi^{\dagger}\Phi$ operators that often
have to be severely fine-tuned to maintain a separation between the
electroweak scale and that of the dark operators.

Finally, though our derivation of the duality severely restricts the form of
the effective interactions, its validity is more general and holds whenever
those states are external. This can be demonstrated as in the massless case
before, by comparing a process $C\rightarrow J$ and $\phi\rightarrow J$. We
start by adding the vertex $C_{\mu\nu\rho}J^{\mu\nu\rho}$ to the parent
Lagrangian. The corresponding amplitudes are%
\begin{subequations}
\begin{align}
\mathcal{M}(C &  \rightarrow J)=\frac{m_{1}}{m_{2}}\varepsilon_{(\lambda
)}^{\mu\nu\rho}J_{\mu\nu\rho}\ ,\ \ \sum_{\lambda}\varepsilon_{(\lambda
)}^{\ast\alpha\beta\gamma}\varepsilon_{(\lambda)}^{\mu\nu\rho}=-3!\left(
\mathcal{I}_{0}^{\alpha\beta\gamma,\mu\nu\rho}-\frac{3}{m_{V}^{2}}%
\mathcal{I}_{2}^{\alpha\beta\gamma,\mu\nu\rho}\right)  \ ,\\
\mathcal{M}(\phi &  \rightarrow J)=-\dfrac{1}{m_{3}}k^{\sigma}\epsilon
_{\mu\nu\rho\sigma}J^{\mu\nu\rho}\ .
\end{align}
For $C\rightarrow J$, the $m_{1}/m_{2}$ comes from the rescaling $C\rightarrow
C\times m_{1}/m_{2}$, necessary after eliminating $\phi$, while for
$\phi\rightarrow J$, the amplitude is derived by expressing $C_{\mu\nu\rho
}J^{\mu\nu\rho}$ in terms of $C_{\mu\nu\rho}=-(m_{2}/m_{3}^{2})\epsilon
_{\mu\nu\rho\sigma}\partial^{\sigma}\phi$ of Eq.~(\ref{CmassEoM}), followed by
the rescaling $\phi\rightarrow\phi\times m_{3}/m_{2}$. With this, summing over
the polarizations:
\end{subequations}
\begin{equation}
\sum_{\lambda}|\mathcal{M}(C\rightarrow J)|^{2}=|\mathcal{M}(\phi\rightarrow
J)|^{2}+3!\frac{k^{2}-m_{S}^{2}}{m_{3}^{2}}J_{\mu\nu\rho}J^{\mu\nu\rho
}\ ,\label{dualmassC}%
\end{equation}
where $m_{S}=m_{1}m_{3}/m_{2}$ is the dark field mass. Thus, on shell, duality
is expected to hold even for multiple external dark states. Notice that this
identity predicts that the squared amplitude for $\bar{\psi}\gamma_{\sigma
}\psi\epsilon^{\mu\nu\rho\sigma}C_{\mu\nu\rho}$ vanishes identically when
$k^{2}=m_{S}^{2}$ since that of $\bar{\psi}\gamma_{\alpha}\psi\partial
^{\alpha}\phi\ $does (a similar prediction was obtained in the massless case
in Sec.~\ref{massless2}). This is a nontrivial result that could not have
been obtained from the symmetry properties of the coupling alone. This can
have striking phenomenological implications whenever $C$ and $\phi$ are
exchanged between such conserved currents, since only the $C$ could produce
visible effects.

\subsection{Equivalence between the massive 1- and 2-form models}

The other massive duality is that between the $p=1$ and $p=2$ fields, derived
from the parent Lagrangian%
\begin{equation}
\mathcal{L}_{\mathrm{parent}}(A,B)=\frac{1}{2}\frac{m_{1}^{2}}{1!}A_{\mu
}A^{\mu}-\frac{m_{2}}{3!}\epsilon^{\mu\nu\rho\sigma}A_{\mu}F_{\nu\rho\sigma
}^{B}-\frac{1}{2}\frac{m_{3}^{2}}{2!}B_{\mu\nu}B^{\mu\nu}\ .
\end{equation}
Again, $A$ or $B$ can be made algebraic and eliminated since under partial
integration,%
\begin{equation}
\frac{m_{2}}{3!}\epsilon^{\mu\nu\rho\sigma}A_{\mu}F_{\nu\rho\sigma}^{B}%
=\frac{1}{2!}\frac{m_{2}}{2!}\epsilon^{\mu\nu\rho\sigma}F_{\rho\sigma}%
^{A}B_{\mu\nu}\ .
\end{equation}
This generates either the massive vector model or the massive 2-form tensor
model:%
\begin{equation}%
\begin{array}
[c]{rl}%
A_{\mu}=\dfrac{m_{2}}{3!m_{1}^{2}}\epsilon_{\mu\nu\rho\sigma}F^{B,\nu
\rho\sigma}: & \mathcal{L}_{\mathrm{parent}}(A(B),B)=\dfrac{1}{2}\dfrac{1}%
{3!}F_{\mu\nu\rho}^{B}F^{B,\mu\nu\rho}-\dfrac{1}{2}\dfrac{1}{2!}\dfrac
{m_{1}^{2}m_{3}^{2}}{m_{2}^{2}}B_{\mu\nu}B^{\mu\nu}\ ,\\
B_{\mu\nu}=-\dfrac{m_{2}}{2!m_{3}^{2}}\epsilon_{\mu\nu\rho\sigma}%
F^{A,\rho\sigma}: & \mathcal{L}_{\mathrm{parent}}(A,B(A))=-\dfrac{1}{2}%
\dfrac{1}{2!}F_{\mu\nu}^{A}F^{A,\mu\nu}+\dfrac{1}{2}\dfrac{m_{1}^{2}m_{3}^{2}%
}{m_{2}^{2}}A_{\mu}A^{\mu}\ .
\end{array}
\label{BmassEoM}%
\end{equation}
Again, after their appropriate rescalings $A\rightarrow A\times m_{3}/m_{2}$
and $B\rightarrow B\times m_{1}/m_{2}$, both fields end up with the same mass
$m_{V}=m_{1}m_{3}/m_{2}$. As mentioned earlier, had we put the BF mixing term
$B_{\mu\nu}F^{A,\mu\nu}$, the same duality would arise but with the Lorenz
kinetic term $\partial^{\nu}B_{\mu\nu}\partial_{\rho}B^{\rho\mu}$, so we do
not consider that case here. For simplicity, initially, we do not include the
pseudoscalar mass term $B_{\mu\nu}\tilde{B}^{\mu\nu}$, but we will correct for
that in a second step below.

Following the same logic as in the previous Section, our goal is to see how
introducing a massive vector via a 2-form field affects the effective theory
once expressed back in terms of the usual vector field. Again, we cannot
simply replace $B_{\mu\nu}\rightarrow\epsilon_{\mu\nu\rho\sigma}%
F^{A,\rho\sigma}/m_{3}$ because the presence of effective interactions affects
the EoM. Instead, let us again keep only the effective interactions linear in
the $B$ field, and encode them into a current $\mathcal{L}_{\mathrm{parent}%
}^{eff}(A,B)=\mathcal{L}_{\mathrm{parent}}(A,B)+B_{\mu\nu}J_{eff}^{\mu\nu}%
/2!$. Integrating $A$ out gives back the 2-form field, but with its
interactions rescaled by $(m_{1}/m_{2})B_{\mu\nu}J_{eff}^{\mu\nu}/2!$.
Integrating $B$ out gives the Proca Lagrangian in terms of $A$, effective
interactions for the Proca field rescaled by $1/m_{3}$, and contact
interactions:%
\begin{equation}
\mathcal{L}_{\mathrm{parent}}^{eff}(A,B(A))=-\frac{1}{2}\frac{1}{2!}%
F_{\rho\sigma}^{A}F^{A,\rho\sigma}+\frac{1}{2}\frac{m_{1}^{2}m_{3}^{2}}%
{m_{2}^{2}}A_{\mu}A^{\mu}+\frac{1}{2!}\dfrac{1}{2!}\dfrac{1}{m_{3}}%
\epsilon^{\mu\nu\rho\sigma}F_{\rho\sigma}^{A}J_{\mu\nu}-\frac{1}{2}\frac
{1}{2!}\dfrac{1}{m_{3}^{2}}J_{\mu\nu}J^{\mu\nu}\ .
\end{equation}
The effective interactions are thus related to those of the $B$ operators as
\begin{subequations}
\label{dualeffBA}%
\begin{align}
\frac{m_{1}}{m_{2}}\bar{\psi}_{L}\sigma^{\mu\nu}\psi_{R}B_{\mu\nu} &
\leftrightarrow\frac{1}{m_{3}}i\bar{\psi}_{L}\sigma^{\mu\nu}\psi_{R}F_{\mu\nu
}^{A}\ ,\\
\frac{m_{1}}{m_{2}}\bar{\psi}_{L}\gamma_{\sigma}\psi_{L}\epsilon^{\mu\nu
\rho\sigma}F_{\mu\nu\rho}^{B} &  \rightarrow\frac{1}{m_{3}}\bar{\psi}%
_{L}\gamma_{\nu}\psi_{L}\partial^{\mu}F_{\mu\nu}^{A}\ ,\\
\frac{m_{1}}{m_{2}}\bar{\psi}_{L}\gamma^{\mu}\mathcal{D}^{\nu}\psi_{L}%
B_{\mu\nu} &  \rightarrow\frac{1}{m_{3}}\bar{\psi}_{L}\gamma^{\mu}%
\mathcal{D}^{\nu}\psi_{L}\tilde{F}_{\mu\nu}^{A}\ ,\\
\frac{m_{1}}{m_{2}}\bar{\psi}_{L}\gamma^{\nu}\psi_{L}\partial^{\mu}B_{\mu\nu}
&  \rightarrow0\ ,
\end{align}
while the dimension-six effective $B$ operators generate dimension-seven $A$
operators that we do not keep. As before, operators with $L\leftrightarrow R$
are understood. Obviously, only gauge invariant operators arise since $B$ is
replaced by the dual field strength $\star F^{A}$. Concerning the couplings to
the other SM fields, concentrating on operators of dimensions less than six,
we have to include the mass mixing with $F_{Y}^{\mu\nu}$ since the
reparametrization used in Sec.~\ref{EffOp2} to rotate it away would mess up
the terms in the parent Lagrangian. Thus, those operators become
\end{subequations}
\begin{align}
\Lambda_{\gamma}\frac{m_{1}}{m_{2}}B_{\mu\nu}F_{Y}^{\mu\nu} &  \leftrightarrow
0\ ,\\
\Lambda_{\gamma}\frac{m_{1}}{m_{2}}B_{\mu\nu}\tilde{F}_{Y}^{\mu\nu} &
\leftrightarrow\frac{\Lambda_{\gamma}}{m_{3}}F_{\mu\nu}^{A}F_{Y}^{\mu\nu}\ .
\end{align}
Provided $m_{2}$ and $m_{3}$ are sufficiently larger than $\Lambda_{\gamma}$,
the kinetic mixing can be strongly suppressed. Further, this does not prevent
the vector field from being light since it suffices for $m_{1}$ to be small to compensate.

The operators involving the $B$ field Lorenz condition have no counterparts in
the $A$ picture, but they do contribute to the contact terms.\ Altogether, the
operators of dimensions up to six are (setting all the Wilson coefficients to
one for clarity, and introducing the scale $\Lambda\neq\Lambda_{\gamma}$ for
nonrenormalizable effective interactions),
\begin{align}
\dfrac{1}{m_{3}^{2}}J_{\mu\nu}J^{\mu\nu} &  =\dfrac{\Lambda_{\gamma}}%
{m_{3}^{2}}\bar{\psi}_{L}\sigma_{\mu\nu}\psi_{R}F_{Y}^{\mu\nu}+\dfrac
{\Lambda_{\gamma}}{m_{3}^{2}}\bar{\psi}_{L}\sigma_{\mu\nu}\psi_{R}\tilde
{F}_{Y}^{\mu\nu}+h.c.\\
&  +\frac{\Lambda_{\gamma}}{\Lambda}\frac{1}{m_{3}^{2}}\bar{\psi}_{L,R}%
\gamma_{\mu}\mathcal{D}_{\nu}\psi_{L,R}F_{Y}^{\mu\nu}+\frac{\Lambda_{\gamma}%
}{\Lambda}\frac{1}{m_{3}^{2}}\bar{\psi}_{L,R}\gamma_{\nu}\psi_{L,R}%
\partial_{\mu}F_{Y}^{\mu\nu}\\
&  +\dfrac{1}{m_{3}^{2}}\bar{\psi}_{L}\sigma_{\mu\nu}\psi_{R}\bar{\psi}%
_{R}\sigma^{\mu\nu}\psi_{L}+\dfrac{1}{m_{3}^{2}}\bar{\psi}_{L}\sigma_{\mu\nu
}\psi_{R}\bar{\psi}_{L}\sigma^{\mu\nu}\psi_{R}+h.c.\ .
\end{align}
If Higgs fields are included, the operators in the third line become
dimension eight and can be discarded while all the others are dimension six.
The presence of the operators in the first line is quite striking. When
coupling the $B$ field to fermions, those have to be included for consistency.
At low energy, barring fine-tuned scenarios, the existence of a dark vector
could then signal itself via shifts in the magnetic and/or electric dipole
operators. Said differently, any fundamental theory leading to this set of
effective $B$ interaction has to also induce dipole operators. Though
increasing $m_{3}$ could make these EDM and MDM corrections small, that would
also suppress all the direct couplings in Eq.~(\ref{dualeffBA}). So, it is
$\Lambda_{\gamma}$ alone that needs to be small to allow for observable effects.

As for the other scenarios, the above correspondences under duality hold more
generally for external dark states. If we consider a $B_{\mu\nu}J^{\mu\nu}$
coupling in the parent Lagrangian, the amplitudes are%
\begin{subequations}
\begin{align}
\mathcal{M}(B &  \rightarrow J)=\frac{m_{1}}{m_{2}}\varepsilon_{(\lambda
)}^{\mu\nu}J_{\mu\nu}\ ,\ \ \sum_{\lambda}\varepsilon_{(\lambda)}^{\ast
\alpha\beta}\varepsilon_{(\lambda)}^{\mu\nu}=2!\left(  \mathcal{I}_{0}%
^{\alpha\beta,\mu\nu}-\frac{2}{m_{V}^{2}}\mathcal{I}_{2}^{\alpha\beta,\mu\nu
}\right)  \ ,\\
\mathcal{M}(A &  \rightarrow J)=\dfrac{1}{m_{3}}\epsilon_{\mu\nu\rho\sigma
}k^{\rho}\varepsilon_{(\lambda)}^{\sigma}J^{\mu\nu}\ ,\ \ \sum_{\lambda
}\varepsilon_{(\lambda)}^{\ast\alpha}\varepsilon_{(\lambda)}^{\mu}=-\left(
\mathcal{I}_{0}^{\alpha,\mu}-\frac{2}{m_{V}^{2}}\mathcal{I}_{2}^{\alpha,\mu
}\right)  \ ,
\end{align}
and
\end{subequations}
\begin{equation}
\sum_{\lambda}|\mathcal{M}(B\rightarrow J)|^{2}=\sum_{\lambda}|\mathcal{M}%
(A\rightarrow J)|^{2}-2!\frac{k^{2}-m_{V}^{2}}{m_{3}^{2}}J_{\mu\nu}J^{\mu\nu
}\ ,\label{BdualA}%
\end{equation}
where $m_{V}=m_{1}m_{3}/m_{2}$ is the dark vector mass. This equation is the
exact analog of Eq.~(\ref{dualmassC}), and shows that on shell, duality is
expected to hold even for multiple external dark states (a similar relation was derived already a long time ago in Ref.~\cite{Sugamoto:1978ft}) .

It is time now to show how duality accommodates the presence of a pseudoscalar
mass term for the $B$ field. For that, let us simply add it to the parent
Lagrangian%
\begin{equation}
\mathcal{L}_{\mathrm{parent}}(A,B)=\frac{1}{2}\frac{m_{1}^{2}}{1!}A_{\mu
}A^{\mu}+\frac{1}{2!}\frac{m_{2}}{2!}\epsilon^{\mu\nu\rho\sigma}F_{\rho\sigma
}^{A}B_{\mu\nu}-\frac{1}{2}\frac{m_{3}^{2}}{2!}B_{\mu\nu}B^{\mu\nu}+\frac
{1}{2}\frac{\tilde{m}_{3}^{2}}{2!}B_{\mu\nu}\tilde{B}^{\mu\nu}\ .
\end{equation}
Eliminating $A$ proceeds as before, and after the rescaling $B\rightarrow
B\times m_{1}/m_{2}$, Eq.~(\ref{Bmass}) predicts the $B$ mass to be%
\begin{equation}
m_{V}^{2}=m_{1}^{2}\frac{m_{3}^{4}+\tilde{m}_{3}^{4}}{m_{3}^{2}m_{2}^{2}}\ .
\label{Bmassdual}%
\end{equation}
Treating instead $B$ as the auxiliary field, the EoM becomes%
\begin{equation}
\left(  m_{3}^{2}\mathcal{I}_{0}^{\mu\nu,\rho\sigma}-\tilde{m}_{3}%
^{2}\mathcal{I}_{3}^{\mu\nu,\rho\sigma}\right)  B_{\rho\sigma}=\frac{m_{2}%
}{2!}\epsilon^{\mu\nu\rho\sigma}F_{\rho\sigma}^{A}\Leftrightarrow B^{\mu\nu
}=m_{2}\frac{m_{3}^{2}\tilde{F}^{A,\mu\nu}-\tilde{m}_{3}^{2}F^{A,\mu\nu}%
}{m_{3}^{4}+\tilde{m}_{3}^{4}}\ . \label{EoMmmt}%
\end{equation}
Plugged back in the Lagrangian, this gives, after the appropriate rescaling of
the $A$ field%
\begin{equation}
\mathcal{L}_{\mathrm{parent}}(A,B(A))=-\frac{1}{4}F_{\mu\nu}^{A}F^{A,\mu\nu
}-\frac{1}{4}\frac{\tilde{m}_{3}^{2}}{m_{3}^{2}}F_{\mu\nu}^{A}\tilde{F}%
^{A,\mu\nu}+\frac{1}{2}m_{1}^{2}\frac{m_{3}^{4}+\tilde{m}_{3}^{4}}{m_{2}%
^{2}m_{3}^{2}}A_{\mu}A^{\mu}\ ,
\label{EoMmmt2}
\end{equation}
which is the Proca Lagrangian for a massive vector field with the same mass as
in Eq.~(\ref{Bmassdual}). This confirms the validity of Eq.~(\ref{Bmass}) in
an independent and somewhat simpler way. The second term in $\mathcal{L}%
_{\mathrm{parent}}$ is an irrelevant Abelian theta term that can be safely discarded.

From here, we can repeat the matching between the effective operators by
adding a $B_{\mu\nu}J^{\mu\nu}$ term in the parent Lagrangian. With now
$B^{\mu\nu}$ having both a $\tilde{F}^{A,\mu\nu}$ and $F^{A,\mu\nu}$
component, the $CP$ properties of all the operators get mixed up. The relative
scaling between their $B$ and $A$ representations is also parametrically more
complicated, being dependent on all the mass parameters, but the main features
discussed previously remain valid so we will not go into those details here.
However, before closing this Section, let us use duality to confirm the
polarization sum appearing in the numerator of the full $B$ propagator in the
presence of the $\tilde{m}_{3}$ parameter, i.e., Eq.~(\ref{PropMtilde}) with
$m=m_{3}m_{1}/m_{2}$ and $\tilde{m}=\tilde{m}_{3}m_{1}/m_{2}$. To this end, we
compare%
\begin{subequations}
\label{mtildePol}%
\begin{align}
\mathcal{M}(B &  \rightarrow J)=\frac{m_{1}}{m_{2}}\varepsilon_{(\lambda
)}^{\mu\nu}J_{\mu\nu}\ ,\ \ \sum_{\lambda}\varepsilon_{(\lambda)}^{\ast
\alpha\beta}\varepsilon_{(\lambda)}^{\mu\nu}=2\left(  \mathcal{I}_{0}%
^{\alpha\beta,\mu\nu}-\frac{2m_{2}^{2}}{m_{3}^{2}m_{1}^{2}}\mathcal{I}%
_{2}^{\alpha\beta,\mu\nu}+\frac{\tilde{m}_{3}^{2}}{m_{3}^{2}}\mathcal{I}%
_{3}^{\alpha\beta,\mu\nu}\right)  \ ,\\
\mathcal{M}(A &  \rightarrow J)=\frac{1}{m_{3}}\left(  \frac{m_{3}^{2}%
\epsilon_{\mu\nu\rho\sigma}k^{\rho}\varepsilon_{(\lambda)}^{\sigma}}%
{\sqrt{m_{3}^{4}+\tilde{m}_{3}^{4}}}-\frac{\tilde{m}_{3}^{2}(k_{\mu
}\varepsilon_{\nu}^{(\lambda)}-k_{\nu}\varepsilon_{\mu}^{(\lambda)})}%
{\sqrt{m_{3}^{4}+\tilde{m}_{3}^{4}}}\right)  J^{\mu\nu}\ .
\end{align}
As before, these equations account for the rescaling necessary to have
canonical kinetic terms. From them, it is immediate to check that%
\end{subequations}
\begin{equation}
\sum_{\lambda}|\mathcal{M}(B\rightarrow J)|^{2}=\sum_{\lambda}|\mathcal{M}%
(A\rightarrow J)|^{2}-2!\frac{k^{2}-m_{V}^{2}}{m_{3}^{4}+\tilde{m}_{3}^{4}%
}\left(  m_{3}^{2}J_{\mu\nu}J^{\mu\nu}+\tilde{m}_{3}^{2}\epsilon^{\mu
\nu\rho\sigma}J_{\mu\nu}J_{\rho\sigma}\right)  \ ,
\end{equation}
with $m_{V}$ given in Eq.~(\ref{Bmassdual}). These descriptions are equivalent
on shell, and their difference off shell precisely matches the contact terms one
could derive by adding the $J^{\mu\nu}$ term in the EoM of Eq.~(\ref{EoMmmt}%
). Finally, it is worth remarking that equating $\mathcal{M}(A\rightarrow J)$
and $\mathcal{M}(B\rightarrow J)$ on shell provides a representation of the
three polarization matrices $\varepsilon_{(\lambda)}^{\mu\nu}$ in terms of
that of a massive vector $\varepsilon_{\mu}^{(\lambda)}$ satisfying the
modified transversality constraint of Eq.~(\ref{GenLorenz}).

\section{Equivalences \`{a} la Stueckelberg}

The massless and massive dualities discussed in the previous sections permit one
to match a parent Lagrangian either into a higher-form effective theory, or
onto its corresponding scalar or vector effective theory. One should realize
though that these are truly different realizations of the dynamics, not a mere
change of variables. This is well illustrated in the case of the Maxwell
theory, in which imposing either $\partial_{\mu}F^{\mu\nu}=J^{\nu}$ or
$\partial_{\mu}\tilde{F}^{\mu\nu}=J^{\nu}$ switches electric and magnetic
charges. To some extent, an exact equivalence is recovered when the dark
fields stay external since the free equations of motion are satisfied. In
practice, equivalence then follows from algebraic identities among
polarization sums, and effective operators for two dual scenarios can be
matched onto each other, up to various rescalings. Those are important, not
least because they allow for suppressing even the renormalizable dimension 3
and 4 couplings.

At the same time, looking at Table~\ref{TableDoF}, we can see that the
dynamics itself should be sufficient to relate fields having the same number
of degrees of freedom, without recourse to parity-violating dualization. For
example, it is well known that a vector field of mass $m_{V}$ is essentially
made of a transverse massless vector field together with a longitudinal scalar
field. This is the essence of the Higgs mechanism. In the Stueckelberg picture,
the vector field becomes massive without breaking gauge invariance thanks to
the presence of a scalar field, in the combination $A_{\mu}-\partial_{\mu}%
\phi/m_{V}$. Whenever $m_{V}$ is much smaller than the typical energy scale of
a given process, this $1/m_{V}$ factor has important phenomenological
consequences: an effective interaction in which the $\phi$ component does not
decouple is strongly enhanced and can become a prime target for experiments.

The Stueckelberg picture is straightforwardly extended to $p$-form fields, and
offers an alternative to duality to explore the relationships between $p$-form
effective interactions (the $p=2$ case has been considered recently in Ref.~\cite{deGracia:2023jky}). In practice, comparing the gauge transformations in
Eq.~(\ref{FieldsGauge}) with the definitions of the field strengths in
Eq.~(\ref{FieldStrength}), one notices that the gauge variation of a $p$-form
field $A\rightarrow A+d\Lambda$ can be compensated by the shift $F^{B+\Lambda
}\rightarrow F^{B}+d\Lambda$ with $B\rightarrow B+\Lambda$ a $p-1$-form field.
Provided the gauge and shift transformations are made coherently, $A-F^{B}$
becomes invariant. For dimensional reasons, the $F$ component should actually
involve a $1/m$ factor, such that the $A$ mass term becomes
\begin{equation}
m^{2}A\wedge\star A\rightarrow m^{2}\left(  A-\frac{1}{m}F^{B}\right)
\wedge\star\left(  A-\frac{1}{m}F^{B}\right)  =m^{2}A\wedge\star
A-2mF^{B}\wedge\star A+F^{B}\wedge\star F^{B}\ .
\end{equation}
Applied to $d=4$, the possible constructions are then (setting $m=1$ for
clarity)
\begin{subequations}
\begin{align}
(A_{\mu}-F_{\mu}^{\phi}) &  :\phi\rightarrow\phi+\Lambda\ ,\ \ A_{\mu
}\rightarrow A_{\mu}+\partial_{\mu}\Lambda\ ,\\
(B_{\mu\nu}-F_{\mu\nu}^{A}) &  :A_{\mu}\rightarrow A_{\mu}+\Lambda_{\mu
}\ ,\ \ B_{\mu\nu}\rightarrow B_{\mu\nu}+\partial_{\mu}\Lambda_{\nu}%
-\partial_{\nu}\Lambda_{\mu}\ ,\\
(C_{\mu\nu\rho}-F_{\mu\nu\rho}^{B}) &  :B_{\mu\nu}\rightarrow B_{\mu\nu
}+\Lambda_{\mu\nu}\ ,\ \ C_{\mu\nu\rho}\rightarrow C_{\mu\nu\rho}%
+\partial_{\mu}\Lambda_{\nu\rho}+\partial_{\nu}\Lambda_{\rho\mu}%
+\partial_{\rho}\Lambda_{\mu\nu}\ .
\end{align}
A fourth possibility involves $D_{\mu\nu\rho\sigma}-F_{\mu\nu\rho\sigma}^{C}$,
but it is rather trivial since $D_{\mu\nu\rho\sigma}$ has no dynamics, and
will not be of any use in the following.

In the next Subsection, we explore in some detail the relationship between
the Stueckelberg construction and the equivalence theorem. We also show how it
provides for another way of organizing the physical and unphysical degrees of
freedom. Then, this will be put to phenomenological use in the following Subsection.

\subsection{Stueckelberg construction and equivalence theorem}

Imagine a coupling $A\wedge\star J\rightarrow A_{\mu_{1}...\mu_{p}}J^{\mu
_{1}...\mu_{p}}$, to which we can schematically associate the squared
amplitude
\end{subequations}
\begin{equation}
\sum_{\lambda}|\mathcal{M}(A\rightarrow J)|^{2}=J_{\mu_{1}...\mu_{p}}%
J_{\nu_{1}...\nu_{p}}^{\dagger}\sum_{\lambda}\varepsilon_{(\lambda)}^{\ast
\mu_{1}...\mu_{p}}\varepsilon_{(\lambda)}^{\nu_{1}...\nu_{p}}\ ,
\label{EquivTh}%
\end{equation}
with
\begin{equation}
\sum_{\lambda}\varepsilon_{(\lambda)}^{\ast\mu_{1}...\mu_{p}}\varepsilon
_{(\lambda)}^{\nu_{1}...\nu_{p}}=(-1)^{p}p!\left(  \mathcal{I}_{0}^{A,\mu
_{1}...\mu_{p},\nu_{1}...\nu_{p}}-\frac{p}{m^{2}}\mathcal{I}_{2}^{A,\mu
_{1}...\mu_{p},\nu_{1}...\nu_{p}}\right)  \ . \label{polsum}%
\end{equation}
For a vector field getting its mass through the Stueckelberg mechanism, it is
well known that the $\mathcal{I}_{2}^{A}$ term above can be interpreted as
that coming from its scalar longitudinal degree of freedom. That part
dominates whenever the energy of the process is large compared to the vector
boson mass $m$. Since the Stueckelberg construction is closely related to the
Abelian Higgs model, this is nothing but a reformulation of the equivalence
theorem for Goldstone bosons. Our goal here is to generalize this to higher
form fields.

Consider thus a generic Stueckelberg model written in terms of $A-F^{B}/m$. The
Lorenz condition required to make the Hamiltonian positive definite is
modified to $d\star A+m\star B=0$ (see e.g. Ref.~\cite{Ruegg:2003ps}). In turn, this
condition can only be fulfilled provided $B$ satisfies the usual Lorenz
condition $d\star B=0$. In a path integral formalism, these conditions are
enforced by adding gauge fixing terms in the Lagrangian (along with fermionic
ghosts which we shall not discuss here). In that context, it is customary to
modify them slightly and adopt the $R_{\xi}$ gauge-fixing term:%
\begin{align}
\mathcal{S}_{A,B} &  =\frac{(-1)^{p}}{2}\int\frac{1}{2}F^{A}\wedge\star
F^{A}-\frac{m^{2}}{2}\left(  A-\frac{1}{m}F^{B}\right)  \wedge\star\left(
A-\frac{1}{m}F^{B}\right)  \nonumber\\
&  \ \ \ \ \ \ \ \ \ \ \ \ \ -\frac{1}{2\xi}(d\star A+(-1)^{p-1}\xi m\star
B)\wedge\star(d\star A+(-1)^{p-1}\xi m\star B)\nonumber\\
&  \ \ \ \ \ \ \ \ \ \ \ \ \ +\frac{1}{2\zeta}d\star B\wedge\star(d\star B)\ .
\end{align}
The advantage of this gauge fixing is to immediately remove the mixing between
the $A$ gauge boson and its $B$ partner, $\mathcal{S}_{A,B}=\mathcal{S}%
_{A}+\mathcal{S}_{B}$ with
\begin{align}
\mathcal{S}_{A} &  =(-1)^{p}\int\frac{1}{2}F^{A}\wedge\star F^{A}-\frac{1}%
{2}m^{2}A\wedge\star A-\frac{1}{2\xi}(d\star A)\wedge\star(d\star A)\ ,\\
\mathcal{S}_{B} &  =(-1)^{p-1}\int\frac{1}{2}F^{B}\wedge\star F^{B}-\frac
{1}{2}\xi m^{2}B\wedge\star B-\frac{1}{2\zeta}d\star B\wedge\star(d\star B)\ .
\end{align}
Thanks to this separation, the full propagators can be read off the kinetic
terms without the need for resummations:%
\begin{align}
\mathcal{P}^{A} &  =i\frac{(-1)^{p}p!}{k^{2}-m^{2}}\left(  \mathcal{I}_{0}%
^{A}-(1-\xi)\frac{p}{k^{2}-\xi m^{2}}\mathcal{I}_{2}^{A}\right)  \ ,\\
\mathcal{P}^{B} &  =i\frac{(-1)^{p-1}(p-1)!}{k^{2}-\xi m^{2}}\left(
\mathcal{I}_{0}^{B}-(1-\zeta)\frac{p-1}{k^{2}-\zeta\xi m^{2}}\mathcal{I}%
_{2}^{B}\right)  \ .
\end{align}

The $\zeta$ gauge parameter is needed to deal with the $\xi=0$ gauge. When
$\xi\neq0$, the $B$ field is massive, the Lorenz condition is automatic, and
we should actually move to the unitary gauge $\zeta\rightarrow\infty$.
Alternatively, one could introduce a $p-2$ Stueckelberg field $C$ to regularize the $B$ mass term, and then maybe a $p-3$ field for the $C$ mass term, and so on down to a scalar field. In practice, this tower of Stueckelberg fields is not needed because the $B$ field always couples in a gauge-invariant way via the $(A-F^{B}/m)$ combination, and the $\mathcal{I}_{2}^{B}$ part never contributes. The only interest of introducing the $\zeta$ parameter is for counting the DoF in various gauges, as we now discuss.

\begin{table}[t]
\centering$%
\begin{tabular}
[c]{c|cclllclll}\hline
Gauge: & $\xi\rightarrow\infty$ &  & \multicolumn{3}{c}{$\xi=0$} &  &
\multicolumn{3}{c}{$\xi=1$}\\\hline
$p$ & \multicolumn{1}{|l}{Massive A} & \multicolumn{1}{l}{$=$} & Massless A &
$+$ & Massless B & \multicolumn{1}{l}{$=$} & Unconstrained A & $-$ & Massive
B\\\hline
$0$ & $1$ & $=$ & \multicolumn{1}{c}{$1$} & \multicolumn{1}{c}{$+$} &
\multicolumn{1}{c}{$0$} & $=$ & \multicolumn{1}{c}{$1$} &
\multicolumn{1}{c}{$-$} & \multicolumn{1}{c}{$0$}\\
$1$ & $3$ & $=$ & \multicolumn{1}{c}{$2$} & \multicolumn{1}{c}{$+$} &
\multicolumn{1}{c}{$1$} & $=$ & \multicolumn{1}{c}{$4$} &
\multicolumn{1}{c}{$-$} & \multicolumn{1}{c}{$1$}\\
$2$ & $3$ & $=$ & \multicolumn{1}{c}{$1$} & \multicolumn{1}{c}{$+$} &
\multicolumn{1}{c}{$2$} & $=$ & \multicolumn{1}{c}{$6$} &
\multicolumn{1}{c}{$-$} & \multicolumn{1}{c}{$3$}\\
$3$ & $1$ & $=$ & \multicolumn{1}{c}{$0$} & \multicolumn{1}{c}{$+$} &
\multicolumn{1}{c}{$1$} & $=$ & \multicolumn{1}{c}{$4$} &
\multicolumn{1}{c}{$-$} & \multicolumn{1}{c}{$3$}\\
$4$ & $0$ & $=$ & \multicolumn{1}{c}{$0$} & \multicolumn{1}{c}{$+$} &
\multicolumn{1}{c}{$0$} & $=$ & \multicolumn{1}{c}{$1$} &
\multicolumn{1}{c}{$-$} & \multicolumn{1}{c}{$1$}\\\hline
\end{tabular}
\ \ $\caption{Stueckelberg separation of the propagating number of degrees of
freedom for a $p$-form field $A$ and its auxiliary $(p-1)$-form field $B$ in four dimensions for the unitary, Landau, and Feynman gauges.}%
\label{TableStuck}%
\end{table}

First, in the unitary gauge $\xi\rightarrow\infty$, only $A$ propagates and we
recover the massive propagator of Eq.~(\ref{Pmassive}). For any $\xi<\infty$,
however, the number of DoF does not match the expected physical number, and
this has to be compensated by the $B$ field. To see this, consider again the
$A\wedge\star J$ interaction, which we make gauge invariant by adding the $B$
field as $(A-F^{B}/m)\wedge\star J$. The associated Feynman rules are%
\begin{equation}
\mathcal{V}_{JA}=\frac{1}{p!}\mathcal{I}_{0}^{A}\ ,\ \ \mathcal{V}_{JB}%
=\frac{i}{(p-1)!}\frac{1}{m}(\mathcal{I}_{0}^{A}\cdot k)\mathcal{I}_{0}^{B}\ ,
\end{equation}
where the dot notation means $((\mathcal{I}_{0}^{A}\cdot k)\mathcal{I}_{0}%
^{B})_{\rho_{1}...\rho_{p-1}}^{\mu_{1}...\mu_{p}}=(\mathcal{I}_{0}^{A}%
)^{\mu_{1}...\mu_{p},\nu_{1}...\nu_{p}}k_{\nu_{1}}(\mathcal{I}_{0}^{B}%
)_{\nu_{2}...\nu_{p},\rho_{1}...\rho_{p-1}}$. An amplitude for $J\rightarrow
A\rightarrow J$ is then accompanied by $J\rightarrow B\rightarrow J$, and
their sum gives%
\begin{align}
\langle J(p)J(-p)\rangle &  =\mathcal{V}_{JA}\mathcal{P}^{A}\mathcal{V}%
_{AJ}+\mathcal{V}_{JB}\mathcal{P}^{B}\mathcal{V}_{BJ}\nonumber\\
&  =\frac{1}{p!}\frac{1}{p!}\left(  \mathcal{P}^{A}+\frac{p^{2}}{m^{2}%
}(\mathcal{I}_{0}^{A}\cdot k)\mathcal{P}^{B}(k\cdot\mathcal{I}_{0}%
^{A})\right)  =\frac{1}{p!}\frac{i(-1)^{p}}{k^{2}-m^{2}}\left(  \mathcal{I}%
_{0}^{A}-\frac{p}{m^{2}}\mathcal{I}_{2}^{A}\right)  \ .\label{FullCurr}%
\end{align}
This result is gauge independent, and in particular, corresponds to that in
the unitary gauge where only $A$ propagates. Notice that $\mathcal{P}^{B}$
contributes only to the $\mathcal{I}_{2}^{A}$ term, and this only via its
$\mathcal{I}_{0}^{B}$ component. This follows from the identity
$(\mathcal{I}_{0}^{A}\cdot k)\mathcal{I}_{0}^{B}(k\cdot\mathcal{I}_{0}%
^{A})=\mathcal{I}_{2}^{A}$, which is essentially a rewriting of the definition
Eq.~(\ref{InvariantDef}), but $(\mathcal{I}_{0}^{A}\cdot k)\mathcal{I}_{2}%
^{B}(k\cdot\mathcal{I}_{0}^{A})=0$ by antisymmetry. As said earlier, this
explains why there is no need to regularize the $B$ mass term.

The unitary gauge can be compared to the counting in the Landau gauge $\xi=0 $
and Feynman gauge $\xi=1$.\ In the former case, both $A$ and $B$ are purely
transverse, even off shell, with a decomposition%
\begin{equation}
C_{p}^{n-1}\left\{  \text{massive }p\text{ field}\right\}  =C_{p}%
^{n-2}\left\{  \text{transverse }p\text{ field}\right\}  +C_{p-1}%
^{n-2}\left\{  \text{transverse }p-1\text{ field}\right\}  \ . \label{DoF1}%
\end{equation}
By contrast, in the Feynman gauge $\xi=1$, $A$ is allowed to propagate all its
$C_{p}^{n}$ degrees of freedom, but the massive $B$ field is tasked with
canceling the spurious ones, so that
\begin{equation}
C_{p}^{n-1}\left\{  \text{massive }p\text{ field}\right\}  =C_{p}^{n}\left\{
\text{unconstrained }p\text{ field}\right\}  -C_{p-1}^{n-1}\left\{
\text{massive }p-1\text{ field}\right\}  \ . \label{DoF2}%
\end{equation}
The situation of Eq.~(\ref{DoF1}) and Eq.~(\ref{DoF2}) is summarized in
Table~\ref{TableStuck}, to be compared to Table~\ref{TableDoF}.

Returning to the equivalence theorem, the polarization sum in
Eq.~(\ref{EquivTh}) clearly matches that of the full propagator in
Eq.~(\ref{FullCurr}).\ This is best interpreted in terms of the propagator
numerators in the Feynman gauge, where $A$ and $B$ have the same mass, so that
$A$ and $B$ can be put on shell via $(x-i\varepsilon)^{-1}=P(1/x)-i\pi
\delta(x)$. In that case, the $\mathcal{I}_{2}^{A}$ component is entirely
generated by the $B$ field, as a consequence of the $(\mathcal{I}_{0}^{A}\cdot
k)\mathcal{I}_{0}^{B}(k\cdot\mathcal{I}_{0}^{A})=\mathcal{I}_{2}^{A}$
identity. This proves that the $1/m$ terms of the polarization sum can indeed
be extracted either by using the Stueckelberg construction or from the
equivalence theorem. The two pictures are totally equivalent\footnote{There is
a caveat here for the $B$ field, because of the $\tilde{m}^{2}B\wedge B$ mass
term. Under the Stueckelberg substitution $B\rightarrow B-F^{A}/m$, the
$R_{\xi}$ trick no longer decouples the $B$ field from $F^{A}$, but leaves a
$B\wedge F^{A}$ coupling quite analog to the $B\wedge F^{\gamma}$ coupling
discussed in Sec.~\ref{EffOp2}. Some form of Dyson resummation appears
necessary to prove the equivalence, but we leave this for a further study.}.

\subsection{Phenomenological comparisons\label{Pheno}}

The goal of this Section is to identify the main phenomenological differences
between a dark photon embedded as a one or 2-form field, with or without
gauge invariance, and between a dark scalar embedded as a 0- or 3-form
field. For ease of reference, we repeat in Table~\ref{TableOps} the most
relevant effective operators for each scenario, now adopting the fermion mass
eigenstate basis instead of the chiral basis of Secs~\ref{EffOp0}
to~\ref{EffOp3}. Though the purpose of the present section is
phenomenological, we will not attempt to draw experimental constraints on the
coefficients of the various operators, but simply identify the main portals
through which the dark states could be looked for. A detailed numerical study
of the impact on the exclusion plot for dark matter searches is certainly
called for, but would require a dedicated study that we leave for a future work.

\begin{table}[t]
\renewcommand{\arraystretch}{1.5}\centering$%
\begin{tabular}
[c]{cccccc}\hline
&  & $\phi$ & $A$ & $B$ & $C$\\\hline
I & a & $-$ & $-$ & $F_{\mu\nu}^{\gamma}\tilde{B}^{\mu\nu}$ & $-$\\\cline{2-6}
& b & $-$ & $F_{\mu\nu}^{\gamma}F^{A,\mu\nu}$ & $-$ & $-$\\\hline
II & a & $\Lambda\Phi^{\dagger}\Phi\phi$ & $\Phi^{\dagger}\overleftrightarrow
{D}_{\mu}\Phi A^{\mu}$ & $-$ & $\Phi^{\dagger}\overleftrightarrow{D}_{\alpha
}\Phi\epsilon^{\mu\nu\rho\sigma}C_{\mu\nu\rho}$\\
&  & $\Phi^{\dagger}\Phi\phi^{2}$ & $\Phi^{\dagger}\Phi A_{\mu}A^{\mu}$ &
$\Phi^{\dagger}\Phi B_{\mu\nu}B^{\mu\nu},\Phi^{\dagger}\Phi B_{\mu\nu}%
\tilde{B}^{\mu\nu}$ & $\Phi^{\dagger}\Phi C^{\mu\nu\rho}C_{\mu\nu\rho}%
$\\\cline{2-6}
& b & $-$ & $-$ & $-$ & $\Phi^{\dagger}\Phi\epsilon^{\mu\nu\rho\sigma
}F_{\mu\nu\rho\sigma}^{C}$\\\hline
III & a & $\bar{\psi}\psi\phi$ & $\bar{\psi}\gamma^{\mu}\psi A_{\mu}$ &
$\bar{\psi}\sigma^{\mu\nu}\psi B_{\mu\nu}$ & $\bar{\psi}\gamma_{\sigma}%
\psi\epsilon^{\mu\nu\rho\sigma}C_{\mu\nu\rho}$\\
&  & $\bar{\psi}\gamma_{5}\psi\phi$ & $\bar{\psi}\gamma^{\mu}\gamma_{5}\psi
A_{\mu}$ & $\bar{\psi}\sigma^{\mu\nu}\psi\tilde{B}_{\mu\nu}$ & $\bar{\psi
}\gamma_{\sigma}\gamma_{5}\psi\epsilon^{\mu\nu\rho\sigma}C_{\mu\nu\rho}%
$\\\cline{2-6}
& b & $\bar{\psi}\gamma^{\mu}\psi F_{\mu}^{\phi}$ & $\bar{\psi}\sigma^{\mu\nu
}\psi F_{\mu\nu}^{A}$ & $\bar{\psi}\gamma_{\sigma}\gamma_{5}\psi
\epsilon^{\mu\nu\rho\sigma}F_{\mu\nu\rho}^{B}$ & $\bar{\psi}\gamma_{5}%
\psi\epsilon^{\mu\nu\rho\sigma}F_{\mu\nu\rho\sigma}^{C}$\\
&  & $\bar{\psi}\gamma^{\mu}\gamma_{5}\psi F_{\mu}^{\phi}$ & $\bar{\psi}%
\sigma^{\mu\nu}\psi\tilde{F}_{\mu\nu}^{A}$ & $\bar{\psi}\gamma_{\sigma}%
\psi\epsilon^{\mu\nu\rho\sigma}F_{\mu\nu\rho}^{B}$ & $\bar{\psi}%
\psi\epsilon^{\mu\nu\rho\sigma}F_{\mu\nu\rho\sigma}^{C}$\\\hline
IV & a & $\bar{\psi}\psi\phi^{2}$ & $\bar{\psi}\psi A_{\mu}A^{\mu}$ &
$\bar{\psi}\psi B_{\mu\nu}B^{\mu\nu},\ \ \bar{\psi}\psi B_{\mu\nu}\tilde
{B}^{\mu\nu}$ & $\bar{\psi}\psi C^{\mu\nu\rho}C_{\mu\nu\rho}$\\
&  & $\bar{\psi}\gamma_{5}\psi\phi^{2}$ & $\bar{\psi}\gamma_{5}\psi A_{\mu
}A^{\mu}$ & $\bar{\psi}\gamma_{5}\psi B_{\mu\nu}B^{\mu\nu},\ \ \bar{\psi
}\gamma_{5}\psi B_{\mu\nu}\tilde{B}^{\mu\nu}$ & $\bar{\psi}\gamma_{5}\psi
C^{\mu\nu\rho}C_{\mu\nu\rho}$\\\cline{2-6}
& b & $\bar{\psi}\psi F_{\mu}^{\phi}F^{\phi,\mu}$ & $\bar{\psi}\psi F_{\mu\nu
}^{A}F^{A,\mu\nu}$ & $\bar{\psi}\psi F_{\mu\nu\rho}^{B}F^{B,\mu\nu\rho}$ &
$\bar{\psi}\psi F_{\mu\nu\rho\sigma}^{C}F^{C,\mu\nu\rho\sigma}$\\
&  & $\bar{\psi}\gamma_{5}\psi F_{\mu}^{\phi}F^{\phi,\mu}$ & $\bar{\psi}%
\gamma_{5}\psi F_{\mu\nu}^{A}F^{A,\mu\nu}$ & $\bar{\psi}\gamma_{5}\psi
F_{\mu\nu\rho}^{B}F^{B,\mu\nu\rho}$ & $\bar{\psi}\gamma_{5}\psi F_{\mu\nu
\rho\sigma}^{C}F^{C,\mu\nu\rho\sigma}$\\\hline
\end{tabular}
\ \ $\renewcommand{\arraystretch}{1}\caption{Dominant operators for one or two
dark states to photons, scalars, and fermions, without or with dark gauge
invariance ($a$ and $b$, respectively). For the first two, only renormalizable
interactions are kept, while for fermions the leading operators in
each class are given.}%
\label{TableOps}%
\end{table}

\subsubsection{Dark photon gauge-breaking couplings}

When embedded into a vector field, and if gauge invariance does not hold,
the dominant operator among those that are renormalizable is $\bar{\psi}%
\gamma^{\mu}\gamma_{5}\psi A_{\mu}$ because its longitudinal component is
enhanced. Under $A\rightarrow A-F^{\phi_{A}}/m_{V}$, it becomes%
\begin{equation}
g_{A}\bar{\psi}\gamma^{\mu}\gamma_{5}\psi A_{\mu}\rightarrow\frac{g_{A}}%
{m_{V}}\bar{\psi}\gamma^{\mu}\gamma_{5}\psi\partial_{\mu}\phi_{A}\ .
\end{equation}
This is an axionlike effective interaction, but at the scale $m_{V}$ instead
of the typical scale in the $10^{9}$ GeV region. It thus requires a strong
suppression of its coupling if $m_{V}$ is light compared to the typical energy
scale of the considered process. By contrast, the $\bar{\psi}\gamma^{\mu}\psi
A_{\mu}$ operator, which also encodes the effect of the kinetic mixing after
the reparametrization of Eq.~(\ref{KmixReparam}), is purely transverse thanks
to the conservation of the current and is not enhanced in the $m_{V}%
\rightarrow0$ limit. The same holds for the $\Phi^{\dagger}\overleftrightarrow
{D}_{\mu}\Phi A^{\mu}$ coupling, whether $\Phi$ is a fundamental scalar or a
low-energy meson (see e.g. Ref.~\cite{Kamenik:2012hn}).

The exact opposite happens in the $B$ case. In the absence of gauge
invariance, the leading operator behaves under $B\rightarrow B-F^{A_{B}}%
/m_{V}$ as%
\begin{equation}
g_{T}\bar{\psi}\sigma^{\mu\nu}\psi B_{\mu\nu}+g_{\tilde{T}}\bar{\psi}%
\sigma^{\mu\nu}\psi\tilde{B}_{\mu\nu}\rightarrow\frac{g_{T}}{m_{V}}\bar{\psi
}\sigma^{\mu\nu}\psi F_{\mu\nu}^{A_{B}}+\frac{g_{\tilde{T}}}{m_{V}}\bar{\psi
}\sigma^{\mu\nu}\psi\tilde{F}_{\mu\nu}^{A_{B}}\ .\label{StueckVB}%
\end{equation}
Again, the starting operator being renormalizable, a strong suppression of
$g_{T,\tilde{T}}$ is required if $m_{V}$ is small. Yet, notice that it is now
the transverse component of $B$, encoded into $F^{A_{B}}$, that entirely
dominates in the $m_{V}\rightarrow0$ limit (if $A_{B}$ is treated as a massive
vector, its longitudinal component cancels out). Also, though similar
transverse operators are present in the 1-form picture, they are necessarily
suppressed by some scale $\Lambda$, with in general $\Lambda\gg m_{V}$.

The consequence of a mixing with the photon is also opposite for the $A$ and
$B$: while purely transverse for the $A$, it get enhanced in the
$m_{V}\rightarrow0$ limit for the $B$. As discussed before, adding
$e\Lambda_{\gamma}F_{\mu\nu}^{\gamma}\tilde{B}^{\mu\nu}$ to the $g_{T,\tilde
{T}}$ tensor operators, a tree-level exchange of the $B$ meson generates EDM
and MDM operators $e\bar{\psi}_{L}\sigma^{\mu\nu}\psi_{R}F_{\mu\nu}^{\gamma}$
and $e\bar{\psi}_{L}\sigma^{\mu\nu}\psi_{R}\tilde{F}_{\mu\nu}^{\gamma}$, which
are tightly bounded. Those scale as $g_{T,\tilde{T}}\Lambda_{\gamma}/m_{V}%
^{2}$, so if $\Lambda_{\gamma}\approx m_{V}$, the bound on $g_{T,\tilde{T}}$
would be so strict that a direct $B$ interaction is unlikely to be ever seen.
We thus arrive at the same conclusion as using duality: $\Lambda_{\gamma}$
needs be very suppressed.

All in all, both scenarios require some level of fine-tuning of their
parameters to be viable. In both cases, the non-gauge invariant couplings have
to be suppressed. This makes the situation for the $B$ field slightly less
appealing since both its coupling to fermions and to the photon break gauge
invariance. Yet, from a phenomenological perspective, it is possible that
$e\Lambda F_{\mu\nu}^{\gamma}\tilde{B}^{\mu\nu}$ is absent because of its odd
parity. If that is the case, it would be worth to search for the dark photon
not only via its vector coupling to matter, but to also probe for tensor interactions.

There is yet another feature to analyze. The $B$ field is unique in that it
can have both a parity conserving and parity-violating mass term, $m^{2}%
B_{\mu\nu}B^{\mu\nu}$ and $\tilde{m}^{2}B_{\mu\nu}\tilde{B}^{\mu\nu}$, with
the physical mass $m_{V}^{2}=(m^{4}+\tilde{m}^{4})/m^{2}$. As discussed
earlier, this does not help with MDM and EDM constraints, and $\Lambda
_{\gamma}$ must still be extremely suppressed even though parity is no longer
of much help since $B_{\mu\nu}\tilde{B}^{\mu\nu}$ and $F_{\mu\nu}^{\gamma
}\tilde{B}^{\mu\nu}$ are both parity odd. Nevertheless, assuming that it is the
case, the phenomenology is then different and the Stueckelberg substitution
fails to capture all the $1/m_{V}$ terms. To be more specific, we know from
Eq.~(\ref{PropMtilde}) that if $B$ is coupled to some current $J$, then%
\begin{equation}
\mathcal{M}(B\rightarrow J)=\varepsilon_{(\lambda)}^{\mu\nu}J_{\mu\nu
}\ ,\ \ \sum_{\lambda}\varepsilon_{(\lambda)}^{\ast\alpha\beta}\varepsilon
_{(\lambda)}^{\mu\nu}=2\left(  \mathcal{I}_{0}^{\alpha\beta,\mu\nu}-\frac
{2}{m^{2}}\mathcal{I}_{2}^{\alpha\beta,\mu\nu}-\frac{\tilde{m}^{2}}{m^{2}%
}\mathcal{I}_{3}^{\alpha\beta,\mu\nu}\right)  \ .\label{BpolS}%
\end{equation}
There is no substitution $B\rightarrow B-xF^{A_{B}}-y\tilde{F}^{A_{B}}$ for
some $x$ and $y$ that would alone generate only the $1/m^{2}$ term (instead,
there exist $x$, $y$ such that $xF^{A_{B}}+y\tilde{F}^{A_{B}}$ alone
reproduces the whole polarization sum, see Eq.~(\ref{mtildePol})).

At first sight, there is no manifest pole as $m_{V}\rightarrow0$, but remember
that $m_{V}\rightarrow0$ is attainable only if both $m\rightarrow0$ and
$\tilde{m}\rightarrow0$. For a given $m_{V}$, the maximum $m\rightarrow m_{V}$
is attained when $\tilde{m}\rightarrow0$, while the maximum $\tilde
{m}\rightarrow m_{V}/\sqrt{2}$ is reached when $m\rightarrow m_{V}/\sqrt{2}$,
but both $m$ and $\tilde{m}$ can be as small as one wishes. Actually, for a
given $m_{V}$ and a value for $m\leqslant m_{V}$, it suffices to take
\begin{equation}
\tilde{m}^{2}=m^{2}\sqrt{m_{V}^{2}/m^{2}-1}\ .\label{mtmV}%
\end{equation}
Thus, the $\mathcal{I}_{3}$ contribution is always smaller than that of
$\mathcal{I}_{2}$, but both get enhanced when $m\rightarrow0$.  One should
understand though that this singularity is totally similar to the $1/m_{V}$
singularities in Eq.~(\ref{StueckVB}) since $m_{V}=m$ when $\tilde{m}=0$.
Further, one should remember that it is $m$ that tames the singularity of the
propagator, i.e., which ensures the $B$ kinetic term can be inverted. In both
cases, the underlying physics abruptly changes when $m\rightarrow0$, which is
thus not attainable, and this shows up phenomenologically as a boost in the
rate when $m$ becomes small compared to the typical energy scale of the
process. All that happens in the presence of the $\tilde{m}$ term is to
decorrelate the physical mass $m_{V}$ from the parameter $m$, but the
underlying physics stay identical. In practice, for the tensor operators of
Eq.~(\ref{StueckVB}), the $\mathcal{I}_{3}$ component cancels out completely
(we will encounter later on a case where the $\mathcal{I}_{3}$ component does
contribute), but the presence of $\tilde{m}$ can make $m$ reach values much
smaller than $m_{V}$, thereby strengthening the bounds on $g_{T,\tilde{T}}$.

\subsubsection{Dark photon gauge-invariant couplings}

If the $A$ and $B$ gauge-invariance hold for the effective interactions, but
still assuming the dark photon gets a small mass term, the situations are
again opposite\footnote{Notice that it is not sufficient to break gauge
invariance softly, as this would still allow for the $B$-$\gamma$ mixing
terms. Instead, we assume that breaking occurs in a secluded sector, and does
not directly affect the $B$ couplings to SM particles.}. When all the
effective interactions are written in terms of field strengths, the extra
piece in the Stueckelberg substitution disappears. In effect, only the degrees
of freedom that would exist if the field was massless can contribute. Thus, if
we look at the effective couplings to fermions, the dominant interactions are
always of dimension five, with the purely transverse $\bar{\psi}\sigma^{\mu
\nu}\psi F_{\mu\nu}^{A}$, $\bar{\psi}\sigma^{\mu\nu}\psi\tilde{F}_{\mu\nu}%
^{A}$ for $A$, and the purely longitudinal $\bar{\psi}\gamma_{\sigma}%
\psi\epsilon^{\mu\nu\rho\sigma}F_{\mu\nu\rho}^{B}$, $\bar{\psi}%
\gamma_{\sigma}\gamma_{5}\psi\epsilon^{\mu\nu\rho\sigma}F_{\mu\nu\rho}^{B}$
for $B$. Whether $\tilde{m}$ is zero or not is irrelevant here since the
$\mathcal{I}_{2,3}$ components of the polarization sum Eq.~(\ref{BpolS})
cancel out, leaving $\mathcal{I}_{0}$ only. 

A major difference though is the fact that the mixing of the photon with $A$ is gauge invariant, but not that with $B$, which is thus now forbidden. Starting with $\varepsilon F^A_{\mu\nu}F^{\gamma,\mu\nu}$, there will then remain dimension-four couplings of the form $\varepsilon e\bar{\psi}\gamma^{\mu}\psi
A_{\mu}$ (and $\varepsilon e\Phi^{\dagger}\overleftrightarrow{D}_{\mu}\Phi A^{\mu}$), from
which $\varepsilon$ is experimentally constrained to be small. In this case,
it is the $B$ embedding that appears more natural, requiring no fine-tuning at all.

The $\bar{\psi}\gamma_{\sigma}\psi\epsilon^{\mu\nu\rho\sigma}F_{\mu\nu\rho
}^{B}$ and $\bar{\psi}\gamma_{\sigma}\gamma_{5}\psi\epsilon^{\mu\nu
\rho\sigma}F_{\mu\nu\rho}^{B}$ couplings produce only longitudinally polarized
$B$, but are not equivalent to axionlike couplings. We cannot use massless
dualities to represent $\epsilon^{\mu\nu\rho\sigma}F_{\mu\nu\rho}%
^{B}\rightarrow\partial^{\sigma}\phi_{B}$ since contact terms would contribute
for $k^{2}\neq0$. To be more specific, let us compare the off-shell squared
amplitude obtained from either
\begin{align}
\mathcal{L}_{eff} &  \supset\frac{g_{V}}{\Lambda}\bar{\psi}_{1}\gamma_{\mu
}\psi_{2}\partial^{\mu}\phi+\frac{g_{A}}{\Lambda}\bar{\psi}_{1}\gamma_{\mu
}\gamma_{5}\psi_{2}\partial^{\mu}\phi+h.c.\ ,\\
\mathcal{L}_{eff} &  \supset\frac{g_{V}}{\Lambda}\bar{\psi}_{1}\gamma_{\mu
}\psi_{2}\epsilon^{\mu\nu\rho\sigma}F_{\mu\nu\rho}^{B}+\frac{g_{A}}%
{\Lambda}\bar{\psi}_{1}\gamma_{\mu}\gamma_{5}\psi_{2}\epsilon^{\mu\nu
\rho\sigma}F_{\mu\nu\rho}^{B}+h.c.\ ,
\end{align}
which are for $\phi$:%
\begin{align}
|\mathcal{M}(\phi(k)\overset{}{\rightarrow}\bar{\psi}_{1}\psi_{2})|^{2} &
=2g_{A}^{2}(m_{1}+m_{2})^{2}(k^{2}-(m_{1}-m_{2})^{2})\nonumber\\
&  \ \ \ +2g_{V}^{2}(m_{1}-m_{2})^{2}(k^{2}-(m_{1}+m_{2})^{2})\ ,
\end{align}
and for $B$:%
\begin{align}
\sum_{\lambda}|\mathcal{M}(B(k)\overset{}{\rightarrow}\bar{\psi}_{1}\psi
_{2})|^{2} &  =2g_{A}^{2}(k^{2}-(m_{1}+m_{2})^{2})(2k^{2}+(m_{1}-m_{2}%
)^{2})\nonumber\\
&  \ \ \ +2g_{V}^{2}(k^{2}-(m_{1}-m_{2})^{2})(2k^{2}+(m_{1}+m_{2})^{2})\ .
\end{align}
Clearly, the vector current contributes even for $m_{1}=m_{2}$, when it is
conserved, for the $B$ field but not for the $\phi$ field, showing that these
two scenarios are intrinsically different. Notice that these two amplitudes
match at $k^{2}=0$ though, as they should from Eq.~(\ref{SpinBphi}). Finally,
it should be stressed that introducing direct couplings to $B$ is not the same
as introducing them via the Stueckelberg component of a 3-form, as in Refs.~\cite{dvali2005threeformgaugingaxionsymmetries,dvali2022strongcpgravity}
. Starting with the couplings $\bar{\psi}\gamma_{\mu}\psi\epsilon^{\mu
\nu\rho\sigma}C_{\mu\nu\rho}$ and $\bar{\psi}\gamma_{\mu}\gamma_{5}%
\psi\epsilon^{\mu\nu\rho\sigma}C_{\mu\nu\rho}$, substituting $C\rightarrow
C-F^{B_{C}}/m$, the $F^{B_{C}}$ accounts for the $\mathcal{I}_{2}$ component
in Eq.~(\ref{polsum}). On shell, the squared amplitude for $C$ is the same as
that of $\phi$. In particular, the current being conserved for $\bar{\psi
}\gamma_{\mu}\psi\epsilon^{\mu\nu\rho\sigma}C_{\mu\nu\rho}$, both the
$\phi\rightarrow\bar{\psi}\psi$ and $C\rightarrow\bar{\psi}\psi$ squared
amplitude vanish. In terms of components, the $\mathcal{I}_{2}$ contribution
in Eq.~(\ref{polsum}) becomes proportional to $k^{2}$, allowing it to cancel
exactly with the $\mathcal{I}_{0}$ part at $k^{2}=m^{2}$. Yet, that
$\mathcal{I}_{2}$ part does not vanish by itself, even on shell. This means
that whenever the $B$ field is fundamentally a vector field with three true
degrees of freedom, its longitudinal component does couple to conserved
fermionic currents.

\subsubsection{Dark scalar couplings}

Let us now compare the $\phi$ and $C$ pictures for a scalar field of mass
$m_{S}$. Notice first that neither of these states mixes with the photon,
simplifying the analysis compared to the dark vector scenario. On the other
hand, the Stueckelberg substitution is not particularly interesting here since
both $\phi$ and $C$ carry a unique degree of freedom. As a result, it does not
exist for $\phi$, and would substitute $C$ by $C-F^{B}/m_{S}$ with $F^{B}$ not
related to $\phi$ (as discussed in the previous Section). So, this cannot help
to single out and characterize possible $1/m_{S}$ enhancements. What we can
use instead are the dualities discussed previously, which relate algebraically
the on shell squared amplitudes for $\phi$ and $C$.

Let us start with the fermionic couplings, which for $\phi$ and $C$ are either
to the vector and axial currents $V,A=\bar{\psi}\gamma_{\sigma}\psi,\bar{\psi
}\gamma_{\sigma}\gamma_{5}\psi$ or to the scalar and pseudoscalar currents
$S,P=\bar{\psi}\psi,\bar{\psi}\gamma_{5}\psi$. Dimensionally, the situation is
inverted for the $C$ and the $\phi$:%
\begin{align}
\mathcal{L}_{eff} &  \supset g_{S}\bar{\psi}_{1}\psi_{2}\phi+g_{P}\bar{\psi
}_{1}\gamma_{5}\psi_{2}\phi\\
&  \ \ \ +\frac{g_{V}}{\Lambda_{\phi}}\bar{\psi}_{1}\gamma^{\mu}\psi_{2}%
F_{\mu}^{\phi}+\frac{g_{A}}{\Lambda_{\phi}}\bar{\psi}_{1}\gamma^{\mu}%
\gamma_{5}\psi_{2}F_{\mu}^{\phi}+h.c.\ ,\\
\mathcal{L}_{eff} &  \supset g_{V}\bar{\psi}_{1}\gamma_{\sigma}\psi
_{2}\epsilon^{\mu\nu\rho\sigma}C_{\mu\nu\rho}+g_{A}\bar{\psi}_{1}%
\gamma_{\sigma}\gamma_{5}\psi_{2}\epsilon^{\mu\nu\rho\sigma}C_{\mu\nu\rho
}\\
&  \ \ \ +\frac{g_{S}}{\Lambda_{C}}\bar{\psi}_{1}\psi_{2}\epsilon^{\mu
\nu\rho\sigma}F_{\mu\nu\rho\sigma}^{C}+\frac{g_{P}}{\Lambda_{C}}\bar{\psi}%
_{1}\gamma_{5}\psi_{2}\epsilon^{\mu\nu\rho\sigma}F_{\mu\nu\rho\sigma}%
^{C}+h.c.\ .
\end{align}
From Eq.~(\ref{dualmassC}), the $S$ and $P$ interactions are dominant for
$\phi$, and related to the corresponding subdominant $C$ interactions via%
\begin{equation}
|\mathcal{M}_{S,P}(C(k)\rightarrow\bar{\psi}_{1}\psi_{2})|^{2}=\frac{k^{2}%
}{\Lambda_{C}^{2}}|\mathcal{M}_{S,P}(\phi(k)\rightarrow\bar{\psi}_{1}\psi
_{2})|^{2}\ .
\end{equation}
The $k^{2}$ factor can be understood by noting that the contact term satisfies
$|\mathcal{M}_{S,P}(\phi(k)\rightarrow\bar{\psi}_{1}\psi_{2})|^{2}=J^{2}$ when
$J=\bar{\psi}_{1}\psi_{2}$ or $\bar{\psi}_{1}\gamma_{5}\psi_{2}$, which is
quite evident from a Feynman rule perspective. Phenomenologically, this means
that the gauge-invariant scalar and pseudoscalar interactions for $C$ are
significantly more suppressed than expected when $C$ is light and on shell.
The opposite holds for the $V$ and $A$ interactions, which are dominant for
the $C$ and suppressed for $\phi$, but related as:%
\begin{equation}
|\mathcal{M}_{V,A}(\phi(k)\rightarrow\bar{\psi}_{1}\psi_{2})|^{2}=\frac
{m_{S}^{2}}{\Lambda_{\phi}^{2}}|\mathcal{M}_{V,A}(C(k)\rightarrow\bar{\psi
}_{1}\psi_{2})|^{2}+\emph{O}\left(  \frac{k^{2}-m_{S}^{2}}{\Lambda_{\phi}^{2}%
}\right)  \ .
\end{equation}
Notice though that for $C$ off shell, the squared amplitudes are intrinsically
different functions of $k^{2}$, the mass squared, and $g_{A,V}$ couplings.

An important peculiarity though is that $\mathcal{M}_{V,A}(\phi(k)\rightarrow
\bar{\psi}\psi)$ is itself related to $\mathcal{M}_{S,P}(\phi(k)\rightarrow
\bar{\psi}\psi)$ since the couplings are by partial integration and use of the
equation of motion. Let us take $\psi_{1}=\psi_{2}$ for simplicity, in which
case
\begin{equation}
\mathcal{M}_{V}(\phi(k)\rightarrow\bar{\psi}\psi)=0\ ,\ \mathcal{M}_{A}%
(\phi(k)\rightarrow\bar{\psi}\psi)=\frac{2m}{\Lambda_{\phi}}\mathcal{M}%
_{P}(\phi(k)\rightarrow\bar{\psi}\psi)\ .
\end{equation}
This means that the dominant $C$ interactions must satisfy%
\begin{align}
|\mathcal{M}_{V}(C(k)  &  \rightarrow\bar{\psi}\psi)|^{2}=\emph{O}\left(
\frac{k^{2}-m_{S}^{2}}{m_{S}^{2}}\right)  \ ,\ \ \\
|\mathcal{M}_{A}(C(k)  &  \rightarrow\bar{\psi}\psi)|^{2}=4\frac{m_{\psi}^{2}%
}{m_{S}^{2}}\frac{\Lambda_{C}^{2}}{k^{2}}|\mathcal{M}_{P}(C(k)\rightarrow
\bar{\psi}\psi)|^{2}+\emph{O}\left(  \frac{k^{2}-m_{S}^{2}}{m_{S}^{2}}\right)
\ .
\end{align}
These are two striking predictions: the on shell vector current coupling gives
no contribution on-shell, while the axial one must be proportional to the
fermion mass squared (it is actually equal to $8g_{A}^{2}m_{\psi}^{2}$).\ This
could not have been expected on the basis of the operators alone.

Concerning the couplings to scalar fields, the main difference with the dark
vector case is that both $\phi$ and $C$ do have linear renormalizable
couplings. Their properties are quite different though. In the $\phi$ picture,
the presence of $\Lambda\Phi^{\dagger}\Phi\phi$ generates tadpoles if $\Phi$
is the Higgs doublet, which is then required to shift the $\phi$ field by a large
constant. In that case, all the other effective interactions better be
shift invariant, otherwise SM particles could all receive large corrections to
their masses and couplings, in particular from the (also renormalizable)
$\bar{\psi}\psi\phi$ and $\bar{\psi}\gamma_{5}\psi\phi$ couplings. So, either
$\Phi^{\dagger}\Phi\phi$ is severely fine-tuned, or these fermionic
interactions are forbidden. By contrast, in the $C$ picture, the tadpole
interaction from $\Phi^{\dagger}\Phi\epsilon^{\mu\nu\rho\sigma}F_{\mu
\nu\rho\sigma}^{C}$ automatically drops out since it involves a derivative. All that remains then is
effective interactions with the $Z$ boson and the physical Higgs field from
\begin{equation}
g_{\Phi}\Phi^{\dagger}\overleftrightarrow{D}_{\alpha}\Phi\epsilon^{\mu
\nu\rho\sigma}C_{\mu\nu\rho}\rightarrow g_{\Phi}g(v_{ew}+h)^{2}Z_{\mu
}\epsilon^{\mu\nu\rho\sigma}C_{\mu\nu\rho}\ ,
\end{equation}
with $g$ the electroweak coupling and $v_{ew}$ the electroweak vacuum expectation value. This includes a parity-odd mixing of $C$
with the $Z$ boson. A detailed analysis is left for a future work. Here,
sticking to a low energy perspective and integrating the $Z$ boson out, this
mixing term combined with the $Z$ couplings to light fermions simply generates
$\mathcal{O}(g_{\Phi})$ corrections to the renormalizable $\bar{\psi}%
\gamma_{\sigma}\psi\epsilon^{\mu\nu\rho\sigma}C_{\mu\nu\rho}$ and
$\bar{\psi}\gamma_{\sigma}\gamma_{5}\psi\epsilon^{\mu\nu\rho\sigma}%
C_{\mu\nu\rho}$ interactions. As such, since the former decouples on shell,
constraints on $g_{A}$ immediately translate into similar constraints on
$g_{\Phi}$.

\subsubsection{Two dark field couplings and differential rates}

Looking at the class IV operators in Table~\ref{TableOps}, the dominant
interactions to pairs of dark states share essentially the same structure. For
gauge-breaking operators, the contraction $A\wedge\star A\rightarrow
A_{\mu_{1}...\mu_{p}}A^{\mu_{1}...\mu_{p}}$ is coupled to either fermions via
$\bar{\psi}\psi$ and $\bar{\psi}\gamma_{5}\psi$, or to scalars via
$\Phi^{\dagger}\Phi$. The only exception is the $B$ field, for which the dark
state can also occur in the pseudoscalar combination $B\wedge B\rightarrow
B_{\mu\nu}\tilde{B}^{\mu\nu}$. Gauge-invariant operators are similar, with
$dA\wedge\star dA\rightarrow F_{\mu_{1}...\mu_{p+1}}F^{\mu_{1}...\mu_{p+1}}$
in place of $A\wedge\star A$, and now the exceptional case of the vector that
also couples via $dA\wedge dA\rightarrow F_{\mu\nu}\tilde{F}^{A,\mu\nu}$.

Though all these couplings are very similar, they do not lead to identical
predictions. To illustrate this, we will consider three-body processes like
$\psi_{1}\rightarrow\psi_{2}XX$ or $\Phi_{1}\rightarrow\Phi_{2}XX$, with $X$ a
given dark state, and compare the differential rates as functions of the $XX$
invariant mass $q^{2}=(k_{1}+k_{2})^{2}$,%
\begin{equation}
\frac{d\Gamma}{dq^{2}}=\frac{1}{2m_{a}}\frac{\lambda\left(  m_{a}^{2}%
,m_{b}^{2},q^{2}\right)  \lambda\left(  q^{2},m_{X}^{2},m_{X}^{2}\right)
}{128\pi^{3}}\frac{1}{s_{a}}\sum_{\lambda,s_{a},s_{b}}\left\vert
\mathcal{M}\left(  a\rightarrow bXX\right)  \right\vert ^{2}\ ,
\label{diffrate}
\end{equation}
where $\lambda^{2}\left(  a,b,c\right)  =a^{2}+b^{2}+c^{2}-2\left(
ab+ac+bc\right)  $, which ranges between $q_{\min}^{2}=4m_{X}^{2}$ and
$q_{\max}^{2}=(m_{a}-m_{b})^{2}$.\ For fermions ($s_{a}=2$), this requires
some flavor violation, but our purpose here is illustrative. For scalars
($s_{a}=1$), similarly, we think of $\Phi_{1}$ and $\Phi_{2}$ as low-energy
pseudoscalar mesons. In this respect, it should be stressed that if the
combination $\Phi^{\dagger}\Phi$ involves the Higgs doublet, then all the
$\Phi^{\dagger}\Phi(A\wedge\star A)$ operators generate electroweak mass terms
for the dark states. As usual, maintaining those states light necessarily
requires some level of fine-tuning there, or one needs to assume the gauge (or
shift) symmetry holds. 

\begin{figure}[t]
\centering\includegraphics[width=0.95\textwidth]{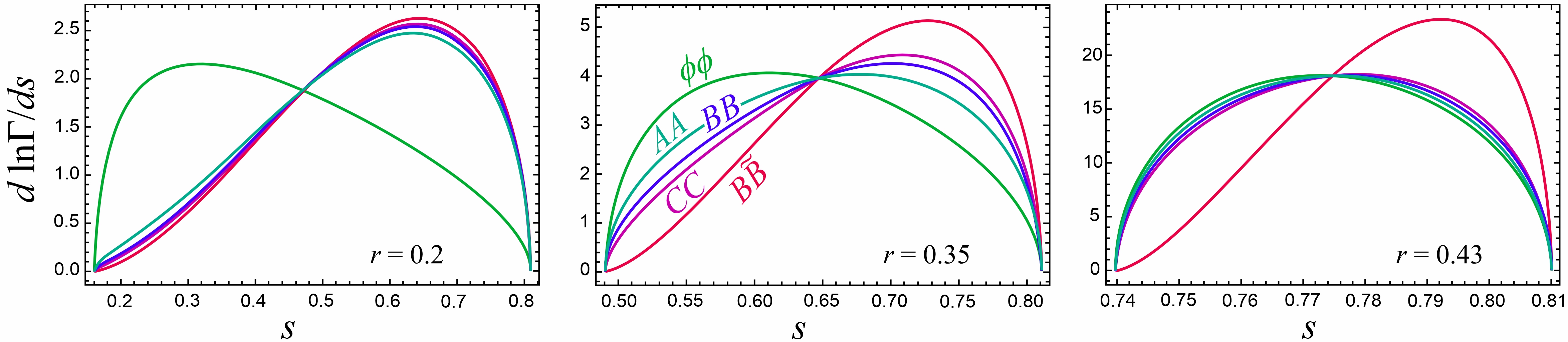}\caption{Normalized
differential rates for $a\rightarrow bXX$, $X=\phi,A,B,C$ as produced via the
$\bar{\psi}_{a}\psi_{b}X\wedge\star X$ and $\bar{\psi}_{a}\psi_{b}B\wedge B$
couplings, see Eq.~(\ref{diffrate}). We take arbitrary units and set $s=q^{2}/m_{a}^{2}$, $m_{b}%
/m_{a}=0.1$, $r=m_{X}/m_{a}$. }%
\label{Fig1}%
\end{figure}

Though completely straightforward, it is instructive to perform the
calculation explicitly. The Feynman rules associated to $A\wedge\star A$ and
$dA\wedge\star dA$, upon proper normalization, are
\begin{subequations}
\label{FRules}%
\begin{align}
\frac{A_{\mu_{1}...\mu_{p}}A^{\mu_{1}...\mu_{p}}}{2p!}  &  \rightarrow
\mathcal{M}_{AA}=\frac{1}{p!}\mathcal{I}_{0}^{\mu_{1}...\mu_{p},\nu_{1}%
...\nu_{p}}\varepsilon_{\mu_{1}...\mu_{p}}^{\lambda_{1}}\varepsilon_{\nu
_{1}...\nu_{p}}^{\lambda_{2}}\ ,\\
\frac{F_{\mu_{1}...\mu_{p+1}}F^{\mu_{1}...\mu_{p+1}}}{2(p+1)!}  &
\rightarrow\mathcal{M}_{FF}=\frac{p+1}{p!}\mathcal{I}_{0}^{\mu_{1}...\mu
_{p+1},\nu_{1}...\nu_{p+1}}k_{1,\mu_{1}}k_{2,\nu_{1}}\varepsilon_{\mu
_{2}...\mu_{p+1}}^{\lambda_{1}}\varepsilon_{\nu_{2}...\nu_{p+1}}^{\lambda_{2}%
}\ .
\end{align}
\end{subequations}
Given the factorized form of the full $a\rightarrow bXX$ amplitudes
$\mathcal{M}_{a\rightarrow bXX}=\mathcal{M}_{a\rightarrow B}\mathcal{M}_{XX}$,
these vertices can be separately squared and summed over polarizations, and we
find
\begin{subequations}
\begin{align}
\sum_{\lambda_{1},\lambda_{2}}|\mathcal{M}_{CC}|^{2}  &  =\frac{(0!)^{2}%
}{m_{S}^{4}}|\mathcal{M}_{F^{\phi}F^{\phi}}|^{2}=1-\frac{q^{2}}{m_{S}^{2}%
}+\frac{q^{4}}{4m_{S}^{4}}\ ,\label{mCC}\\
\sum_{\lambda_{1},\lambda_{2}}|\mathcal{M}_{BB}|^{2}  &  =\frac{(1!)^{2}%
}{m_{V}^{4}}\sum_{\lambda_{1},\lambda_{2}}|\mathcal{M}_{F^{A}F^{A}}%
|^{2}=3-\frac{2q^{2}}{m_{V}^{2}}+\frac{q^{4}}{2m_{V}^{4}}\ ,\label{mBB}\\
\sum_{\lambda_{1},\lambda_{2}}|\mathcal{M}_{AA}|^{2}  &  =\frac{(2!)^{2}%
}{m_{V}^{4}}\sum_{\lambda_{1},\lambda_{2}}|\mathcal{M}_{F^{B}F^{B}}%
|^{2}=3-\frac{q^{2}}{m_{V}^{2}}+\frac{q^{4}}{4m_{V}^{4}}\ ,\label{mAA}\\
|\mathcal{M}_{\phi\phi}|^{2}  &  =\frac{(3!)^{2}}{m_{S}^{4}}\sum_{\lambda
_{1},\lambda_{2}}|\mathcal{M}_{F^{C}F^{C}}|^{2}=1\ . \label{mphiphi}%
\end{align}
\label{mAAFF}
\end{subequations}
The vertices $B_{\mu\nu}\tilde{B}^{\mu\nu}$ and $F_{\mu\nu}^{A}\tilde
{F}^{A,\mu\nu}$ can be treated similarly. Their Feynman rules are identical to
that in Eq.~(\ref{FRules}) but for $\mathcal{I}_{0}\rightarrow\mathcal{I}_{3}%
$, and we find
\begin{equation}
\sum_{\lambda_{1},\lambda_{2}}|\mathcal{M}_{B\tilde{B}}|^{2}=\frac{(1!)^{2}%
}{m_{V}^{4}}\sum_{\lambda_{1},\lambda_{2}}|\mathcal{M}_{F^{A}\tilde{F}^{A}%
}|^{2}=-\frac{2q^{2}}{m_{V}^{2}}+\frac{q^{4}}{2m_{V}^{4}}\ . \label{mBBt}%
\end{equation}
In principle, the $B_{\mu\nu}\tilde{B}^{\mu\nu}$ and $F_{\mu\nu}^{A}\tilde
{F}^{A,\mu\nu}$ contributions should be added to that of $B_{\mu\nu}B^{\mu\nu
}$ and $F_{\mu\nu}^{A}F^{A,\mu\nu}$ at the amplitude level, but producing the
dark pairs in different states, they do not interfere. 

The $\mathcal{M}_{a\rightarrow b}$ amplitude, once squared and appropriately summed and averaged, is $((m_{1}+m_{2})^{2}-T^{2})$, $((m_{1}-m_{2})^{2}-T^{2})$, and $1$
for $\bar{\psi}_{1}\psi_{2}$, $\bar{\psi}_{1}\gamma_{5}\psi_{2}$, and
$\Phi_{1}^{\dagger}\Phi_{2}$, respectively. The different mass dimensionality
comes from that of the vertices, with $\Phi^{\dagger}\Phi(A\wedge\star A)$ of
dimension four, but $\bar{\psi}\psi(A\wedge\star A)$ of dimension five, so
there is an implicit $\Lambda^{-2}$ factor involved for fermions. The same
holds for relating the $A\wedge\star A$ and $F^{A}\wedge\star F^{A}$ squared amplitudes.

Clearly, the massive dualities discussed before are at play to explain the
relationships between $|\mathcal{M}_{AA}|^{2}$ and $|\mathcal{M}_{FF}|^{2}$,
see in particular Eqs.~(\ref{CmassEoM}) and~(\ref{dualmassC}) for $\phi-C$,
and Eqs.~(\ref{BmassEoM}) and~(\ref{BdualA}) for $A-B$. Yet, importantly, the
four $|\mathcal{M}_{AA}|^{2}$ do predict different kinematics, and this is
then reflected in the corresponding differential rates, see Fig.~\ref{Fig1}.
It is not the same to produce two dark scalars via either their $\phi$ or
$C$ representations, or to produce two dark photons via either their $A$ or
$B$ representations. As apparent in Fig.~\ref{Fig1}, all but the $\phi$
normalized differential rates tend to the same curve when $m_{V}\rightarrow0$
because the $q^{4}$ component then dominates, and all but the $B\tilde{B}$
normalized differential rates coincide when $m_{V}\rightarrow(m_{a}-m_{b})/2$
because the constant term then dominates.

\begin{figure}[t]
\centering\includegraphics[width=0.35\textwidth]{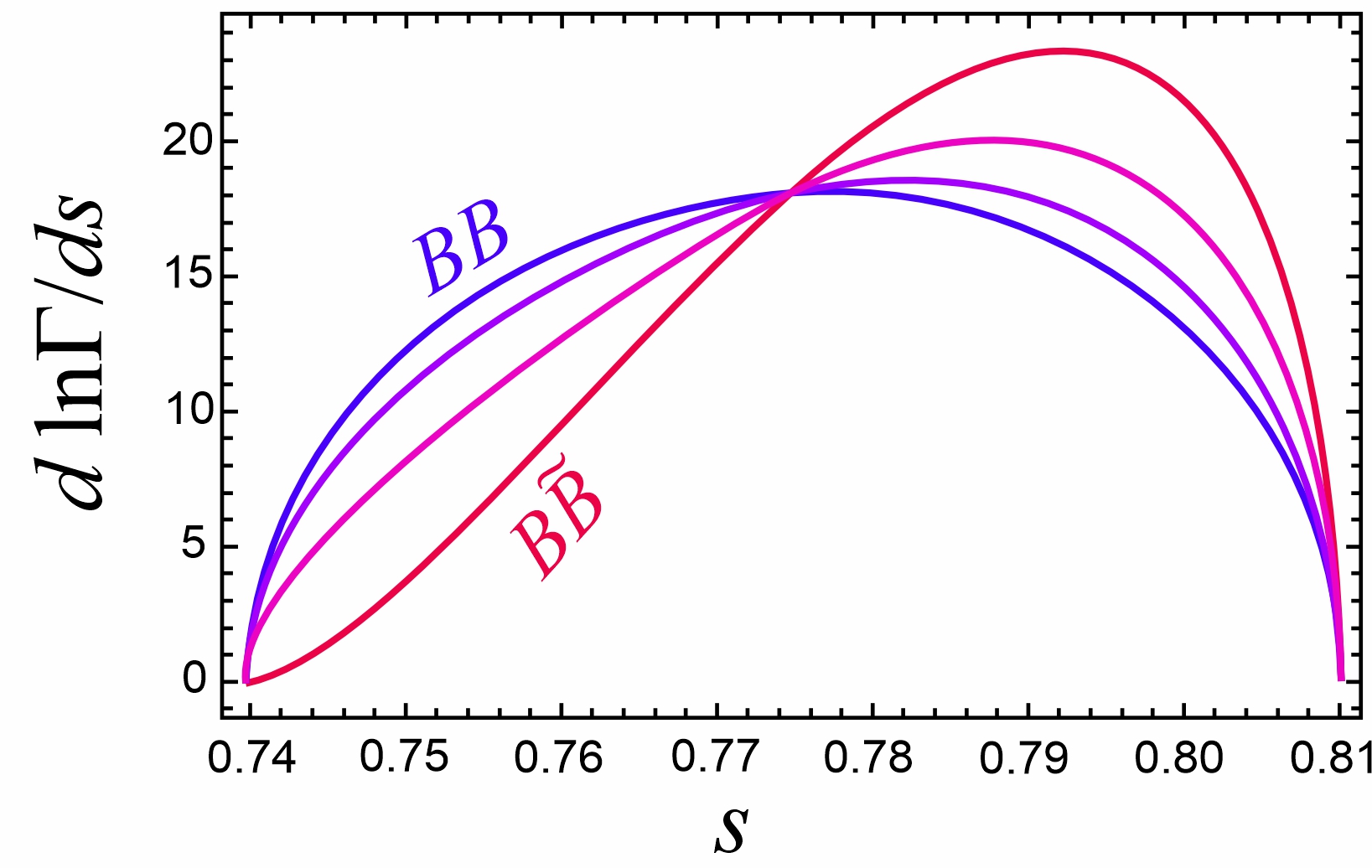}\caption{Normalized
differential rates for $a\rightarrow bBB$ via $g_{V}\bar{\psi}_{a}\psi
_{b}B_{\mu\nu}B^{\mu\nu}+g_{A}\bar{\psi}_{a}\psi_{b}B_{\mu\nu}\tilde{B}^{\mu\nu}$, with $(g_{V},g_{A})=(1,0)$,
$(1,2)$, $(1,9)$, and $(0,1)$, for the kinematical situation depicted in
Fig.~\ref{Fig1} for $r=0.43$.}%
\label{Fig2}%
\end{figure}

For $B$, there is also the possibility to turn on the $\tilde{m}^{2}B_{\mu\nu
}\tilde{B}^{\mu\nu}$ mass term, bringing a $\mathcal{I}_{3}$ component in the
polarization sum, Eq.~(\ref{BpolS}). In its presence, the $a\rightarrow bBB$
and $a\rightarrow bB\tilde{B}$ amplitudes start to interfere. Introducing
$g_{A,V}$ couplings for the operators involving $B_{\mu\nu}B^{\mu\nu}$ and
$B_{\mu\nu}\tilde{B}^{\mu\nu}$, respectively, we find%
\begin{align}
\sum_{\lambda_{1},\lambda_{2}}|\mathcal{M}_{g_{V}BB+g_{A}B\tilde{B}}|^{2} &
=g_{V}^{2}\left(  12-12\frac{m_{V}^{2}}{m^{2}}+3\frac{m_{V}^{4}}{m^{4}}%
-2\frac{m_{V}^{2}}{m^{2}}\frac{q^{2}}{m^{2}}+\frac{q^{4}}{2m^{4}}\right)
\nonumber\\
&  +g_{A}^{2}\left(  -12+12\frac{m_{V}^{2}}{m^{2}}-2\frac{m_{V}^{2}}{m^{2}%
}\frac{q^{2}}{m^{2}}+\frac{q^{4}}{2m^{4}}\right)  \nonumber\\
&  +12g_{V}g_{A}\sqrt{\frac{m_{V}^{2}}{m^{2}}-1}\left(  \frac{m_{V}^{2}}%
{m^{2}}-2\right)  \ ,\label{mBBBBt}%
\end{align}
where $\tilde{m}$ has been expressed as in Eq.~(\ref{mtmV}). We recover
Eqs.~(\ref{mBB}) and~(\ref{mBBt}) if $m\rightarrow m_{V}$, i.e., when
$\tilde{m}\rightarrow0$. In this expression, $m$ is essentially free, so the
total rate becomes singular if $m\rightarrow0$.\ As explained after
Eq.~(\ref{mtmV}), the nature of that singularity is totally similar to the
$1/m_{V}$ singularities in Eqs.~(\ref{mBB}) and~(\ref{mBBt}).

Though not immediately apparent, the massive duality of Eq.~(\ref{EoMmmt}) is
still satisfied, and one can check that
\begin{align}
\sum_{\lambda_{1},\lambda_{2}}|\mathcal{M}_{g_{V}BB+g_{A}B\tilde{B}}|^{2} &
=\left(  \frac{g_{V}(m^{4}-\tilde{m}^{4})-2g_{A}m^{2}\tilde{m}^{2}}{m^{4}%
}\right)  ^{2}\sum_{\lambda_{1},\lambda_{2}}|\mathcal{M}_{BB}|^{2}\nonumber\\
&  +\left(  \frac{g_{A}(m^{4}-\tilde{m}^{4})+2g_{V}m^{2}\tilde{m}^{2}}{m^{4}%
}\right)  ^{2}\sum_{\lambda_{1},\lambda_{2}}|\mathcal{M}_{B\tilde{B}}|^{2}\ ,
\end{align}
provided $m_{V}^{2}=(m^{4}+\tilde{m}^{4})/m^{2}$, with the $BB$ and
$B\tilde{B}$ squared amplitudes in Eqs.~(\ref{mBB}) and~(\ref{mBBt}). Thus,
the $a\rightarrow b(BB+B\tilde{B})$ differential rate is the sum of the
$a\rightarrow bBB$ and $a\rightarrow bB\tilde{B}$ differential rate with just
the right coefficients to eliminate all the $1/m_{V}$ factors, leaving only
the $1/m$ factors apparent in Eq.~(\ref{mBBBBt}). Phenomenologically though,
the range of shapes of the normalized differential rates simply runs from that
of pure $BB$ to pure $B\tilde{B}$, as one could already obtain with arbitrary
$g_{A}$ and $g_{V}$ but $\tilde{m}=0$, see Fig.~\ref{Fig2}. All one needs to
remember is that effectively, a $B\tilde{B}$ component in the rate can come
either from the direct $a\rightarrow bB\tilde{B}$ coupling, or from the
$\tilde{m}$ mass term.

To close this Section, let us remark that the present analysis could easily be adapted to the pair-creation processes $\psi_{1}\psi_{2}\rightarrow XX$ or $\Phi_{1}\Phi_{2}\rightarrow XX$ for $X = \phi, A,B,C$. As a function of the center of mass energy, their cross section would again behave differently because of Eqs.~(\ref{mAAFF}) and~(\ref{mBBBBt}).

\section{Conclusion}

In this paper, we have analyzed in details the theoretical frameworks in which
a dark scalar $\phi$ is represented by a rank-three antisymmetric field $C_{\mu\nu\rho}$, and the dark photon $A_{\mu}$ by a rank-two antisymmetric field $B_{\mu\nu}$. Though well-known dualities relate $\phi$ to $C_{\mu\nu\rho}$, and $A_{\mu}$ to $B_{\mu\nu}$, those are not equivalent
phenomenologically once interactions are turned on. Starting with a more
theoretical note, our main findings are

\begin{itemize}
\item Minimal bases of effective operators involving $\phi$, $A_{\mu}$,
$B_{\mu\nu}$, $C_{\mu\nu\rho}$, or even the fourth rank tensor $D_{\mu\nu
\rho\sigma}$, singly or in pairs, and SM particles have been constructed,
including operators with up to two extra derivatives. Those for $\phi$ and
$A_{\mu}$ were known, but the others have never been systematically derived.
Though it is not phenomenologically useful to go to such high orders, several
interesting features emerged. First, surprisingly, these bases end up
involving relatively few operators. Even though the number of ways to contract
the Lorentz indices quickly becomes huge, the antisymmetry of these states
permits one to relate many operators, sometimes through quite intricate
reductions. As a result, one peculiar feature is that neither $B$ nor $C$ can
have dimension-six couplings with fermions only. A second feature is that the
$C$ basis is actually related to that for $A$. Provided all the effective
operators involving the Lorenz conditions are kept, their operators are in
one-to-one correspondence. For $B$, under the same condition, the operator
basis has to be self-dual, i.e., operators at each order must only get
reorganized under $B_{\mu\nu}\rightarrow\epsilon_{\mu\nu\rho\sigma}%
B^{\rho\sigma}$.

\item We have systematically studied the impact of adding to the mass term
$m^{2}B_{\mu\nu}B^{\mu\nu}$ a pseudoscalar component $\tilde{m}%
^{2}B_{\mu\nu}\tilde{B}^{\mu\nu}$. By treating this term nonperturbatively,
we derived the modified physical mass of the $B$ field, $m_{B}^{2}%
=(m^{4}+\tilde{m}^{4})/m^{2}$, its full propagator, and its polarization sum.
Also, these results were checked by rederiving them independently via an
extension of the duality formalism, see Eqs.~(\ref{EoMmmt}) and~(\ref{EoMmmt2}). Phenomenologically, this mass
term not only alters the physical mass, but it also impacts observables, as we
showed explicitly for a generic pair production process, see Fig.~\ref{Fig2}.
A peculiar feature though is that the polarization sum keeps its pole in
$1/m^{2}$ because even when $\tilde{m}\neq0$, the $B$ kinetic term remains
noninvertible when $m\rightarrow0$. This means that even a not-so-light $B$
field could see its rate strongly enhanced if $m$ becomes very small.

\item Even if strictly speaking, dualities do not hold for interacting
theories, we found that they remain as a powerful tool at the phenomenological
level, where they show up as sum rules for polarization vectors and tensors,
see Eq.~(\ref{dualmassC}) and (\ref{BdualA}). As long as the dark states stay external, these sum rules are universally valid thanks to their algebraic nature. They can even accommodate for the pseudoscalar mass term for the $B$ field, see Eq.~(\ref{mtildePol}).

\item All these higher-rank tensor fields have an Abelian gauge symmetry when
massless. As such, they appear particularly suited to Stueckelberg
representations, and this opens the way to phenomenological interpretations on
the basis of the equivalence theorem, even though we did not try to implement
true Higgs mechanisms for these fields. Though these constructions have
appeared before, it seems a systematic study along that line has not. Yet, we
found that introducing $R_{\xi}$ gauge-fixing for a generic antisymmetric
tensor field sheds new light on their physical degrees of freedom, and their
polarization sums.
\end{itemize}

Phenomenologically, once allowing for higher rank fields, we actually
identified two different points of view to proceed, depending on the assumed
status of dualities. Specifically, these effective frameworks are

\begin{itemize}
\item The parent Lagrangian formalism can be promoted to a construction
mechanism to derive specific effective theories. In this case, the equivalence
holds between the $\phi-C$ or $A-B$ embeddings but in a very special way.
First, starting with fully generic effective interactions for the higher rank
fields, only specific effective interactions can be present in the lower-rank
forms. In particular, the shift symmetry is built in for the dark scalar, while
the dark vector necessarily couples in a gauge-invariant way. Second, the
parent Lagrangian formalism somewhat decouples the scales of the effective
operators in each picture. Technically, it introduces three scales $m_{1}$,
$m_{2}$ and $m_{3}$, with the operators with higher-rank (lower-rank) fields
tuned by $m_{1}/m_{2}$ ($1/m_{3}$), respectively, while its mass remains set
at $m_{dark}=m_{1}m_{2}/m_{3}$. For the scalar theory, these two features
permit one to circumvent stability problems from mixing of $\phi$ with the Higgs
field. For the dark photon, it provides a natural mechanism to suppress the
renormalizable and gauge-invariant kinetic mixing. Another consequence is the
need for contact terms to relate these effective theories. These take the form
of effective interactions among SM fields only, scaling as $1/m_{3}^{2}$,
which can be looked for in totally different settings like, e.g., at colliders
or in low-energy observables.

\item Instead of the parent Lagrangian approach, one can also start directly
with an effective theory written in terms of $B_{\mu\nu}$ for the dark photon,
or $C_{\mu\nu\rho}$ for the dark scalar, without any recourse to duality
arguments. The leading effective operators are then simply different than when
using the $A_{\mu}$ or $\phi$ picture, with some advantages and disadvantages.
Specifically, whatever the picture, some effective interactions need to be
tiny to pass experimental bounds. This in general requires some additional
assumptions on the unknown UV physics. In this context, imposing a $B$ gauge
invariance on the effective operators (but allowing for a mass term) proves
particularly powerful. In this approach, even if duality is no longer called
in to somehow translate the effective theory back into the usual $A$ or $\phi$
picture, it is still present in the form of the polarization sum
relations. Those imply in particular that processes involving $B$ and $C$
must have different momentum dependences compared to that involving $A$ and
$\phi$. Since an external $B$ field behaves essentially as a vector field
strength, and $C$ essentially as a derivatively coupled scalar, extra
enhancements in the form of energy scale over dark mass are expected. To illustrate this effect in a simple setting, we compared the generic pair production processes for each scenario, where this enhancement shows up in the differential rates in terms of the dark state invariant mass, see Fig.~\ref{Fig1}.
\end{itemize}
The stage is set for further theoretical and phenomenological studies. For the former, some questions remain on possible renormalizable UV completion for the higher form effective theories. In particular, the presence of some gauge invariance, and then the mechanism at the origin of the mass term(s), would need to be elucidated. For the latter, all the tools are ready for detailed analyses. First, in a dark matter context and low-energy searches, it would be very interesting to see how interpreting a dark scalar as a $C$ field, or a dark photon as a $B$ field, would alter the available parameter space. Second, more generally, $B$ and $C$ fields could show up in unexpected places. For instance, throughout this work we took the point of view that the $B$ or $C$ field should be light, but nothing in the formalism prevents them from arising at the TeV scale or above. In that case, they could be looked for at colliders, where their different kinematical behaviors would provide rather unique signatures.

\subsubsection*{Acknowledgments:}

This research is supported by the IN2P3 Master project \textquotedblleft
Axions from Particle Physics to Cosmology\textquotedblright, and from the
French National Research Agency (ANR) in the framework of the
\textquotedblleft GrAHal\textquotedblright project no. ANR-22-CE31-0025.

\appendix

\section{Differential forms}
\label{sec:appendix}

The language of differential forms is particularly well suited to the
formulation of gauge theories. The present Appendix collects in the next
Section all the relevant definitions. It is intended more as a repository of
useful relations and conventions rather than a pedagogical introduction. For
that, we refer e.g. to Ref.~\cite{Baez:1995sj} or any other book on differential geometry. In the following section, the formalism is applied to
the vector field, showing explicitly how the usual equations are recovered.

\subsection*{Definitions and conventions}

Differential 1-forms are in essence gradients, in the sense that to
$\partial f/\partial x=f^{\prime}(x)$, one can associate the 1-form
$\omega(x)=f^{\prime}(x)dx$ such that integrating $\omega$ gives back the
function $f$. Including the integration measure makes this definition
independent of the coordinate system. 0-forms are just functions that can
be evaluated (\textquotedblleft integrated\textquotedblright) at any given
point within the range of $f$.

To generalize this definition to higher dimensions while taking care of
possible orientations of the integration region, the basic operation is the
\textit{wedge product}. For a $p$-form $\omega$ and a $q$-form $\eta$, this
product is the $p+q$-form constructed from the antisymmetric product:
\begin{equation}
(\omega\wedge\eta)_{\mu_{1}...\mu_{p+q}}=\frac{(p+q)!}{p!q!}\omega_{\lbrack
\mu_{1}...\mu_{p}}\eta_{\mu_{p+1}...\mu_{p+q}]}\ ,
\end{equation}
where $[...]$ denotes the \textit{normalized} antisymmetrization, hence the
$(p+q)!$ prefactor. From the definition, one can show the graded commutativity
$\omega\wedge\eta=(-1)^{pq}\eta\wedge\omega$ and distributivity over addition
$\omega\wedge(\eta+\varphi)=\omega\wedge\eta+\omega\wedge\varphi$.
Recursively, this permits one to define higher forms out of 1-forms, with, e.g.,
the wedge product of $k$ 1-forms being a $k$-form. A natural basis for
$k$-forms is then $dx^{\mu_{1}}\wedge...\wedge dx^{\mu_{k}}$, in which a
generic $k$-form is%
\begin{equation}
\omega=\frac{1}{k!}\omega_{\mu_{1}...\mu_{k}}dx^{\mu_{1}}\wedge...\wedge
dx^{\mu_{k}}\ .
\end{equation}
In $n$-dimensional spacetime, a $k$-form $\omega$ has $C_{k}^{n}=n!/k!(n-k)!$
independent components. A $k$-form can be integrated over a $k$-dimensional
space, while the $n$-form in $n$ dimensions has one component since it is
necessarily proportional to the \textit{volume form}:
\begin{equation}
\operatorname*{vol}=\frac{1}{n!}\epsilon_{\mu_{1}...\mu_{n}}dx^{\mu_{1}}%
\wedge...\wedge dx^{\mu_{n}}\equiv d^{n}x\ .
\end{equation}
We will work in flat space throughout, so the tensor $\epsilon_{\mu_{1}%
...\mu_{n}}=\varepsilon_{\mu_{1}...\mu_{n}}\ $with $\varepsilon_{\mu_{1}%
...\mu_{n}}$ the usual constant algebraic antisymmetric symbol for which we
set $\varepsilon_{0,...,n-1}=+1$, while $\epsilon^{\mu_{1}...\mu_{n}%
}=-\varepsilon^{\mu_{1}...\mu_{n}}$ given our metric signature $(+1,-1,-1,-1)$%
. This implies that
\begin{equation}
\epsilon_{\rho_{1}...\rho_{k}\mu_{1}...\mu_{n-k}}\epsilon^{\rho_{1}...\rho
_{k}\nu_{1}...\nu_{n-k}}=-k!\delta_{\mu_{1}...\mu_{n-k}}^{\nu_{1}...\nu_{n-k}%
}\ ,\label{AppId}%
\end{equation}
with the \textit{generalized Kronecker symbol} is defined as%
\begin{equation}
\delta_{\mu_{1}...\mu_{k}}^{\nu_{1}...\nu_{k}}=\sum_{\sigma}%
\operatorname{sign}(\sigma)\delta_{\mu_{1}}^{\sigma(\nu_{1})}...\delta
_{\mu_{k}}^{\sigma(\nu_{k})}\ ,
\end{equation}
where summation is over all the $\sigma$ permutations of the $k$ indices
$\nu_{1}$ to $\nu_{k}$. Also, $\delta_{\mu_{1}...\mu_{k}}^{\nu_{1}...\nu_{k}%
}\omega_{\nu_{1}...\nu_{k}}=k!\omega_{\mu_{1}...\mu_{k}}$ and $\delta_{\mu
_{1}...\mu_{k}}^{\rho_{1}...\rho_{k}}\delta_{\rho_{1}...\rho_{k}}^{\nu
_{1}...\nu_{k}}=k!\delta_{\mu_{1}...\mu_{k}}^{\nu_{1}...\nu_{k}}$. With this,
we can also express the volume integration measure as $dx^{\mu_{1}}%
\wedge...\wedge dx^{\mu_{n}}=\varepsilon^{\mu_{1}...\mu_{n}}d^{n}x$.

The \textit{Hodge dual} of a $k$-form is defined as%
\begin{equation}
\omega_{\nu_{1}...\nu_{k}}\rightarrow(\star\omega)_{\mu_{1}...\mu_{n-k}}%
=\frac{1}{k!}\epsilon_{\nu_{1}...\nu_{k}\mu_{1}...\mu_{n-k}}\omega^{\nu
_{1}...\nu_{k}}\ ,
\end{equation}
satisfying $\star(\star\omega)=(-1)^{k(n-k)-1}\omega$. With this, one can
introduce the \textit{inner product} $\langle\omega,\eta\rangle$ of two $k$-forms $\omega$ and $\eta$, as the volume form%
\begin{equation}
\langle\omega,\eta\rangle\equiv\omega\wedge\star\eta=\frac{1}{k!}\omega
_{\mu_{1}...\mu_{k}}\eta^{\mu_{1}...\mu_{k}}d^{n}x\ .
\end{equation}
By symmetry, $\langle\omega,\eta\rangle=\langle\eta,\omega\rangle$, and
$\langle\omega,\eta\rangle=-\langle\star\eta,\star\omega\rangle$ for a
negative-signature metric.

Another way to recursively construct higher forms is by differentiation. The
\textit{exterior derivative} of a $k$-form $\omega$ is the $k+1$-form obtained
after properly antisymmetrizing the partial derivative%
\begin{equation}
d\omega=\frac{1}{k!}\partial_{\mu_{1}}\omega_{\mu_{2}...\mu_{k+1}}dx^{\mu_{1}%
}\wedge dx^{\mu_{2}}\wedge...\wedge dx^{\mu_{k+1}}\ ,
\end{equation}
or $(d\omega)_{\mu_{1}...\mu_{k}\mu_{k+1}}=(k+1)\partial_{\lbrack\mu_{1}%
}\omega_{\mu_{2}...\mu_{k+1}]}$, where $[...]$ is normalized. The important
properties of $d$ are
\begin{equation}
d(\omega+\eta)=d\omega+d\eta\ ,\ \ d(\omega\wedge\eta)=d\omega\wedge
\eta+(-1)^{p}\omega\wedge d\eta\ \ ,\ \ \ d^{2}\omega=0\ .
\end{equation}
A form such that $d\omega=0$ is said to be a \textit{closed form}, while one
such that $\omega=d\eta$ is said to be an \textit{exact form}. All exact forms
are closed, but the converse is true only locally. In terms of $k$-forms,
\textit{Stokes theorem }takes the form%
\begin{equation}
\int_{M}d\omega=\int_{\partial M}\omega\ ,
\end{equation}
for some $k+1$ dimensional space with $k$-dimensional boundary $\partial M$.

To combine exterior derivative with the Hodge dual, one defines the
\textit{codifferential} of a $k$-form as the $k-1$-form obtained via%
\begin{equation}
\delta=(-1)^{n(k+1)}\star d\star\ \rightarrow(\delta\omega)_{\mu_{1}%
...\mu_{k-1}}=-\partial^{\alpha}\omega_{\alpha\mu_{1}...\mu_{k-1}}\ .
\label{deltaDiv}%
\end{equation}
It has fewer properties than $d$ but still $\delta^{2}\omega=0$. The
codifferential is the adjoint of the exterior derivative, since from the
definition we have $\star\delta=(-1)^{k}d\star$ and $\delta\star
=(-1)^{k+1}\star d$ when acting on a $k$ form. Explicitly, if $\omega$ is a
$k-1$-form and $\eta$ a $k$-form such that $\omega\wedge\star\eta$ vanishes on
the integration volume boundary, then
\begin{equation}
\int d(\omega\wedge\star\eta)=0=\int d\omega\wedge\star\eta-\int\omega
\wedge\star\delta\eta\ .
\end{equation}

The exterior derivative and codifferential can be combined as $\delta
d=(-1)^{n(k+2)}\star d\star d$ and $d\delta=(-1)^{n(k+1)}d\star d\star$, both
producing $k$-forms when acting on a $k$-form. Of special interest is their
sum, called the \textit{Laplace-Beltrami operator }$\Delta=(\delta
+d)^{2}=\delta d+d\delta$. It is positive-definite and such that $\Delta
\omega=0$ is attained for $d\omega=\delta\omega=0$, in which case $\omega$ is
said to be a \textit{harmonic form}. Notice finally that $\Delta$ commutes
with the Hodge dual, $\star\Delta=\Delta\star$.

\subsection*{Application to Maxwell and Proca theories}

To see all the definitions in action in a simple setting, it may be useful to
rederive known results for vector fields. In that case, the field $A=A_{\mu
}dx^{\mu}$ and the current $j=J_{\mu}dx^{\mu}$ are 1-forms. The field
strength is the 2-form defined as $F=dA$:%
\begin{equation}
F=\frac{1}{2!}F_{\mu\nu}dx^{\mu}\wedge dx^{\nu}=\partial_{\mu}A_{\nu}dx^{\mu
}\wedge dx^{\nu}\ .
\end{equation}
The Hodge dual $\star F$ is then%
\begin{equation}
\star F=\frac{1}{2}\left(  \frac{1}{2}\epsilon_{\alpha\beta\mu\nu}%
F^{\alpha\beta}\right)  dx^{\mu}\wedge dx^{\nu}\ \ ,
\end{equation}
and is often denoted $\star F=\tilde{F}$. The homogeneous Maxwell's equations
immediately follow from $dF=d^{2}A=0$ since $d^{2}=0$. Equivalently, $\star
dF=0$ translates into the Bianchi identity $\epsilon^{\beta\mu\nu\rho}%
\partial_{\mu}F_{\nu\rho}=0$. The inhomogeneous equations correspond to the EoM
derived from the Maxwell action,%
\begin{equation}
S_{\text{\textrm{EM}}}=\int-\frac{1}{2}F\wedge\star F-A\wedge\star
j=\int\left(  -\frac{1}{4}F_{\mu\nu}F^{\mu\nu}-A_{\mu}J^{\mu}\right)
d^{n}x\ .
\end{equation}
Notice that $F\wedge\star F$ is quadratic in $F$ since the wedge product is
symmetric. The action must be stationary against small variations
$A\rightarrow A+\delta_{A}$, which to linear order imposes the vanishing of%
\begin{equation}
0=\delta S_{\text{\textrm{EM}}}=\int-d\delta_{A}\wedge\star F-\delta_{A}%
\wedge\star j=\int\delta_{A}\wedge(-d\star F-\star j)\ .
\end{equation}
The total derivative is discarded since $F$ is assumed to vanish at infinity,
and $d(\delta_{A}\wedge\star F)=d\delta_{A}\wedge F-\delta_{A}\wedge d\star
F=0$ since $\delta_{A}$ is a 1-form. Thus, the equation of motion is%
\begin{equation}
d\star F+\star j=0\Leftrightarrow\delta F+j=0\Leftrightarrow(\partial^{\mu
}F_{\mu\nu}-J_{\nu})dx^{\nu}=0\ ,
\end{equation}
where $\star(\star j)=j$ while $\delta F=-\partial^{\mu}F_{\mu\nu}dx^{\nu}$
from Eq.~(\ref{deltaDiv}). Acting with $d$ on the EoM also implies $\delta
j=0$ since $d^{2}\star F=0$ automatically, i.e., the current must be conserved
$\partial^{\mu}J_{\mu}=0$. With this, the action $S_{\text{\textrm{EM}}}$ is
invariant under the gauge transformations $A\rightarrow A+d\Lambda$ for
$\Lambda$ a 0-form. Indeed, $F=dA$ is automatically invariant, while the
source term varies as%
\begin{equation}
\delta S_{\text{\textrm{EM}}}=\int-d\Lambda\wedge\star j=\int\Lambda\wedge
d\star j=0\ ,
\end{equation}
upon integrating by part and discarding the surface term, over which $j$ is
supposed to vanish. In terms of gauge fields, the EoM takes the form%
\begin{equation}
(\Delta-d\delta)A=-j\rightarrow(\square g^{\mu\nu}-\partial^{\mu}\partial
^{\nu})A_{\nu}=0\ .
\end{equation}
Here, it is customary to enforce the Lorenz condition to fix the gauge,
$\delta A=0$, that is, $\partial^{\mu}A_{\mu}=0$, so that the EoM collapses to
$\Delta A=0$. Notice that this leaves a residual gauge freedom, $A\rightarrow
A+d\Lambda^{\prime}$ with $\Lambda^{\prime}$ such that $\delta d\Lambda
^{\prime}=0$. This can again be written as $(\Delta-d\delta)\Lambda^{\prime
}=0$, but since $\Lambda^{\prime}$ is a 0-form, we immediately get
$\Delta\Lambda^{\prime}=0$ and the residual gauge freedom is harmonic.

For the Proca Lagrangian, $F=dA$ and $dF=0$ still hold, but the equation of
motion is modified by the presence of the mass term,%
\begin{equation}
S_{\text{\textrm{Proca}}}=S_{\text{\textrm{EM}}}+\int\frac{1}{2}m^{2}%
A\wedge\star A=S_{\text{\textrm{EM}}}+\int\frac{m^{2}}{2}A_{\mu}A^{\mu}%
d^{4}x\ \ ,
\end{equation}
from which the EoM $-\star d\star F+m^{2}~A=j$ is derived. In this case, if
the current is conserved, $m^{2}~d\star A=d\star j$, and the Lorenz condition
$\delta A=0$ emerges automatically. In terms of field, the EoM is then the
usual Klein-Gordon equation, $(-\Delta+m^{2})A=j$.

To close this Appendix, let us stress that once written in differential form,
all the equations of the present section remain essentially valid for higher
form fields. The definition of the fields and field strengths, the action,
EoM, gauge freedom, or mass term are, up to trivial signs, identical.

\bibliographystyle{BiblioStyle}
\bibliography{references}

\end{document}